\documentclass[useAMS,usenatbib]{mn2e}

\usepackage{tabularx}
\usepackage{graphicx}
\usepackage{txfonts}
\usepackage{longtable}
\usepackage{hhline}\usepackage{arydshln}
\usepackage{multirow}
\usepackage{lscape}
\usepackage{array}

\newcommand{\hersc}{{\it Herschel}}

\newcommand{\spitz}{{\it  Spitzer}}

\newcommand{\aap}{A\&A}

\newcommand{\apjl}{ApJ}
\newcommand{\apjs}{ApJS}

\newcommand{\lsun}{$L_\odot$}
\newcommand{\msun}{$M_\odot$}

\newcommand{\mic}{$\mu$m}

\newlength{\pointwidth}
\begin{document}

 \title[ Calibration of L$_{TIR}$ from Herschel bands]{Calibration of the total infrared luminosity of nearby galaxies from Spitzer and Herschel bands. }
  
  \author[Galametz et al.]
{\parbox{\textwidth}{M. Galametz$^{1}$\thanks{e-mail: mgalamet@ast.cam.ac.uk}, 
R. C. Kennicutt$^{1}$, 
D. Calzetti$^{2}$,
G. Aniano$^{3,4}$,
B. T. Draine$^{3}$, 
M. Boquien$^{5}$,
B. Brandl$^{6}$,
K. V. Croxall$^{7,8}$,
D. A. Dale$^{9}$,
C. W. Engelbracht$^{10,11}$,
K. D. Gordon$^{12}$,
B. Groves$^{13}$,
C.-N. Hao$^{14}$,
G. Helou$^{15}$,
J. L. Hinz$^{10}$,
L. K. Hunt$^{16}$,
B. D. Johnson$^{17}$,
Y. Li$^{2}$,
E. Murphy$^{18}$,
H. Roussel$^{17}$,
K. Sandstrom$^{13}$,
R. A. Skibba$^{10,19}$ and
F. S. Tabatabaei$^{13}$
}\vspace{0.5cm}\\
\parbox{\textwidth}{
$^{1}$Institute of Astronomy, University of Cambridge, Madingley Road, Cambridge CB3 0HA, UK\\
$^{2}$ Department of Astronomy, University of Massachusetts, Amherst, MA 01003, USA\\
$^{3}$Department of Astrophysical Sciences, Princeton University, Princeton, NJ 08544, USA\\
$^{4}$Institut d'Astrophysique Spatiale, b\^{a}timent 121, Universit\'{e} Paris-Sud 11, CNRS UMR~8617, 91405 Orsay, France \\
$^{5}$Laboratoire d'Astrophysique de Marseille, Universit\'e Aix-Marseille, CNRS UMR~7326, 13388 Marseille Cedex 13, France \\
$^{6}$Leiden Observatory, Leiden University, P.O. Box 9513, 2300 RA Leiden, The Netherlands\\
$^{7}$Department of Physics and Astronomy, Mail Drop 111, University of Toledo, 2801 West Bancroft Street, Toledo, OH 43606, USA\\
$^{8}$Department of Astronomy, The Ohio State University, 140 West 18th Avenue, Columbus, OH 43210, USA\\
$^{9}$Department of Physics $\&$ Astronomy, University	of Wyoming, Laramie, WY 82071, USA\\
$^{10}$Steward Observatory, University of Arizona, Tucson, AZ 85721, USA\\
$^{11}$Raytheon Company, 1151 East Hermans Road, Tucson, AZ 85756, USA\\
$^{12}$Space Telescope Science Institute, 3700 San Martin Drive, Baltimore, MD 21218, USA\\
$^{13}$Max-Planck-Institut f\"ur Astronomie, K\"onigstuhl 17, D-69117 Heidelberg, Germany\\
$^{14}$Tianjin Astrophysics Center, Tianjin Normal University, Tianjin 300387, China\\
$^{15}$NASA Herschel Science Center, IPAC, California Institute of Technology, Pasadena, CA 91125, USA\\
$^{16}$INAF - Osservatorio Astrofisico di Arcetri, Largo E. Fermi 5, 50125 Firenze, Italy\\
$^{17}$Institut d'Astrophysique de Paris, Universit\'{e} Pierre et Marie Curie, CNRS UMR~7095, 75014 Paris, France  \\
$^{18}$Observatories of the Carnegie Institution for Science, Pasadena, CA 91101, USA \\
$^{19}$Center for Astrophysics and Space Sciences, Department of Physics, University of California, 9500 Gilman Dr., San Diego, CA 92093, USA}
}

\maketitle{}


 \begin{abstract}

When combined with infrared observations with the \spitz\ telescope (3 to 160 \mic), the \hersc\ {\it Space Observatory} now fully samples the thermal dust emission up to 500 \mic\ and enables us to better estimate the total infrared-submm energy budget (L$_{TIR}$) of nearby galaxies. We present new empirical calibrations to estimate resolved and integrated total infrared luminosities from \spitz\ and \hersc\ bands used as monochromatic or combined tracers. We base our calibrations on resolved elements of nearby galaxies (3 to 30 Mpc) observed with \hersc. We perform a resolved SED modelling of these objects using the \citet{Draine_Li_2007} dust models and investigate the influence of the addition of SPIRE measurements in the estimation of L$_{TIR}$. We find that using data up to 250 \mic\ leads to local L$_{TIR}$ values consistent  with those obtained with a complete coverage (up to 500 \mic) within $\pm$10$\%$ for most of our resolved elements. We then study the distribution of energy in the resolved SEDs of our galaxies. The bulk of energy (30-50$\%$) is contained in the [70-160 \mic] band. The [24-70 \mic] fraction decreases with increasing metallicity. The [160-1100 \mic] submillimeter band can account for up to 25$\%$ of the L$_{TIR}$ in metal-rich galaxies.
We investigate the correlation between TIR surface brightnesses/luminosities and monochromatic \spitz\ and \hersc\ surface brightnesses/luminosities. The three PACS bands can be used as reliable monochromatic estimators of the L$_{TIR}$, the 100 \mic\ band being the most reliable monochromatic tracer. There is also a strong correlation between the SPIRE 250 \mic\ and L$_{TIR}$, although with more scatter than for the PACS relations. We also study the ability of our monochromatic relations to reproduce integrated L$_{TIR}$ of nearby galaxies as well as L$_{TIR}$ of z$\sim$1-3 sources. Finally, we provide calibration coefficients that can be used to derive TIR surface brightnesses/luminosities from a combination of \spitz\ and \hersc\ surface brightnesses/fluxes and analyse the associated uncertainties.
 \end{abstract}

\begin{keywords}
galaxies: ISM --
     		ISM: dust --
		submillimeter: galaxies
\end{keywords}

 \maketitle


\section{Introduction} 

Interstellar dust obscures our view of the star formation sites in galaxies. Indeed, 30 to 50$\%$ of the starlight emission is thermally reprocessed by dust, and re-emitted at infrared (IR) wavelength \citep{Draine2003,Tielens2005}. This wavelength regime enables us to directly investigate the dust physics and indirectly probe the star formation activity obscured by dust within galaxies and is thus crucial to understand how galaxies evolve and recycle their interstellar material. The total bolometric IR emission L$_{TIR}$ constitutes the emission of all the dust-enshrouded stellar populations (but can also include emission from Active Galaxy Nucleus or AGN) and is one of most reliable tracers of the star formation obscured by dust. Several studies have thus derived calibrations of the star formation rate (SFR) based on L$_{TIR}$ \citep{Kennicutt1998,Perez-Gonzalez_2006,Kennicutt2009,Kennicutt_Evans_2012}.

L$_{TIR}$ can be estimated by combining multi-wavelength observations sampling the thermal dust emission from mid-IR to submillimeter wavelengths and integrating the emission directly or using realistic dust models to interpolate the data.
Unfortunately, many galaxies do not benefit from a complete sampling of their Spectral Energy Distribution (SEDs), which prevents the modelling of their SEDs and thus a correct estimate of their L$_{TIR}$. 
Previous works have thus provided calibrations of the L$_{TIR}$ using monochromatic IR wave bands or a combination of IR wave bands. For instance, \citet{Sanders_Mirabel1996} or \citet{Sanders2003} provided a relation to derive the L$_{TIR}$ of luminous IR galaxies using the 4 {\it IRAS} (Infrared Astronomical Satellite) filters at 12, 25, 60 and 100 \mic. \citet{Dale_Helou_2002} updated this relation using a combination of the three \spitz/MIPS wavelengths (24, 70 and 160 \mic), matching their modelled L$_{TIR}$ with very good accuracy. More recently, \citet{Boquien2010} used the \citet{Dale_Helou_2002} relation to estimate the L$_{TIR}$ from other \spitz\ bands, the 8 and 24 \mic\ bands in particular.

The good resolution of the two IR-submillimeter instruments PACS and SPIRE onboard the \hersc\ {\it Space Observatory} opens a new window on how to quantify L$_{TIR}$ at local scale. The resolutions of the SPIRE instrument match that of the MIPS instrument of the \spitz\ {\it Space telescope}, the predecessor of \hersc. Indeed, the resolution of SPIRE at 250 \mic\ ($\sim$18\arcsec) is similar to that of \spitz/MIPS 70 \mic\ and the resolution at 500 \mic\ ($\sim$36\arcsec) is similar to that of MIPS 160 \mic. Furthermore, \hersc\ data enable a more complete coverage of the peak of the thermal dust emission and of the submm slope of nearby galaxies up to 500 \mic, allowing a refinement of our measurements of the L$_{TIR}$. 

Combining \spitz\ and \hersc\ bands, \citet{Boquien2011} derived resolved estimators of the total infrared brightness in the galaxy M33. 
In this study, we aim to similarly model resolved L$_{TIR}$ for a wider sample of galaxies, investigate the relation between L$_{TIR}$ and different \hersc\ bands, study how those relations evolve with the galaxy characteristics as well as provide recipes to obtain reliable L$_{TIR}$ predictions from a large choice of single or combined wavelengths. As previously mentioned, many authors have studied relations between IR monochromatic fluxes and L$_{TIR}$ as well. Recipes like those provided by \citet{Sanders_Mirabel1996} and \citet{Dale_Helou_2002} are often used in the literature as a proxy for the derivation of L$_{TIR}$. This estimated L$_{TIR}$ is then used to derive calibrations from monochromatic luminosities \citep[][among others]{Boquien2010,Elbaz2011}. One of the advantages of the approach we follow in this study is that we now have access to the whole coverage of the thermal dust IR emission in our nearby objects with \hersc. Our L$_{TIR}$ will be directly modelled using the IR observations and a realistic dust SED model, which limits uncertainties and biases linked with previous calibrations.

We perform this analysis using the \hersc\ data of $\sim$60 nearby galaxies observed as part of the KINGFISH \citep[Key Insights on Nearby Galaxies: A Far-Infrared Survey with Herschel;][]{Kennicutt2011} programme. The paper is organised as follows. We present the sample, \spitz\ and \hersc\ data in Section 2. As we aim to derive resolved estimators of L$_{TIR}$, we would like to use the highest resolution available. The first step of this study is thus to determine the best compromise between resolution and sufficient constraint on the L$_{TIR}$ estimates (Section 3). We then analyse the distribution of the total infrared energy with wavelength on a local basis in Section 4. We present a calibration of the TIR surface brightnesses/luminosities using single \spitz\ or \hersc\ bands in Section 5 as well as calibrations from a combination of various bands in Section 6. Because the metallicity and the hardness of the radiation field are parameters that strongly affect the far-IR emission and the range in dust temperature from galaxy to galaxy, throughout the paper, we investigate how our relations and their reliability evolve with global or local galaxy properties.


\section{A multi-wavelength mapping}

\subsection{The sample}

We obtain the \hersc\ data (PACS and SPIRE maps) as part of the \hersc\ key programme KINGFISH. This sample provides a unique opportunity to study the relation between Herschel bands and total IR luminosities. The sample comprises 61 galaxies, probing a wide range of galaxy types (from elliptical to irregular galaxies) and various star-formation activities, from active star-forming regions to more quiescent ISM, with global SFRs ranging from 10$^{-3}$ to 7 \msun\ yr$^{-1}$ (\citet{Howell2010} even estimate a SFR of 23 \msun~yr$^{-1}$ for the luminous IR galaxy NGC~2146). It also includes galaxies hosting low-luminosity AGN. The KINGFISH galaxies are located between 3 and 31 Mpc, leading to ISM resolution elements of 0.2 to 2.6 kpc at the resolution of SPIRE 250 \mic\ (FWHM of the PSF: 18\arcsec), the resolution at which we work in the following study (see Section 3 for justification). The KINGFISH galaxies also probe various metallicities. We use the metallicities tabulated by \citet{Kennicutt2011} who provide two metallicities per galaxy, one derived from the theoretical calibration of \citet{Kobulnicky2004}, the other from the empirical calibration of \citet{Pilyugin2005}. Here we use the latter; with this calibration, oxygen abundances (defined as 12+log(O/H)) range from 7.54 for the low-metallicity galaxy DDO~154 to 8.9 for the galaxy NGC~3077. We note that metallicity gradients are observed in some of the KINGFISH galaxies \citep{Moustakas2010} but only few gradients are currently well constrained. In this paper, we adopt the same metallicity in each resolution element of a given galaxy, equal to that determined globally for the galaxy using the \citet{Pilyugin2005} calibration. 

Previous and on-going studies are also analysing global and local SED models of galaxies of the KINGFISH sample using the \hersc\ data. \citet{Dale2012} present the \hersc\ far-IR and sub-millimeter photometry of the KINGFISH survey as well as integrated SED models of these objects from which total dust masses are in particular derived. \citet{Skibba2011} also compare the global emission from dust and from stars in the same sample and analyse how the dust-to-stellar flux ratio varies with properties such as morphology, L$_{TIR}$ or metallicity. Using local modified blackbody models for a sample of the KINGFISH galaxies, \citet{Galametz2012} investigate the physical properties (temperature, emissivity) of the cold dust phase and associated uncertainties. Finally, \citet{Aniano2012} present a pixel-by-pixel SED modelling and a mapping of the dust and radiation field properties for the two spirals NGC~628 and NGC~6946 using the \citet{Draine_Li_2007} dust models. This local modelling will be extended to the whole KINGFISH sample in Aniano et al. (in prep). \\

\subsection{Herschel maps}

The \hersc/PACS instrument \citep{Poglitsch2010} provides maps with FWHMs of the PSFs of 5\farcs76$\times$5\farcs46, 6\farcs69$\times$6\farcs89 and 12\farcs13$\times$10\farcs65 at 70, 100 and 160 \mic\ respectively for the chosen scan speed (20\arcsec/s). Observations of the KINGFISH galaxies with this instrument were obtained with 15\arcmin\ long cross-scans (perpendicular scans). 
From raw data to Level 1, the processing of PACS data follows the main steps of the recommended standard procedure for steps of pointing association, conversion to physical units or flat-fielding. Glitches are removed using a second-level deglitching method based on a comparison of individual readouts with a reference sky value at the same position that allows us to detect outlier values. We refer to \citet{Kennicutt2011} for more details on the initial processing of the data within the \hersc\ Interactive Processing Environment (HIPE, version 8). 

We use the Scanamorphos technique (version 16.9) to process the data from these Level~1 data, correct for 1/f noise and project the pixel timelines in the sky in order to build the final maps. Scanamorphos in particular subtracts the brightness drifts caused by the low-frequency noise using the redundancy built in the observations \citep{Roussel2012}. The final pixel sizes of our PACS data are 1\farcs4, 1\farcs7 and 2\farcs85 at 70, 100 and 160 \mic\ respectively. The PACS calibration uncertainties are $\sim$5 $\%$\footnote{http://herschel.esac.esa.int/Docs/PACS/html/pacs$\_$om.html}.

The \hersc/SPIRE instrument \citep{Griffin2010} provides maps with FWHMs of the PSFs 18\farcs3$\times$17\arcsec, 24\farcs7$\times$23\farcs2 and 37\arcsec$\times$33\farcs4 at 250, 350 and 500 \mic\ respectively. Observations were obtained in scan mode. The data reduction was performed from raw data with the HIPE environment. We refer to the KINGFISH Data Products Delivery User's Guide\footnote{http://herschel.esac.esa.int/UserReducedData.shtml} for details on the data reduction. The SPIRE maps used in this study are built with a nearest-neighbor projection on sky and averaging of the time ordered data. The final pixel sizes of our SPIRE data are 6\arcsec, 10\arcsec\ and 14\arcsec\ at 250, 350 and 500 \mic\ respectively. Calibration uncertainties are estimated to be $\sim$7$\%$ for the three wave bands\footnote{http://herschel.esac.esa.int/Docs/SPIRE/html/spire$\_$om.html}. 

We refer to \citet{Kennicutt2011}, \citet{Engelbracht2010} and \citet{Sandstrom2010} for more details on the KINGFISH sample, the observation strategy and the different steps of the data processing. We do not include the KINGFISH galaxies DDO~154, DDO~165, Holmberg~I and NGC~1404, since they are barely detected with \hersc\ \citep[upper limits in the global flux catalogue of][]{Dale2012}. We also note that PACS observations for the galaxy NGC~584 are contaminated by emission from Jupiter. This galaxy is thus also excluded from the following analysis.

\subsection{Spitzer maps}

Most of the KINGFISH galaxies have been observed with \spitz/IRAC and MIPS as part of the SINGS programme \citep[Spitzer Infrared Nearby Galaxies Survey;][]{Kennicutt2003}. Four galaxies of the sample are drawn from other \spitz\ surveys: IC~342, NGC~5457 (M101), NGC~2146 and NGC~3077. IRAC observes at 3.6, 4.5, 5.8 and 8 \mic\ with PSF FWHMs of 1\farcs7, 1\farcs7, 1\farcs9 and 2\arcsec\ respectively. The IRAC images are reduced using the SINGS Fifth Data Delivery pipeline\footnote{http://data.spitzer.caltech.edu/popular/sings/20070410$\_$enhanced$\_$v1/Documents/ sings$\_$fifth$\_$delivery$\_$v2.pdf}. Maps are multiplied by 0.91, 0.94, 0.66 and 0.74 at 3.6, 4.5, 5.8 and 8 \mic\ respectively to account for extended-source flux calibration\footnote{http://irsa.ipac.caltech.edu/data/SPITZER/docs/irac/iracinstrumenthandbook/}. MIPS observes at 24, 70 and 160 \mic\ with FWHMs of the PSFs of 6\arcsec, 18\arcsec\ and 40\arcsec\ respectively. Because of their lower resolution compared to PACS 160 \mic\ maps, we do not use the MIPS 160 \mic\ maps in the following study. Galaxies of the LVL (Local Volume Legacy) survey are reduced using the LVL pipeline\footnote{http://irsa.ipac.caltech.edu/data/SPITZER/LVL/LVL$\_$DR5$\_$v5.pdf}. Galaxies that are not part of the LVL survey were re-processed using the LVL reduction technique for consistency.

\section{Influence of SPIRE data on the L$_{TIR}$ maps}

We aim to derive L$_{TIR}$ estimators using \spitz\ and \hersc\ bands (as monochromatic or combined tracers). To make the most of the good resolution of \hersc\ and work at the highest resolution available, we first investigate in this section how SPIRE wavelengths influence the determination of the L$_{TIR}$. We thus derive L$_{TIR}$ maps using data between 3.6 and 160 \mic, between 3.6 and 250 \mic\ or between 3.6 and 500 \mic\ to quantify the difference in the global and local L$_{TIR}$ values. 

\subsection{Obtaining the L$_{TIR}$ maps} 

In a first step, we subtract the sky background of \spitz\ and \hersc\ maps using a plane-subtraction technique (see \citealt{Aniano2012} and Aniano et al., in prep).
We then use the convolution kernels provided by \citet{Aniano2011} to convolve the \spitz\ and \hersc\ maps to three different resolutions:

\vspace{4pt}
{\it - PACS 160 \mic\ resolution}~:~We convolve the IRAC 3.6 \mic, 4.5 \mic, 5.8 \mic\ and 8.0 \mic, the MIPS 24 \mic, the PACS 70 \mic\ and PACS 100 \mic\ maps to the resolution of PACS 160 \mic\ (FWHM $\sim$ 12") and regrid them to a common pixel size of 4" (original pixel size of the PACS 160 \mic\ image),

\vspace{4pt}
{\it - SPIRE 250 \mic\ resolution}~:~We convolve the IRAC 3.6 \mic, 4.5 \mic, 5.8 \mic\ and 8.0 \mic, the MIPS 24 \mic, the PACS 70 \mic, 100 \mic\ and 160 \mic\ maps to the resolution of SPIRE 250 \mic\ (FWHM $\sim$ 18") and regrid them to a common pixel size of 6" (original pixel size of the SPIRE 250 \mic\ image),

\vspace{4pt}
{\it - SPIRE 500 \mic\ resolution}~:~We convolve the IRAC 3.6 \mic, 4.5 \mic, 5.8 \mic\ and 8.0 \mic, the MIPS 24 and 70 \mic, the PACS 70 \mic, 100 \mic\ and 160 \mic, the SPIRE 250 and 350 \mic\ maps to the resolution of SPIRE 500 \mic\ (FWHM $\sim$ 36") and regrid them to a common pixel size of 14" (original pixel size of the SPIRE 500 \mic\ image). \\

Using our maps convolved at three different resolutions - and consequently providing three different coverages - we perform local SED fits using the \citet{Draine_Li_2007} (hereafter [DL07]) dust models in order to match the observed fluxes in each resolution element. We refer to \citet{Aniano2012} for a full description of the pre-data treatment (convolution or background subtraction steps, production of the uncertainty maps for each bands etc.) and the resolved SED modelling process (description of the model, assumption on parameters etc.). The SED modeling uses \spitz\ and \hersc\ bandpasses directly and alleviates the need for colour-corrections. In most regions, stellar emission dominates the emission of the two first IRAC bands (3.6 and 4.5 \mic) while dust dominates observations beyond 4.5 \mic\ and at least up to 500 \mic\ \citep[see][ for instance]{Engelbracht2008}. 
To account for thermal dust emission only, we subtract the contribution of stellar emission to the short wavelengths during the modelling process. We approximate the stellar emission at $\lambda$ $>$ 3 \mic\ by scaling a blackbody function, using a representative photospheric temperature of 5000K \citep{Bendo2006,Draine2007}.

L$_{TIR}$ measures the total dust emission and is obtained by integrating the SED in a $\nu$-L$_{\nu}$ space. 
In this paper, we define L$_{TIR}$ as:
\begin{equation}
L_{TIR} = \int_{3~\mu m}^{1100~\mu m}L_{\nu}~d\nu
\label{eq1}
\end{equation}

We thus integrate the SEDs from 3 to 1100 \mic\ to obtain the L$_{TIR}$ in each resolved element. We note that our resolved SED models have a logarithmically-spaced wavelength grid of $\sim$350 values from 3 to 1100 \mic. We use the IDL function {\it INT$\_$TABULATED} (5-point Newton-Cotes formula) to perform the integration. We restrict ourselves to resolved elements that {\it 1)} are located within the elliptical apertures used by \citet{Dale2012} to perform the global photometry, {\it 2)} do not contain contamination from foreground stars or known background galaxies along the line-of-sight and {\it 3)} have a 3-$\sigma$ detection in all the bands used for the modelling. This leads to L$_{TIR}$ maps of our KINGFISH galaxies obtained at three different resolutions (and for three different SED coverages).  \\

In the rest of this paper, we use the nomenclature: 

\vspace{4pt}
{\it - L$_{TIR~P160}$}~:~the L$_{TIR}$ modelled with data constraining the SED from 3.6 to 160 \mic\ (at PACS 160 \mic\ resolution),

\vspace{4pt}
{\it - L$_{TIR~S250}$}~:~the L$_{TIR}$ modelled with data constraining the SED from 3.6 to 250 \mic\ (at SPIRE 250 \mic\ resolution),

\vspace{4pt}
{\it - L$_{TIR~S500}$}~:~the L$_{TIR}$ modelled with data constraining the SED from 3.6 to 500 \mic\ (at SPIRE 500 \mic\ resolution). We consider our L$_{TIR~S500}$ maps as our ``reference'' maps, since resolved SEDs were modelled with the most complete coverage of the thermal dust emission. \\

\begin{table}
\centering \small
\caption{\large Integrated L$_{TIR}$ values for different SED coverages.}
\begin{tabular}{ccccc}
\hline
\hline
			&&&&\\
			&  Integrated &Integrated &&\\
Name 		& L$_{TIR~SINGS}$ & L$_{TIR~S500}$ 	& {\Large $\frac{L_{TIR~P160}}{L_{TIR~S500}}$}	&	{\Large $\frac{L_{TIR~S250}}{L_{TIR~S500}}$}\\
			&(\lsun)&	(\lsun)	& &		\\
			&&&&\\
\hline
			&&&&\\
DDO~053  	&1.24$\times$10$^{7}$&8.83 $^{\pm1.6}$$\times$10$^{6}$		& 1.08	&1.04\\
Holmberg~II 	&7.08$\times$10$^{7}$&6.09 $^{\pm0.5}$$\times$10$^{7}$ 		& 0.98	&0.96\\
IC~342 		&-&1.47 $^{\pm0.1}$$\times$10$^{10}$ 		& 1.06	&1.03\\
IC~2574  		&2.04$\times$10$^{8}$&1.69 $^{\pm0.2}$$\times$10$^{8}$ 		& 0.98	&0.99\\
M81~dwB 	&-&3.54 $^{\pm1.0}$$\times$10$^{6}$ 		& 1.02	&1.00\\
NGC~0337  	&1.25$\times$10$^{10}$&1.06 $^{\pm0.1}$$\times$10$^{10}$ 		& 1.04	&1.06\\
NGC~0628  	&7.78$\times$10$^{9}$&6.89 $^{\pm0.2}$$\times$10$^{9}$ 		& 1.04	&1.02\\
NGC~0855  	&3.91$\times$10$^{8}$&3.55 $^{\pm0.2}$$\times$10$^{8}$ 		& 1.02	&1.01\\
NGC~0925  	&4.18$\times$10$^{9}$&3.68 $^{\pm0.4}$$\times$10$^{9}$ 		& 0.97	&0.98\\
NGC~1097  	&4.56$\times$10$^{10}$&4.16 $^{\pm0.4}$$\times$10$^{10}$ 		& 0.99	&1.01\\
NGC~1266  	&2.63$\times$10$^{10}$&2.45 $^{\pm0.4}$$\times$10$^{10}$ 		& 1.04	&0.98\\
NGC~1291  	&2.76$\times$10$^{9}$&2.70 $^{\pm1.1}$$\times$10$^{9}$ 		& 0.93	&0.97\\
NGC~1316  	&6.83$\times$10$^{9}$&5.95 $^{\pm0.3}$$\times$10$^{9}$ 		& 1.08	&1.02\\
NGC~1377  	&-&1.26 $^{\pm0.1}$$\times$10$^{10}$ 		& 1.02	&1.00\\
NGC~1482 	&5.05$\times$10$^{10}$&4.53 $^{\pm0.9}$$\times$10$^{10}$ 		& 0.97	&1.03\\
NGC~1512  	&3.51$\times$10$^{9}$&3.43 $^{\pm0.3}$$\times$10$^{9}$ 		& 1.05	&1.04\\
NGC~2146  	&-&1.30 $^{\pm0.1}$$\times$10$^{11}$ 		& 0.98	&0.95\\
NGC~2798  	&4.15$\times$10$^{10}$&3.36 $^{\pm0.5}$$\times$10$^{10}$ 		& 0.97	&1.02\\
NGC~2841 	&1.11$\times$10$^{10}$&9.22 $^{\pm1.2}$$\times$10$^{9}$ 		& 1.04	&1.03\\
NGC~2915  	&4.19$\times$10$^{7}$&3.36 $^{\pm0.4}$$\times$10$^{7}$ 		& 1.14	&1.05\\
NGC~2976  	&8.46$\times$10$^{8}$&7.57 $^{\pm0.7}$$\times$10$^{8}$ 		& 1.00	&1.00\\
NGC~3049  	&3.65$\times$10$^{9}$&3.21 $^{\pm0.4}$$\times$10$^{9}$ 		& 1.06	&1.08\\
NGC~3077  	&-&7.31 $^{\pm0.5}$$\times$10$^{8}$ 		& 0.98	&1.02\\
NGC~3184  	&1.07$\times$10$^{10}$&8.45 $^{\pm1.3}$$\times$10$^{9}$ 		& 1.03	&1.02\\
NGC~3190  	&6.61$\times$10$^{9}$&5.96 $^{\pm1.0}$$\times$10$^{9}$ 		& 1.05	&1.02\\
NGC~3198 	&8.83$\times$10$^{9}$&7.27 $^{\pm0.5}$$\times$10$^{9}$ 		& 1.02	&1.02\\
NGC~3265 	&2.83$\times$10$^{9}$&2.49 $^{\pm0.3}$$\times$10$^{9}$ 		& 1.03	&1.02\\
NGC~3351  	&7.81$\times$10$^{9}$&6.86 $^{\pm1.0}$$\times$10$^{9}$ 		& 1.03	&1.03\\
NGC~3521  	&3.24$\times$10$^{10}$&3.15 $^{\pm0.2}$$\times$10$^{10}$ 		& 1.04	&1.03\\
NGC~3621  	&7.90$\times$10$^{9}$&6.80 $^{\pm0.6}$$\times$10$^{9}$ 		& 1.03	&1.01\\
NGC~3627 	&2.66$\times$10$^{10}$&2.53 $^{\pm0.3}$$\times$10$^{10}$ 		& 1.00	&1.01\\
NGC~3773 	&6.69$\times$10$^{8}$&5.04 $^{\pm1.4}$$\times$10$^{8}$ 		& 1.07	&1.06\\
NGC~3938 	&1.87$\times$10$^{10}$&1.57 $^{\pm0.2}$$\times$10$^{10}$ 		& 1.02	&1.02\\
NGC~4236  	&4.68$\times$10$^{8}$&3.85 $^{\pm0.5}$$\times$10$^{8}$ 		& 1.05	&1.04\\
NGC~4254  	&4.21$\times$10$^{10}$&3.57 $^{\pm0.2}$$\times$10$^{10}$ 		& 1.03	&1.02\\
NGC~4321  	&3.38$\times$10$^{10}$&2.88 $^{\pm0.2}$$\times$10$^{10}$ 		& 1.02	&1.01\\
NGC~4536  	&2.12$\times$10$^{10}$&2.06 $^{\pm0.2}$$\times$10$^{10}$ 		& 0.98	&1.05\\
NGC~4559  	&3.01$\times$10$^{9}$&2.51 $^{\pm0.3}$$\times$10$^{9}$ 		& 0.97	&1.00\\
NGC~4569  	&1.41$\times$10$^{10}$&1.37 $^{\pm0.3}$$\times$10$^{10}$ 		& 1.05	&1.04\\
NGC~4579  	&1.20$\times$10$^{10}$&9.22 $^{\pm0.5}$$\times$10$^{9}$ 		& 1.06	&1.06\\
NGC~4594  	&3.42$\times$10$^{9}$&3.03 $^{\pm0.3}$$\times$10$^{9}$ 		& 1.03	&1.03\\
NGC~4625  	&6.19$\times$10$^{8}$&5.20 $^{\pm0.6}$$\times$10$^{8}$ 		& 1.00	&1.01\\
NGC~4631  	&2.54$\times$10$^{10}$&2.04 $^{\pm0.2}$$\times$10$^{10}$ 		& 0.98	&1.00\\
NGC~4725 	&7.33$\times$10$^{9}$&6.22 $^{\pm0.1}$$\times$10$^{9}$ 		& 1.01	&1.03\\
NGC~4736 	&6.00$\times$10$^{9}$&5.68 $^{\pm1.0}$$\times$10$^{9}$ 		& 0.98	&1.01\\
NGC~4826  	&4.05$\times$10$^{9}$&3.60 $^{\pm0.3}$$\times$10$^{9}$ 		& 1.02	&1.01\\
NGC~5055  	&2.10$\times$10$^{10}$&1.75 $^{\pm0.1}$$\times$10$^{10}$ 		& 1.05	&1.03\\
NGC~5398 	&3.68$\times$10$^{8}$&2.98 $^{\pm0.2}$$\times$10$^{8}$ 		& 0.99	&1.07\\
NGC~5408  	&1.90$\times$10$^{8}$&1.63 $^{\pm0.2}$$\times$10$^{8}$ 		& 1.05	&1.05\\
NGC~5457 	&-&1.87 $^{\pm0.2}$$\times$10$^{10}$ 		& 1.03	&1.02\\
NGC~5474 	&5.34$\times$10$^{8}$&4.26 $^{\pm0.5}$$\times$10$^{8}$ 		& 1.02	&1.01\\
NGC~5713  	&3.40$\times$10$^{10}$&3.10 $^{\pm0.2}$$\times$10$^{10}$ 		& 1.02	&1.03\\
NGC~5866  	&4.52$\times$10$^{9}$&4.55 $^{\pm0.4}$$\times$10$^{9}$ 		& 1.02	&0.99\\
NGC~6946 	&3.42$\times$10$^{10}$&3.30 $^{\pm0.2}$$\times$10$^{10}$ 		& 1.06	&1.03\\
NGC~7331  	&4.75$\times$10$^{10}$&4.18 $^{\pm0.2}$$\times$10$^{10}$ 		& 1.05	&1.02\\
NGC~7793 	&2.07$\times$10$^{9}$&1.67 $^{\pm0.1}$$\times$10$^{9}$ 		& 0.98	&0.99\\
			&&&\\
			\hline
\end{tabular}
\label{Global_TIR}
\end{table}

   \begin{figure*}
   \centering
    \includegraphics[height=15cm]{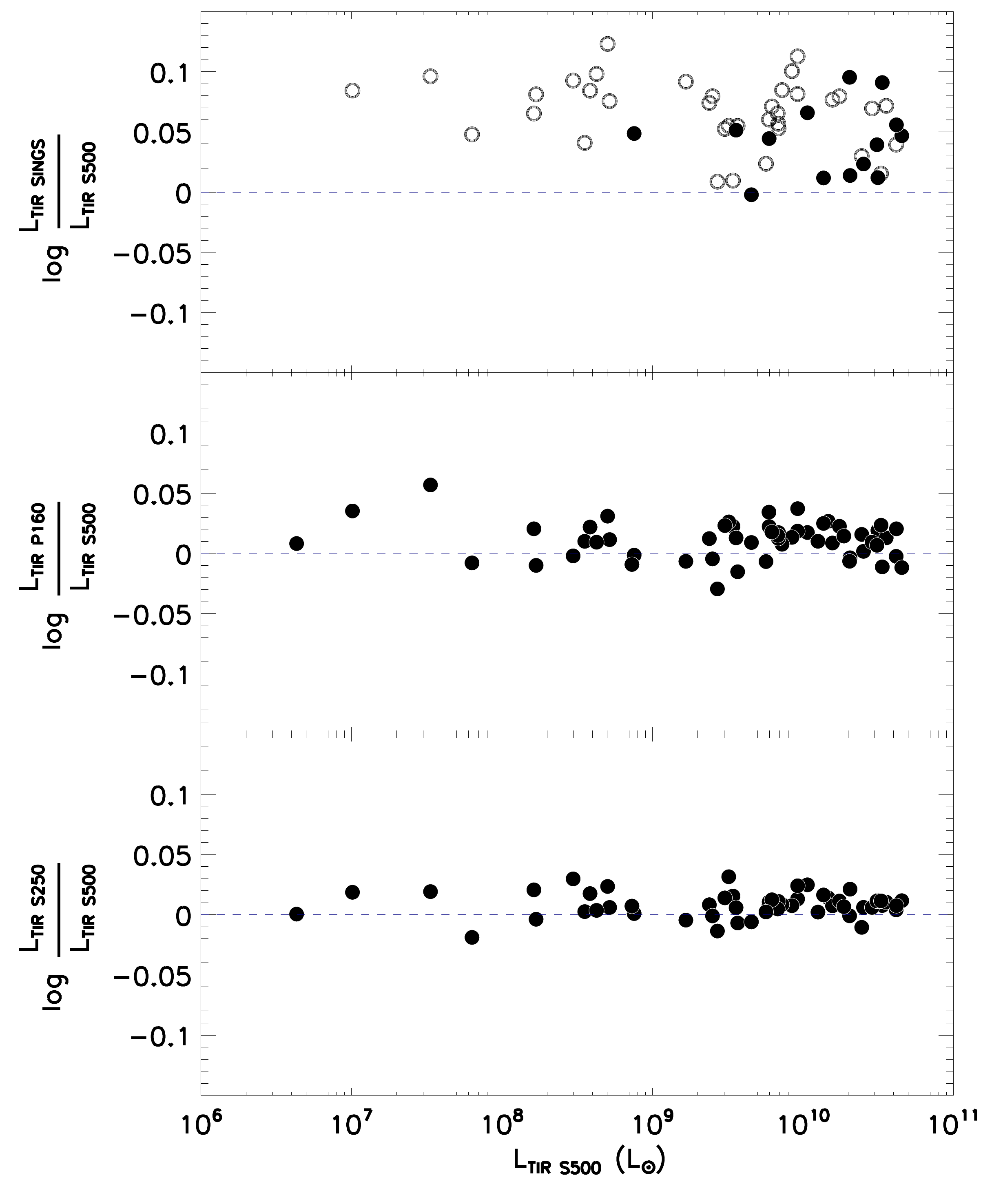} \\
    \caption{Integrated L$_{TIR~SINGS}$ from \citet{Draine2007} (top), L$_{TIR~P160}$ (middle) or L$_{TIR~S250}$ (bottom) compared to our reference L$_{TIR~S500}$. Integrated L$_{TIR}$ are expressed in \lsun.  For the top panel, we indicate with filled circles the luminosities L$_{TIR~SINGS}$ derived including SCUBA data at 850 \mic\ in the SED fitting and with empty circles the galaxies for which SCUBA data were not available.}
    \label{Global_LTIR_diffres}
    \end{figure*}

{\it Submm excess or AGN contribution - } 
Ground-based data and \hersc\ observations at submm wavelengths have helped us to better investigate the properties of the coldest phases of dust. A flattening of the submm slope or an excess compared to fits performed without submm data is very often detected in metal-poor galaxies \citep[][among others]{Galliano2003,Galliano2005,Marleau2006, Galametz2009,OHalloran2010, Bot2010_2}. Various hypotheses have been investigated to explain this excess: emission from a shielded cold dust reservoir \citep{Galliano2005,Galametz2009}, temperature-emissivity dependence of dust grains \citep{Meny2007,Paradis2010}, ``spinning dust" emission \citep{Murphy2010,Planck_collabo_2011_MagellanicClouds} and recently magnetic dipole radiation from magnetic nanoparticles \citep{Draine2012}. In the following study, we do not include data at wavelengths greater than 500 \mic\ (wavelength at which submm excess starts to be detected) but we include modelled emission out to 1100 \mic\ in our estimations. Our estimations of L$_{TIR}$ would not take the submm excess into account, if any. Nevertheless, as quantified further in this paper (Section 4), we expect thermal dust emission above 500 \mic\ to be negligible in the bolometric infrared energy budget of our galaxies, especially for low-metallicity environments where the excess is usually detected. In addition, a large number of the KINGFISH galaxies show nuclear emission indicating excitation by a non stellar continuum but no dominant AGN (except NGC~1316). We do not expect the AGN contribution to significantly affect our estimates of L$_{TIR}$.

   \begin{figure*}
   \centering    
   \begin{tabular} {cc}     
         \vspace{-15pt}               
      {\large \bf NGC~628} & \\
      & \includegraphics[width=12.5cm]{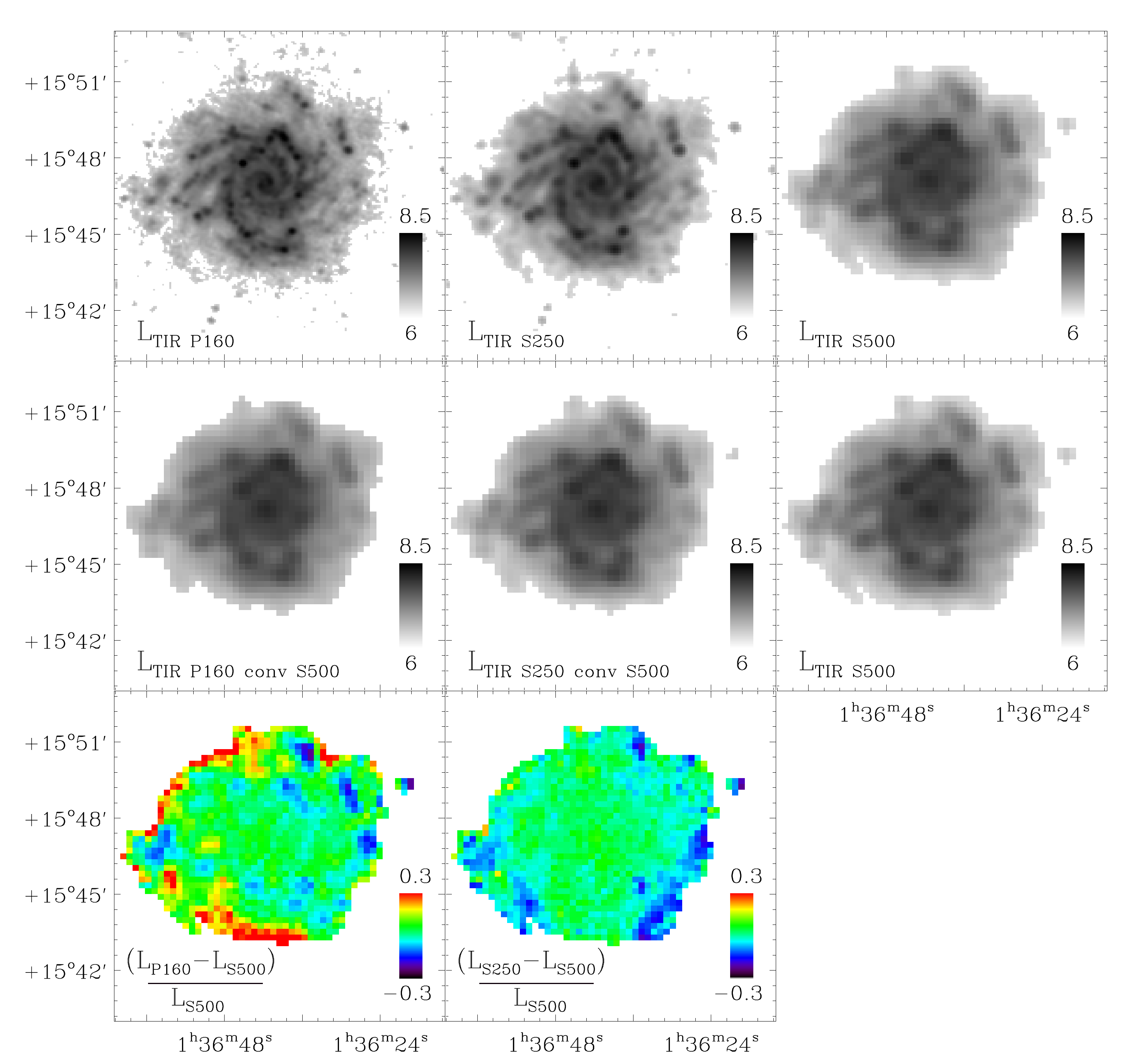}  \\
            \vspace{-15pt}
        {\large \bf NGC~6946} & \\
       &  \includegraphics[width=12.5cm]{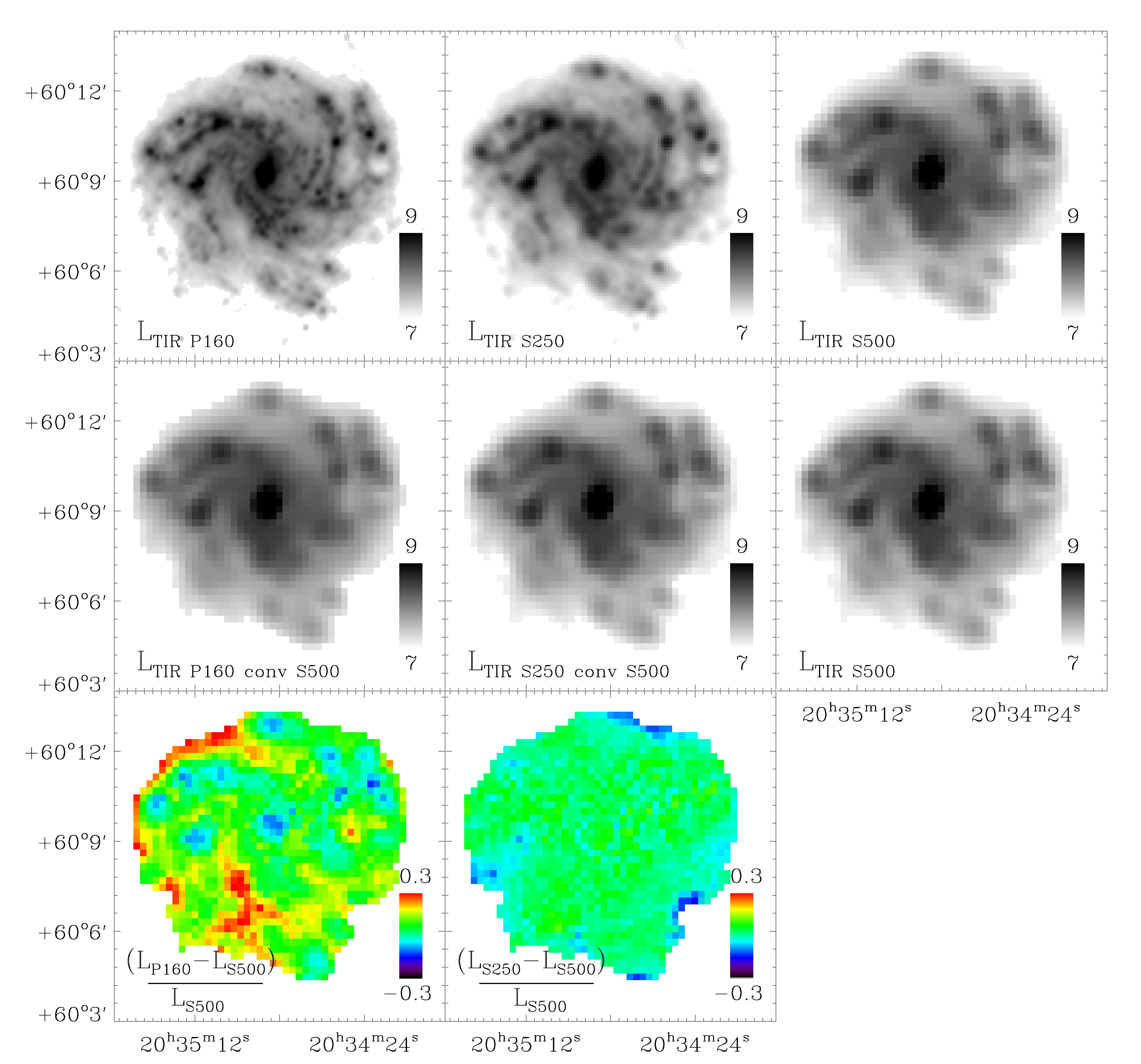}  \\
                    \end{tabular}
      \caption{Comparison of L$_{TIR}$ maps of NGC~628 and NGC~6946 for different resolutions. For each galaxy: Top: L$_{TIR~P160}$, L$_{TIR~S250}$ and L$_{TIR~S500}$ maps (\lsun~kpc$^2$, log scale). Middle: L$_{TIR~P160}$ and L$_{TIR~S250}$ maps convolved to the SPIRE 500 \mic\ resolution (\lsun~kpc$^2$, log scale). The last column is unchanged. Bottom: Relative difference between the convolved L$_{TIR~P160}$ or L$_{TIR~S250}$ maps and the reference L$_{TIR~S500}$ map.}
    \label{LTIR_diffres_maps}
     \end{figure*}     

   \begin{figure*}
   \centering
    \includegraphics[width=15cm]{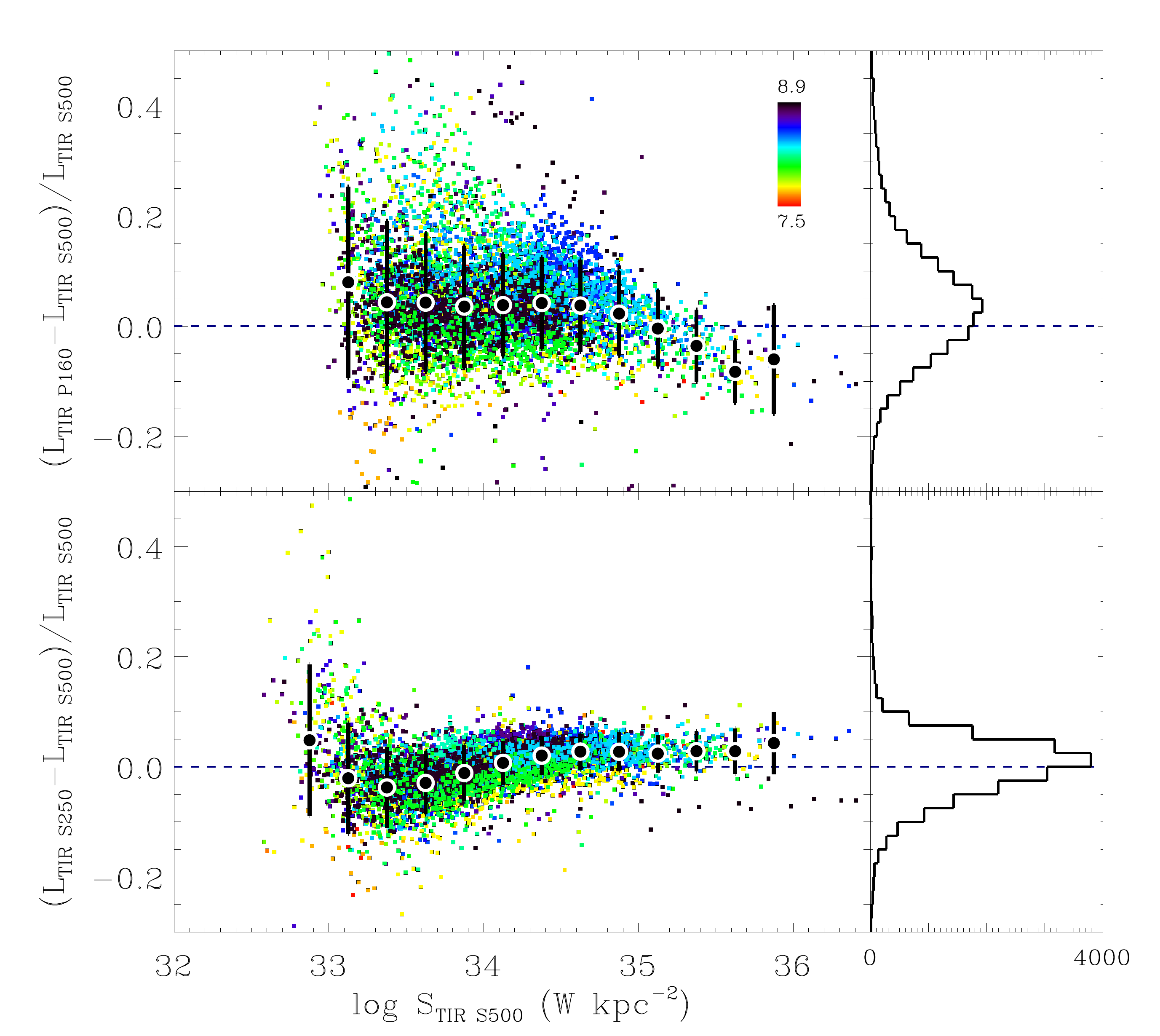} \\
    \caption{Resolved differences between L$_{TIR~P160}$ or L$_{TIR~S250}$ and the reference L$_{TIR~S500}$ as a function of the TIR brightness S$_{TIR~S500}$.  Galaxies are metallicity coded: purple for metal-rich, red for low-metallicity (colour scale in terms of 12+log(O/H) in the top panel). For each galaxy, we sort and average the data 3-by-3 for clarity. The black points indicate the median of the complete distribution per TIR brightness bins with 1-$\sigma$ error bars.  A histogram of the relative difference for the full sample of pixels is shown on the right of each plot.}
    \label{LTIR_diffres_perpixel}
    \end{figure*}

\subsection{Comparison of the integrated L$_{TIR}$} 

L$_{TIR}$ being a linear quantity, we do not expect strong differences between integrated L$_{TIR}$ obtained from a SED model fitting the galaxy as one big pixel or the L$_{TIR}$ we obtained by summing the resolved L$_{TIR}$. To check this assumption, we derive the integrated L$_{TIR}$ using the two techniques. We find that for each galaxy, the two integrated L$_{TIR}$ values differ by 5$\%$ at most whatever the wavelength coverage, which is comparable to the error bars on these quantities. For the rest of the paper, integrated L$_{TIR}$ values are thus obtained by summing the resolved L$_{TIR}$.

We want to investigate how integrated L$_{TIR}$ values vary when we include SPIRE data in the SED modelling. We thus derive integrated L$_{TIR}$ by summing the resolved L$_{TIR}$ in the \citet{Dale2012} apertures obtained using a [3-160\mic] coverage, a [3-250\mic] coverage and a [3-500\mic]. 
Table~\ref{Global_TIR} lists the integrated L$_{TIR}$ values of our sample for the three different coverages. Global errors are obtained by summing the resolved uncertainties estimated during the SED modelling process, with a median offset of $\sim$12$\%$ for the integrated L$_{TIR~S500}$. We normalise the integrated L$_{TIR~P160}$ and L$_{TIR~S250}$ to L$_{TIR~S500}$ for comparison. As part of the SINGS project, \citet{Draine2007} modelled most of the KINGFISH galaxies using \spitz\ fluxes (thus an SED coverage up to 160 \mic) and the [DL07] dust models and derive global dust luminosities (hereafter L$_{TIR~SINGS}$). They also include SCUBA fluxes at 850 \mic, when available, for a small subsample of their objects. In this paper, we use the distances provided by \citet{Kennicutt2011}, some of them being different from those used in the study of \citet{Draine2007}. We thus first rescale the L$_{TIR~SINGS}$ values to the distances we choose to use and add these corrected values to Table~\ref{Global_TIR} for comparison.

Figure~\ref{Global_LTIR_diffres} illustrates how the integrated L$_{TIR~SINGS}$ (top), L$_{TIR~P160}$ (middle) or L$_{TIR~S250}$ (bottom) compare with the integrated L$_{TIR}$ obtained in our complete coverage case L$_{TIR~S500}$. For the top panel, filled circles indicate when SCUBA data at 850 \mic\ were used in the fit to determine L$_{TIR~SINGS}$, empty circles when SCUBA data were not available. Our integrated L$_{TIR~S500}$ estimates are close to the L$_{TIR~SINGS}$ estimates within 10-15$\%$, even if systematically lower than the values derived from \spitz\ data only. Comparing the top and middle panel, we observe differences (a shift) between integrated L$_{TIR~SINGS}$ and L$_{TIR~P160}$, both obtained using the same coverage up to 160 \mic. As previously mentioned, resolution effects are not sufficient to explain such a shift between SINGS values and our values. The difference is thus probably related to the use of PACS data in the fit compared to the MIPS data used previously \citep[see][for discussion on PACS-MIPS photometry disagreement]{Aniano2012}. Integrated L$_{TIR}$ values obtained with our three different coverages (L$_{TIR~P160}$, L$_{TIR~S250}$, L$_{TIR~S500}$) are very similar. The agreement is within 8$\%$ between the integrated L$_{TIR~P160}$ or L$_{TIR~S250}$ and the reference L$_{TIR~S500}$ (except for NGC~2915 for which L$_{TIR~P160}$ is higher by 14$\%$).  \\

\subsection{Comparison of the resolved L$_{TIR}$}

We now probe the variations in the L$_{TIR}$ estimates driven by the different coverages on a resolved scale. We convolve the L$_{TIR~P160}$ and L$_{TIR~S250}$ maps of the KINGFISH galaxies to the resolution of SPIRE 500 \mic\ and compare them to our reference L$_{TIR~S500}$ map. 
Figure~\ref{LTIR_diffres_maps} gives an example of this comparison for the two spiral galaxies NGC~628 and NGC~6946. The top panels show the L$_{TIR}$ maps at the original resolution (PACS 160, SPIRE 250 and SPIRE 500 \mic\ from left to right) in \lsun~kpc$^{-2}$. The middle panels show these maps convolved to the SPIRE 500 \mic\ resolution (the last column is unchanged). The bottom panels finally compare the convolved L$_{TIR~P160}$ or L$_{TIR~P250}$ maps with the reference L$_{TIR~S500}$ map. We remind the reader that we restrict our study to resolved elements with a 3-$\sigma$ detection in PACS 160 and SPIRE 250 and 500 \mic\ bands. We can still distinguish the structure of NGC~628 and NGC~6946 on the bottom left panels that show the difference between the L$_{TIR~P160}$ maps and the reference maps L$_{TIR~S500}$. In both galaxies, we observe that the absence of submm (SPIRE) data leads to an underestimation of the L$_{TIR}$ in bright regions (blue structures in Figure~\ref{LTIR_diffres_maps}) of up to 10-15$\%$, and an overestimation in the outer part of the galaxies, so at low-surface brightnesses (red structures), of up to 30$\%$. The L$_{TIR}$ maps obtained with data up to SPIRE 250 and with data up to SPIRE 500 \mic\ are, on the contrary, very similar, as illustrated by the small difference residuals of those maps (bottom middle panels). 

Figure~\ref{LTIR_diffres_perpixel} gathers, for the complete sample, the resolved relative differences between L$_{TIR~P160}$ (top) or L$_{TIR~S250}$ (bottom) with the reference L$_{TIR~S500}$ as a function of the TIR surface brightness S$_{TIR~S500}$ (in W~kpc$^{-2}$, log scale). Galaxies are metallicity-coded, from low-metallicity in red to high-metallicity in dark purple colour. Discrepancies between resolved L$_{TIR~P160}$ and L$_{TIR~S500}$ can be significant as illustrated in Fig.~\ref{LTIR_diffres_perpixel} (top). L$_{TIR}$ is under-predicted when modelled with data up to PACS 160 \mic\ in high surface brightness regions (drop of the median for log S$_{TIR~S500}$ $>$ 34.5 W~kpc$^{-2}$), and over-predicted in low-surface brightness regions, similar to the trends observed for NGC~628 and NGC~6946 (Fig.~\ref{LTIR_diffres_maps}).
Using data up to 250 \mic\ resolution, we can observe a slight under-prediction of L$_{TIR}$ at 33 $<$ log S$_{TIR~S500}$ $<$ 34 (in W~kpc$^{-2}$) and an over-prediction at log S$_{TIR~S500}$ $>$ 34 but L$_{TIR~S250}$ and L$_{TIR~S500}$ are nevertheless close (within $\pm$10$\%$ for most of our resolved elements). For both resolution, determining the L$_{TIR}$ with accuracy seems difficult at very low S$_{TIR}$ ($<$ 33 W~kpc$^{-2}$), due to the large uncertainties on the flux measurements in those regions. 
In conclusion, using data up to 250 \mic\ seems to be the best compromise between sufficiently constraining the submm slope in order to obtain a L$_{TIR}$ consistent with that obtained with a complete coverage of the dust thermal emission, while still keeping a good working resolution. \\

{\it We choose to work at the resolution of SPIRE 250 \mic\ for the rest of this study}. 
At this resolution, resolved L$_{TIR}$ have uncertainties of 10-15$\%$ on average in the resolved elements we select. In the Appendix (Figure \ref{LTIR_Maps}), we show the surface brightness S$_{TIR}$ maps obtained at SPIRE 250 \mic\ resolution for the full KINGFISH sample. Maps are in \lsun~kpc$^{-2}$ (log scale). We remind the reader that the FWHM of the SPIRE 250 PSF is $\sim$18\arcsec. Our final maps have a pixel size of 6\arcsec, which corresponds to ISM elements of 88 pc for the closest galaxy of the sample Holmberg~II (3.05 Mpc) and 890 pc for the furthest galaxy of the sample NGC~1266 (30.6 Mpc).


\section{The infrared to submm distribution of luminosities}

For normal star-forming galaxies, \citet{Dale_Helou_2002} studied the distribution of the infrared energy budget, namely how much energy emerges in various wavebands, and investigate its dependence on the star formation activity. \citet{Dale2009} also present monochromatic-to-bolometric ratios and analyse their dependence with f$_{\nu}$(70 \mic) / f$_{\nu}$ (160 \mic) ratios and morphologies. In a similar fashion, we analyse the distribution of the total infrared energy with wavelength for the KINGFISH sample, now using our \hersc\ data on a resolved scale. We derive the fractions of the L$_{TIR}$ emitted in 4 different wavebands ([3-24 \mic], [24-70 \mic], [70-160 \mic] and [160-1100 \mic]): we integrate the local SED models over the stated limits and divide these values by the resolved L$_{TIR}$.

   \begin{figure}
   \centering
    \includegraphics[width=9cm]{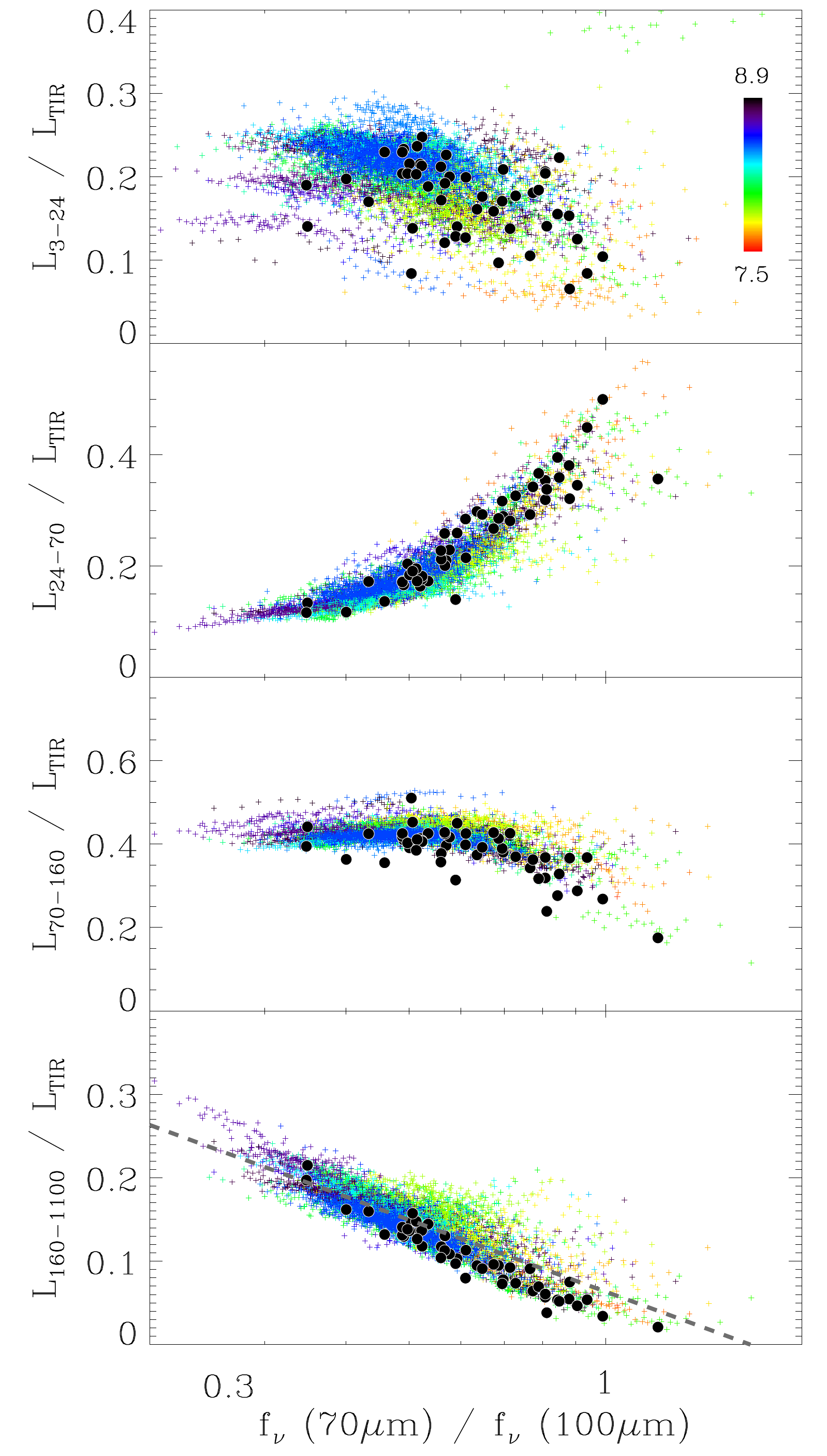} \\
    \caption{Resolved fraction of L$_{TIR}$ emitted in different wavebands ([3-24], [24-70], [70-160] and [160-1100]) as a function of the f$_{\nu}$(70 \mic)/f$_{\nu}$(100 \mic) far-IR colour. Galaxies are metallicity coded: purple for metal-rich, red for metallicity-poor (colour scale in the top panel in terms of 12+log(O/H)). For each galaxy, we sort and average the resolved element 5-by-5 for clarity. Global luminosity fractions are overlaid with black circles. A log-linear fit to the data in the bottom panel is shown by the grey dashed line.}
    \label{Distribution_IRluminosities}
    \end{figure}

Figure~\ref{Distribution_IRluminosities} displays these fractions as a function of the 70-to-100 \mic\ flux density ratio (hereafter 70/100 colour), which parameterises various star formation activities and is also a good proxy for dust temperatures. This ratio also correlates well with the starlight intensity. Galaxies are metallicity coded, from dark purple for metal-rich to red for metal-poor. We overlay integrated values for comparison. We obtain averaged 70/100 colours (x-axis) by summing the 70 and 100 \mic\ flux densities of each resolution element then dividing them. We then sum the resolved luminosities in our 4 different bands (thus L$_{3-24}$, L$_{24-70}$, L$_{70-160}$ or L$_{160-1100}$) and divide these integrated luminosities in each band by the integrated L$_{TIR~S500}$ (Table~\ref{Global_TIR}). This enables us to obtain the integrated fractions (y-axis).

The thermal dust emission peaks in the [70-160 \mic] band for all our galaxies. We thus naturally observe that the bulk of IR energy is contained in that wavelength range. This [70-160 \mic] fraction is homogeneous across the sample and accounts for $\sim$40$\%$ of the L$_{TIR}$. The [24-70 \mic] fraction ranges from 0.1 to 0.4, increasing with decreasing metallicity (or increasing 70/100 colour). Since low-metallicity environments usually contain warmer dust \citep{Hunter1989,Dale2005}, the [24-70 \mic] fraction is indeed expected to significantly contribute to the total IR emission in those objects. 

While the [3-24 \mic] band contributes to $\sim$20$\%$ on average to the L$_{TIR}$ for the sample, there is a large scatter in this fraction. This is likely associated with the contribution of Polycyclic Aromatic Hydrocarbons (PAH) emission. Several factors drive the scatter in the contribution of PAHs to the [3-24 \mic] band. Low-metallicity galaxies usually show weak emission from PAHs \citep{Engelbracht2005,Jackson2006} and PAHs are also known to be sensitive to the hardness of the radiation field \citep{Madden2006,Galliano2003, Galliano2005}. The PAH size distribution also tends to be different at low metallicity, but whether the small PAHs are destroyed in the harsh conditions or whether they dominate because of different formation processes is still not clear \citep{Hunt2010,Sandstrom2012}. 
The paucity of PAHs in low-metallicity environments could also be due to a delayed injection of carbon dust by AGB stars \citep{Dwek1998,Galliano_Dwek_Chanial_2008}. 
This low PAH emission is responsible for the weak [3-24 \mic] fraction in our metal-poor galaxies. Some metal-rich galaxies (NGC~5866, NGC~4594, NGC~1316) also present low [3-24\mic] resolved fractions (resolved elements with L$_{3-24}$/L$_{TIR}$ $<$15$\%$ and f$_{\nu}$(70 \mic)/f$_{\nu}$(100 \mic)$<$0.5). Those peculiar galaxies show only little dust emission relative to their stellar emission, as commented by \citet{Draine2007}.

   \begin{figure*}
    \centering
    \begin{tabular}{c}
    \includegraphics[width=13cm]{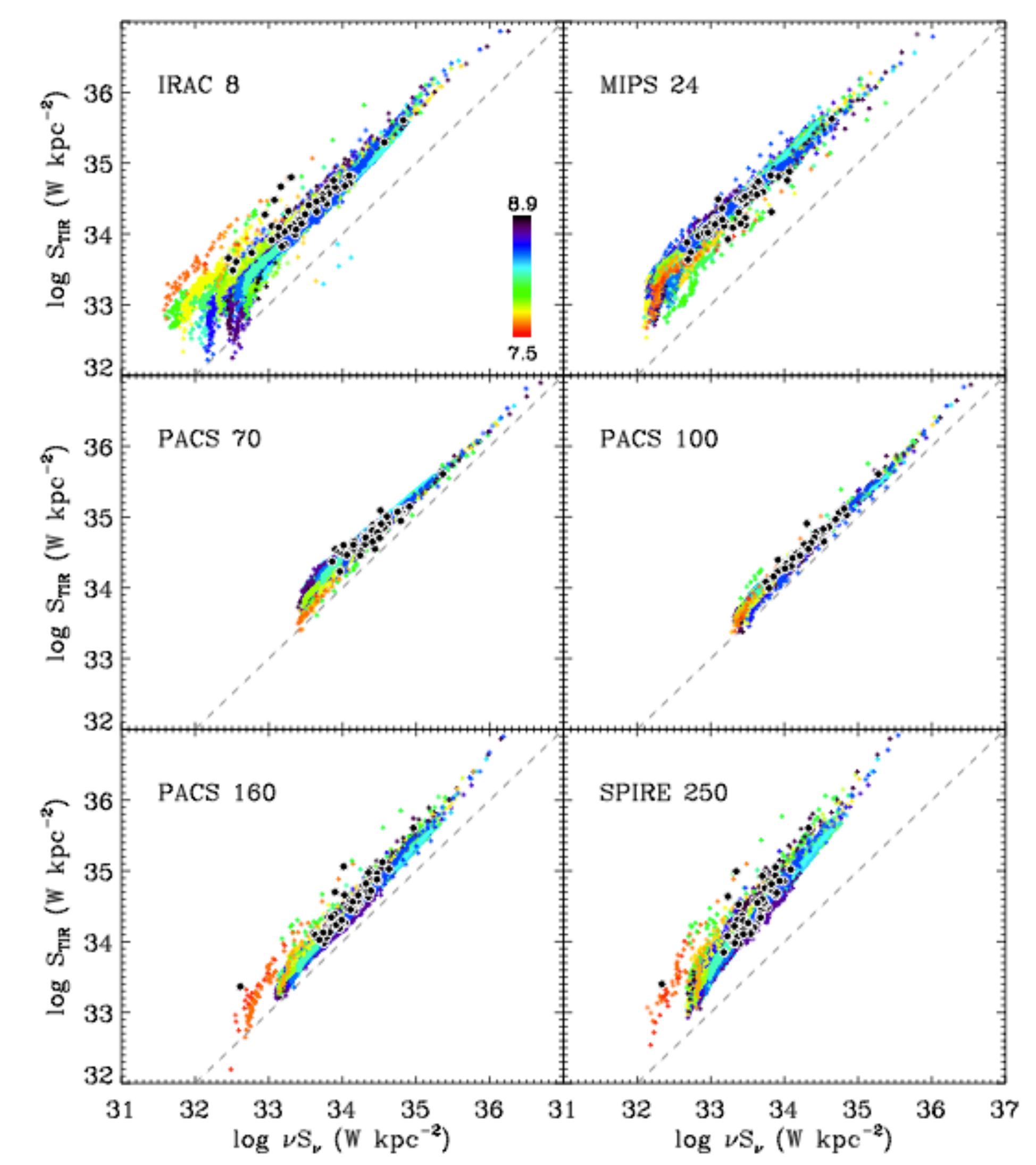}\\
    \end{tabular}
    \caption{TIR surface brightness as a function of the individual bands (IRAC~8 \mic, MIPS~24 \mic, PACS~70 \mic, PACS~100 \mic, PACS~160 \mic\ and SPIRE~250 \mic). Galaxies are colour-coded by metallicity (colour scale in the top left panel expressed as 12 + log(O/H)). We overlay global surface brightnesses with black circles. Dashed lines indicate the 1:1 relation.}
    \label{LTIRvsSB_separatewave}
    \end{figure*}

We observe a very clear trend of the submillimeter [160-1100 \mic] fraction with the 70/100 colour or the metallicity. The [160-1100 \mic] band accounts for a few $\%$ for low-metallicity galaxies to up to 25$\%$ of the total infrared luminosity budget for metal-rich environments, consistent with the previous studies of \citet{Dale2001} and \citet{Dale_Helou_2002}. A log-lin fit of our data leads to the relation: L$_{160-1100}$/L$_{TIR}$ = -0.28 log(f$_{\nu}$(70\mic)/f$_{\nu}$(100\mic)) + 0.07.

While not shown, we also quantified the [500-1100 \mic] fraction. This band contributes no more than 0.6$\%$ in the resolved elements of our sample, and only up to $\sim$0.2$\%$ for the most metal-poor galaxies. Thus, even if present, we do not expect a submm excess to significantly modify our conclusions in this work. \\


\section{Spitzer and Herschel bands as L$_{TIR}$ monochromatic calibrators}

In this section, we want to study the resolved relationships linking the individual \spitz\ and \hersc\ bands with the L$_{TIR}$. A calibration using the resolved elements of the complete sample of KINGFISH galaxies has been derived. However, we also analyse the relations for individual galaxies in order to get a handle on the scatter driven by the variety of our sample and study how individual relations change with global galaxy characteristics or local ISM conditions (metallicity, 70/100 colour for instance). This will help us to unify the picture initiated by \citet{Boquien2011} for M33 using a wider sample of galaxies, more representative of the diversity of the local Universe. We thus first analyse the individual and global relations between monochromatic far-IR {\it surface brightnesses} and TIR {\it surface brightnesses} in Sections 5.1 and 5.2. We also derive empirical calibrations of L$_{TIR}$ from monochromatic far-IR {\it luminosities} in Section 5.3. We analyse the validity of our monochromatic calibrations for the KINGFISH sample in Section 5.4. Finally, we investigate in Section 5.5 the goodness of our calibrations for a wider range of environments, including high-redshift galaxies.

\subsection{A qualitative view of the relations}

We display the resolved TIR surface brightnesses (S$_{TIR}$) of our galaxies as a function of the IRAC 8 \mic, MIPS 24 \mic, PACS 70, 100 and 160 \mic\ and SPIRE 250 \mic\ brightnesses (in W~kpc$^{-2}$, log scale, 3-$\sigma$ detection) in Fig.~\ref{LTIRvsSB_separatewave}. 
To study how metallicity influences those monochromatic relations, galaxies are colour-coded by metallicity (expressed as 12+log(O/H)), from metal-poor in red to metal-rich in dark purple colour. We overlay integrated values for comparison. Since we restrict our L$_{TIR}$ maps to pixels with a sufficient signal-to-noise in each \hersc\ bands, we could be missing flux in the faint outskirts of our objects. We prefer to use the global \spitz\ and \hersc\ fluxes \citep[from][]{Dale2007,Dale2012} instead. These integrated fluxes are divided by the area covered by our selected pixels to obtain the average global monochromatic brightnesses in W~kpc$^{-2}$. In order to remove the stellar contribution from the integrated 8 \mic\ flux densities of \citet{Dale2007}, we apply the recipe of \citet{Marble2010}, namely f$_8^{stellar}$/f$_{3.6}$$\sim$24$\%$. This estimate is close to the value derived in \citet{Helou2004} using the recipe of Starburst 99 (23.2$\%$). We consider that the stellar contribution to the integrated 24 \mic\ flux densities is negligible. \\

We observe that : 

\vspace{4pt}
{\it -} the relation between the IRAC 8 \mic\ surface brightness and S$_{TIR}$ changes from galaxy to galaxy, as already shown by \citet{Calzetti2007_2}, and strongly depends on the metallicity of the galaxy. For a given S$_{TIR}$, the resolved 8 \mic\ brightnesses are systematically lower in metal-poor objects. This trend is consistent, as discussed in Section 4, with the decrease of PAH emission often observed in low metallicity environments,

\vspace{4pt}
{\it - } MIPS 24 \mic\ is also a decent tracer of S$_{TIR}$, but a significant scatter can be observed from galaxy-to-galaxy, as already observed by \citet{Calzetti2007_2},

\vspace{4pt}
{\it - } the relation between the brightness in the 70 \mic\ band and S$_{TIR}$ also seems to be slightly dependent on metallicity. For a given S$_{TIR}$, we observe higher 70 \mic\ brightnesses in low-metallicity galaxies. The mid (MIR) to far-infrared (FIR) part of the SEDs of dwarf galaxies are known to be elevated compared to normal spiral galaxies, since they contain more small grains and hotter dust (\citealt{Engelbracht2005}, \citealt{Galliano2005} or the global SEDs of KINGFISH low-metallicity objects in \citealt{Dale2012}), which could explain this trend. The lowest metallicity still appears as an outlier in the PACS 160 and SPIRE 250 \mic\ panel.

\vspace{4pt}
{\it -} the \hersc\ bands (PACS 70, 100 and 160 \mic\ bands, as well as SPIRE 250 \mic\ band to a lesser extent) appear to be very good monochromatic indicators of S$_{TIR}$ as suggested by the tight correlations obtained in Fig.~\ref{LTIRvsSB_separatewave}. \\

We show the resolved S$_{TIR}$ (in W~kpc$^{-2}$, log scale) as a function of the different monochromatic surface brightnesses (in W~kpc$^{-2}$, log scale)  for each galaxy of the KINGFISH sample in the Appendix (Figure~\ref{LTIRvsSB}). We restrict this calibration to pixels with a 3-$\sigma$ detection in the individual bands. We remind the reader that stellar contribution to the 8 \mic\ and 24 \mic\ bands is removed using the stellar emission modelled during the SED fitting process. As mentioned before, this contribution is minor at 24 \mic, with a median contribution of 0.67$\%$ over the pixels fulfilling our 3-$\sigma$ criterion. 

\subsection{Quantitative analysis}

For each galaxy and for the complete sample, we derive the calibration coefficients (a$_i$, b$_i$) such as:

\begin{equation}
log~S_{TIR} = a_i~log~S_i + b_i 
\label{eq2}
\end{equation}

\noindent where S$_{TIR}$ refers to the TIR brightness, S$_i$ the brightness in a given \spitz\ or \hersc\ band {\it i} (from IRAC 8 \mic\ to SPIRE 250 \mic), both in W~kpc$^{-2}$, and a$_i$ and b$_i$ respectively the slope and the intercept of the fit. 

Here and for the rest of Section 5, we choose to work in log-log space to account for non-linearities between the L$_{TIR}$ and the monochromatic emissions. We refer to \citet{Boquien2011} for a description of non-linearity effects in the relations between TIR surface brightnesses and monochromatic surface brightnesses. We perform the regressions using a least-squares-bisector algorithm (function {\it sixlin} of the Astrolibrary of IDL) and use a ``jack-knife" technique to quantify the goodness-of-fit and provide calibration coefficients with conservative errors. Indeed, we randomly select N=1/10 of the resolved elements used for a given calibration\footnote{The choice of N does not influence our final results as long as the subsample still contains enough points to be representative of the relation. Coefficients start to differ by a few percent if we choose N$<$1/100}, perform the regression for that subsample, save the coefficients and repeat this procedure 10000 times. The calibration coefficients provided in this paper are therefore the median of the coefficient distributions and errors on the coefficients are obtained from the standard deviations. \\

   \begin{figure*}
   \centering
    \includegraphics[width=17cm]{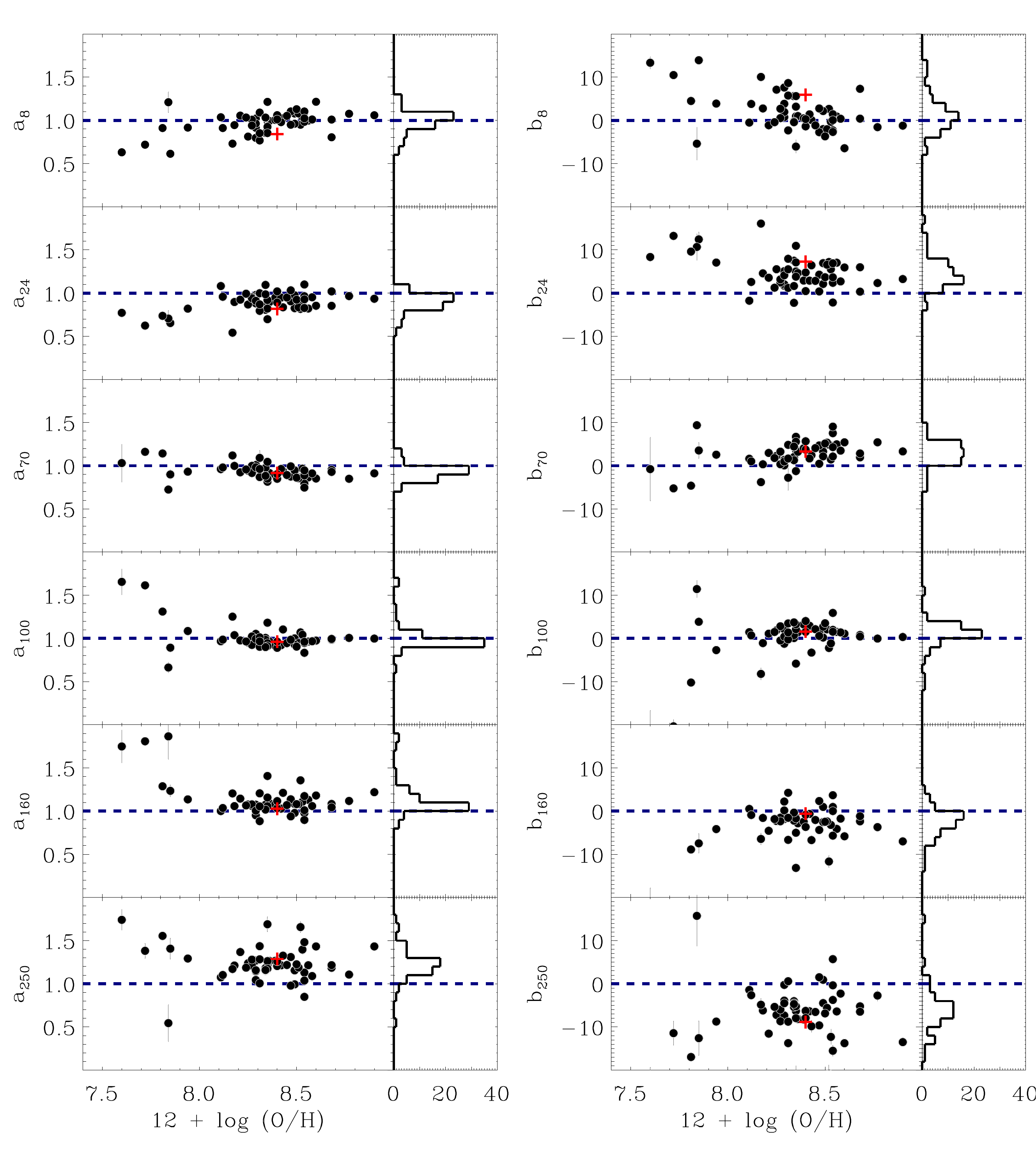} \\
    \vspace{-10pt}
    \caption{Calibration coefficients to convert monochromatic brightnesses from \spitz\ or \hersc\ bands into S$_{TIR}$ plotted as a function of metallicity expressed as 12+log(O/H). The relation is : log~S$_{TIR} $= a$_i$~log~S$_i$ + b$_i$ with S$_i$ in W~kpc$^{-2}$. Horizontal dashed lines indicate a$_i$=1 (linear relation) and b$_i$=0. We overlay the coefficients derived by \citet{Boquien2011} for M33 with red crosses and join an histogram of the coefficients on the right hand side of each plot. }
    \label{Coeffs_fits}
    \end{figure*}
   
 \begin{table*}
\caption{\large Calibration coefficients to predict the TIR brightness/luminosity from monochromatic \spitz\ and \hersc\ brightnesses/luminosities. }
\label{Coeff_total_Mono}
 \centering
  \begin{tabular}{ccccccccc}
\hline
\hline
&\\
{\large Surface brightnesses}&  \\
&\\
\cline{1-6}
&											&waveband &a$_{i}$	   &	b$_{i}$	&  Scatter (dex) \\
\cline{2-6}
&\\
{\large log~S$_{TIR}$= a$_i$~log~S$_i$ + b$_i$} 			&&8    & 0.869$\pm$0.007 & 5.127$\pm$0.220		&0.137\\
S$_{TIR}$, S$_i$ in W~kpc$^{-2}$						&&24  & 0.919$\pm$0.003& 3.786$\pm$0.106		&0.095\\
(Eq.~\ref{eq2}) 										&&70  &0.931$\pm$0.003 & 2.749$\pm$0.087		&0.081\\
												&&100&0.974$\pm$0.002 & 1.137$\pm$0.066		&0.050\\
												&&160&1.043$\pm$0.004 & -1.151$\pm$0.152 	&0.090\\
												&&250&1.148$\pm$0.006 & -4.180$\pm$0.218 	&0.133\\
&\\												
\hline
\hline					
&\\							
{\large Luminosities}&  \\
&\\
\cline{1-6}
&											&waveband &a$_{i}$	   &	b$_{i}$  &  Scatter (dex) \\
\cline{2-6}
&\\
{\large log~L$_{TIR}$= a$_i$~log~$\nu$L$_{\nu}$(i) + b$_i$} 		&& 8   &0.929$\pm$0.005&1.135$\pm$0.031		&0.148 \\
L$_{TIR}$, $\nu$L$_{\nu}$(i) in \lsun\ 						&&24  &0.954$\pm$0.002& 1.336$\pm$0.013		&0.100 \\	
(Eq.~\ref{eq3}) 											&&70  &0.973$\pm$0.002&0.567$\pm$0.013		&0.086 \\
													&&100&1.000$\pm$0.001&0.256$\pm$0.008		&0.052 \\
													&&160&1.024$\pm$0.003&0.176$\pm$0.018		&0.090 \\
													&&250&1.060$\pm$0.004&0.451$\pm$0.023		&0.136 \\
\hline
\end{tabular}
\end{table*} 

We indicate the calibration coefficients obtained for the whole sample in Table~\ref{Coeff_total_Mono}. We also tabulate the coefficients a$_i$ and b$_i$ obtained for each galaxy in the Appendix in Table~\ref{Coeff_Indiv_mono}. We add the scatter around the best fit (in dex) in the last column of the table.
In order to better assess the results for individual galaxies, we plot in Fig.~\ref{Coeffs_fits} the individual coefficients a$_i$ and b$_i$ derived using Eq.~\ref{eq2} (and tabulated in Table~\ref{Coeff_Indiv_mono}) as a function of metallicity (expressed as 12+log(O/H)). For galaxies with average 12+log(O/H) $>$ 8.0, the MIR 8 and 24 \mic\ brightnesses are linear estimators of  the S$_{TIR}$, as also found in \citet{Zhu2008}. The three PACS bands are very reliable estimators of the S$_{TIR}$. The slope of the relation is, on average, $<$1.0, $\sim$1.0 and $>$1.0 for PACS 70, 100 and PACS 160 \mic\ respectively, consistent with the slope we derive gathering the resolved elements of the whole sample and tabulated in Table~\ref{Coeff_total_Mono}). We observe that the scatter for the MIPS 24 \mic\ and PACS relations is small, with a minimum scatter for the PACS 100 \mic\ relation.

For our low-metallicity galaxies, we observe that the 8 \mic\ emission tends to be a sub-linear estimator of the S$_{TIR}$. This result is consistent with the trend observed for the resolved elements of low-metallicity galaxies in Fig.~\ref{LTIRvsSB_separatewave}. Low-metallicity objects also seem to show steeper relations between PACS 100 and 160 \mic\ monochromatic brightnesses and S$_{TIR}$. This means that for a fixed PACS brightness, the TIR emission of low-metallicity environments will be higher than a normal spiral galaxy. We nevertheless highlight the difficulty in performing a pixel-by-pixel calibration from PACS data in some of our metal-poor galaxies due to poor statistics and a possible lack of detection at low-surface brightnesses that could bias the calibrations toward steeper relations. 
 
We finally note that residuals from the best fit to Eq.~\ref{eq2} for each galaxy are smaller than the residuals from the same fit performed on the integrated values of our galaxies. This could favour global parameters rather than local parameters as a driver of scatter in the relationship between monochromatic and TIR surface brightnesses. \\

   \begin{figure*}
   \centering
    \includegraphics[width=13cm]{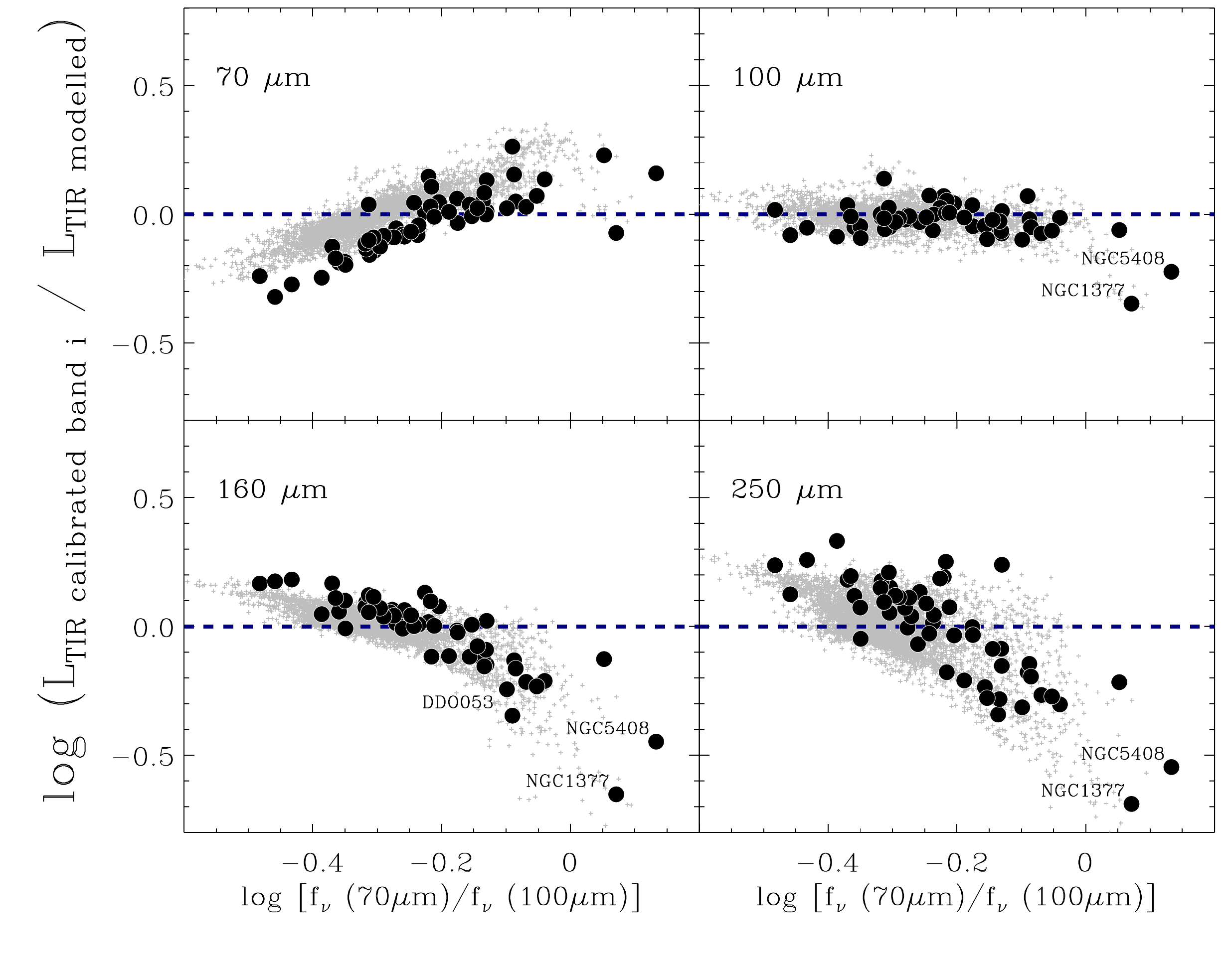} \\
    \caption{Comparison between resolved and integrated L$_{TIR}$ calibrated from \hersc\ monochromatic fluxes L$_{TIR~band~i}$ (with i = 70, 100, 160 or 250 \mic) with modelled L$_{TIR~S500}$ plotted as a function of the f$_{\nu}$(70\mic)/f$_{\nu}$(100\mic) far-IR colour. The calibration relation is of the form log~L$_{TIR~band~i}$= a$_i$~log~$\nu$L$_{\nu}$(i) + b$_i$ with luminosities in \lsun. We report the calibration coefficients in Table~\ref{Coeff_Indiv_mono} (last line). We show the integrated values with black circles and the resolved elements with grey points. For each galaxy, we sort and average resolved element 10-by-10 for clarity.}
    \label{Comp_mono}
    \end{figure*}

{\it Comparison with M33 - }
 \citet{Boquien2011} derived calibration coefficients linking the monochromatic \spitz\ and \hersc\ surface brightnesses to S$_{TIR}$ for the Scd galaxy M33, using Eq.~\ref{eq2}. We overlay the monochromatic coefficients they derive with red crosses in Fig.~\ref{Coeffs_fits} for comparisons with our results for the KINGFISH galaxies. We adopt an oxygen abundance of 12+log(O/H)=8.4 \citep[from][]{Massey1998} for M33.  For galaxies with 12+log(O/H) $>$ 8.0, the calibration coefficients derived in our study and in \citet{Boquien2011} for M33 are very similar. We note that \citet{Boquien2011} found that for the galaxy M33, the MIR 8 and 24 \mic\ brightnesses seem to be sub-linear estimators of S$_{TIR}$. To check if the difference could be linked with the signal-to-noise threshold we choose, we restrict our study to resolved elements above a higher (5$\sigma$) brightness threshold. This does not strongly modify our calibration coefficients for the 8 and 24 \mic\ relations. A superimposition of the resolved elements of M33 with galaxies of our sample sharing similar properties (metallicity, global L$_{TIR}$, SFR) as M33 confirm the difference. Small differences could also arise as a result of differences in the regression methods. We thus apply the same linear regression (bisector algorithm) to the M33 data used by \citet{Boquien2011} and obtain calibration coefficients a$_8$=0.88, b$_8$=4.69, a$_{24}$=0.84 and b$_{24}$=6.61, very similar to the values found in their paper. The calibration coefficients of M33 are within the ranges of values found for our sample. Small differences could possibly due to {\it 1)} a different treatment in the data reduction / background subtraction at those wavelengths, {\it 2)} a real difference in the 8 or 24 \mic\ vs TIR surface brightness relations compared to KINGFISH galaxies with similar average oxygen abundance, {\it 3)} the large uncertainties on these average oxygen abundances, M33 having a metallicity gradient for instance \citep{Magrini2007}.

 \subsection{Calibration from monochromatic luminosities} 
 
Integrated luminosities are sometimes the only measure we can access, in particular for high-redshift observations. Studies of non resolved objects thus require calibrations using luminosities rather than surface brightnesses. Gathering the resolved elements of the complete sample, we derive calibrations coefficients similar to those of Eq.~\ref{eq2} but linking L$_{TIR}$ and 24, 70, 100, 160 or 250 \mic\ monochromatic luminosities. Eq.~\ref{eq2}  can thus be re-written as:
\begin{equation}
log~L_{TIR} = a_i~log ~\nu L_{\nu}(i) + b_i 
\label{eq3}
\end{equation}

\noindent with now L$_{TIR}$ the TIR luminosity, $\nu$L$_{\nu}$(i) the flux in a given \spitz\ or \hersc\ band {\it i} and both L$_{TIR}$ and $\nu$L$_{\nu}$(i) are in \lsun. As before, we use our ``jack-knife" technique to quantify errors on the calibration coefficients (see Section 5.2 for details on the technique) and report the coefficients calibrating L$_{TIR}$ from monochromatic luminosities in Table~\ref{Coeff_total_Mono}. The scatter around the best fit (in dex) is also provided in the table.

If we were to derive a similar calibration for individual galaxies, the individual slopes (a$_i$) would not be modified by the change of units compared to those derived in Section 5.2. Most of the conclusions of Section 5.2 still apply for the calibrations derived in this section. We note that the coefficients we derive for the 24 \mic\ calibration (a$_{24}$=0.954, b$_{24}$=1.336) are consistent with those derived by \citet{Rieke2009} within their error bars (namely a$_{24}$=0.920, b$_{24}$=1.183). They estimate L$_{TIR}$ from IRAS fluxes using the recipe of \citet{Sanders2003}. \\

 \subsection{Validity of the calibrations for KINGFISH galaxies}
 
We aim to test the robustness of the empirical calibrations derived in Section 5.3 - namely their ability to reproduce resolved and integrated luminosities - and analyse the intrinsic biases of our predictions. We thus compare the resolved and integrated L$_{TIR}$ calibrated from the monochromatic luminosities at 70, 100, 160 or 250 \mic\ \footnote{Because the PACS wavebands are close to the peak of the SEDs, they are expected to provide more robust calibrations of the L$_{TIR}$ than the 8 and 24 \mic\ wavebands. The 8 \mic\ calibration is also strongly dependent on the metallicity.} and the coefficients tabulated in Table~\ref{Coeff_total_Mono} versus the resolved and integrated L$_{TIR}$ derived with a proper SED modelling.  Figure~\ref{Comp_mono} illustrates the comparison as a function of the 70/100 colour. For integrated values, we use the global flux densities of \citet{Dale2012} to predict integrated L$_{TIR}$ and compare them to our reference L$_{TIR~S500}$ (Table.~\ref{Global_TIR}). Results are identical for integrated or resolved L$_{TIR}$. The shift between resolved and integrated values (grey points / filled black circles) may be linked with the fact that {\it (i)} the calibration tree used in the data reduction of \hersc\ observations has changed since the \citet{Dale2012} study and {\it (ii)} the background subtraction technique used in both studies are slightly different.

{\it PACS 70 - }The 70 \mic\ band provides a reasonably good monochromatic estimate of the L$_{TIR}$ (difference $<$ 50$\%$ for most of them). The L$_{TIR}$ of the lower IR luminosity objects (L$_{TIR}$ $<$ 3 $\times$ 10$^8$ \lsun) is usually overestimated by the 70 \mic\ monochromatic calibration. Those objects are mostly low-metallicity galaxies (DDO~053, M81DwB, HolmbergII, IC~2574, NGC~2915) that usually present warmer temperatures, which is consistent with the overestimation we observe. We also observe a strong correlation of the goodness of the 70 \mic\ calibration with the 70/100 colour.

{\it PACS 100 - }The 100 \mic\ band offers the best monochromatic estimator, with very little scatter and 53 out of 55 galaxies within $\sim$30$\%$ (as shown in NGC~6946 by Tabatabaei et al, submitted to A\&A). We remind the reader that uncertainties on the resolved L$_{TIR}$ are of the order of 10-15$\%$. We thus observe that many galaxies have predictions that match the modelled L$_{TIR}$ within these error bars. The main outlier is NGC~1377. NGC~1377 shows a significant excess in its infrared-to-radio ratio and is thought to be a rare local nascent starburst probably powered by accretion through a recent merger \citep{Roussel2006}. The galaxy presents a very hot IR SED peaking around 60 \mic\ ($\nu$f$_{\nu}$ units) and the PACS 100 \mic\ observation thus already belongs to the Rayleigh-Jeans slope of the SED of this object \citep{Dale2012}. This hot SED compared to those of the other galaxies of the sample explains why we underestimate its L$_{TIR}$ when PACS 100 \mic\ is used as a monochromatic indicator (inability to capture the starburst component). The same explanation applies for the star-bursting dwarf irregular NGC~5408 whose IR emission peaks at $\sim$70 \mic\ \citep{Dale2012}. Given the fact that the 70 \mic\ band corresponds to the peak of the global SED for NGC~1377, using the 70 \mic\ flux density combined with the coefficients of the 100 \mic\ calibration lead to a much better estimate of the L$_{TIR}$ (L$_{TIR~calibrated}$ / L$_{TIR~modelled}$ = 0.76).

{\it PACS 160 - }Like the 70 \mic\ calibration, the goodness of our 160 \mic\ calibration depends on the 70/100 colour. The hot SED of NGC~1377 or NGC~5408 can explain why our 160 \mic\ monochromatic calibration underestimates the integrated L$_{TIR}$ for these objects. DDO~053 is also an outlier for the PACS 160 \mic\ calibration. DDO~053 shows a very low integrated PACS 160 \mic\ flux compared to the PACS 100 and SPIRE 250 \mic\ fluxes \citep[see][]{Dale2012}. Using the MIPS 160 \mic\ flux of this galaxy estimated in \citet[]{Dale2007} (0.5 Jy in lieu of 0.25 Jy for PACS 160 \mic) would lead to a better agreement of the calibrated TIR value with the modelled integrated L$_{TIR}$. 

{\it SPIRE 250 - }We finally observe a larger uncertainty in the predictions derived using the 250 \mic\ monochromatic calibration (still within a factor of 2 for most of our objects) with, as expected, a strong correlation of the goodness of the calibration with the 70/100 colour similar to that of the 160 \mic\ calibration. Our calibration will thus  probably underestimate the L$_{TIR}$ for hot objects and overestimate the L$_{TIR}$ for cold objects. \\

We conclude that PACS 100 \mic\ luminosities can be safely used as monochromatic estimators of the L$_{TIR}$, even if it should be used with caution for strong starburst environments. We note the small errors on the parameters derived and small scatter around the relation. Our 70 and 160 \mic\ calibrations also lead to reasonably good estimates of the L$_{TIR}$ (within 50$\%$ for most resolved elements or entire galaxies). Here again we caution their use for very cold or very hot SEDs. Calibrations using the 250 \mic\ luminosity alone have larger uncertainties. We observe that predictions deviate from modelled values if f$_{70}$$\ge$f$_{100}$ or if f$_{70}$/f$_{100}$$<$0.4. Calibrations using combined fluxes should be favored if more than one PACS/SPIRE band is available (see Section 6).\\

   \begin{figure*}
   \centering
    \includegraphics[width=17cm]{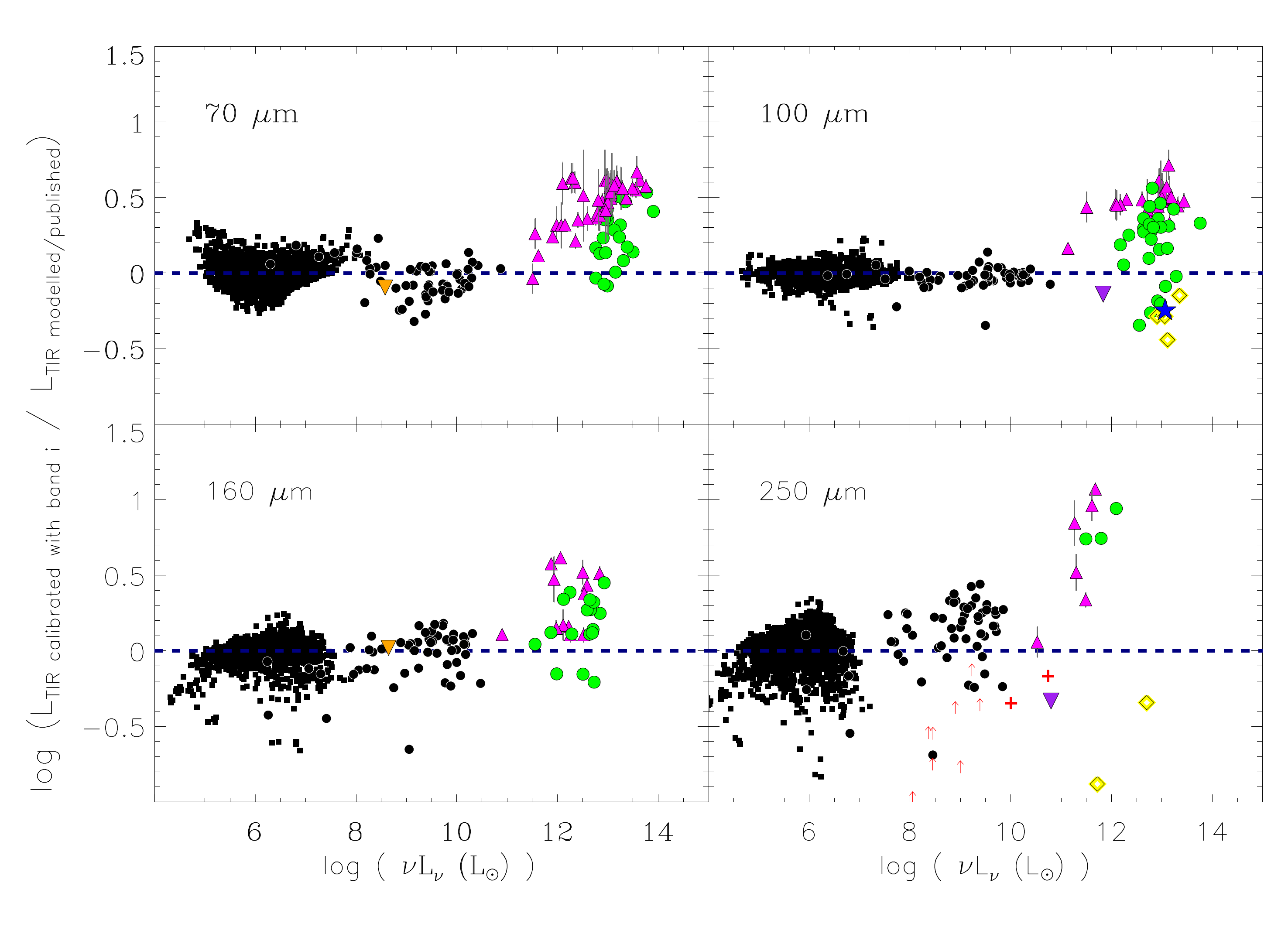} \\
        \vspace{-20pt}
    \caption{Comparison between L$_{TIR}$ (modelled in this paper or published in the literature) of various objects and their L$_{TIR}$ predicted from our monochromatic calibrations (and their rest-frame 70, 100, 160 and 250 \mic\ luminosities). Black points indicate the resolved elements of the KINGFISH galaxies. For each galaxy, we sort and average resolved elements 20-by-20 for clarity. We overplot integrated measures of our KINGFISH sample with black circles.  Orange upside down triangles indicate the environment of the gamma-ray burst GRB~980425 from \citet{LeFloch2012}. Magenta triangles indicate submillimeter galaxies from \citet{Magnelli2012}. We also overlay the uncertainties on their modelled L$_{TIR}$. Green circles indicate z$\sim$2 dust obscured galaxies from \citet{Melbourne2012}. Yellow diamonds indicate WISE-Selected Hyper-luminous Galaxies from \citet{Wu2012}. The blue star shows an Hyper-Luminous Infrared Galaxy from \citet{Eisenhardt2012}. Red crosses and upward arrows (lower limits) indicate local active galaxies from Pereira-Santaella et al (in prep). The upside down purple triangle finally shows Arp~220 studied \citet{Rangwala2011}. }
    \label{Comp_other}
    \end{figure*}

In the present paper, we estimate calibrations of the L$_{TIR}$ from observations in the [8-250 \mic] band, but calibrations at longer wavebands would be useful to understand galaxy properties of nearby or high-redshift objects observed in submm and millimeter, from ground-based telescopes in particular (SCUBA-2, LABOCA, ALMA). In Appendix A, we present and discuss monochromatic calibrations derived for longer wavelengths from {\it 1)} 350 and 500 \hersc\ data {\it 2)}  850 and 1000 \mic\ model predictions.

\subsection{Predictions for near and high-redshift sources}

We now analyse L$_{TIR}$ predictions  for a wider range of environments using our monochromatic calibrations. 
Figure~\ref{Comp_other} compares the L$_{TIR}$ modelled in this paper or published in the literature of various objects with their L$_{TIR}$ predicted from our monochromatic calibrations (and their rest-frame 70, 100, 160 and 250 \mic\ luminosities). Black points indicate the resolved elements of the KINGFISH galaxies, with integrated values overlaid with black circles. The horizontal dashed line indicates when predictions from our calibrations match modelled or published L$_{TIR}$. We add various nearby and high-redshift objects for comparison.

\subsubsection{Nearby sources}

\citet{LeFloch2012} characterise the close environment of the gamma-ray burst GRB~980425, located 36 Mpc away. They provide 70 and 160 \mic\ fluxes for the GRB host (respectively 230 and 615 mJy). Using our monochromatic calibrations, we predict an integrated L$_{TIR}$ of 8.17$^{\pm0.6}$ $\times$ 10$^8$ and  1.07$^{\pm0.1}$ $\times$ 10$^9$ \lsun\ from the 70 and 160 \mic\ fluxes respectively. Those estimates are in very good agreement with the integrated L$_{TIR}$ they derive using standard empirical libraries of galaxy templates (1.02 $\times$ 10$^9$ \lsun). We overlay the environment of GRB~980425 in Fig.~\ref{Comp_other} (70 and 160 \mic\ panels, upside-down orange triangles).

\citet{Rangwala2011} present SPIRE-FTS \hersc\ observations of the nearby ultra-luminous infrared galaxy Arp~220, located at 77 Mpc, and provide a SPIRE 250 \mic\ continuum flux of 30.1 Jy for this galaxy. Using a single-temperature modified blackbody to fit the global SED from 15 \mic\ up to SPIRE bands (including ISOPHOT, IRAS, ISO-LWS, SPIRE and SCUBA data), they estimate a L$_{TIR}$ of 1.77 $\times$ 10$^{12}$ \lsun. From our 250 \mic\ monochromatic calibration, we estimate the integrated L$_{TIR}$ to be 7.95$^{\pm1.3}$ $\times$ 10$^{11}$ \lsun, thus a factor of two lower than their modelled value (see Figure~\ref{Comp_other}, 250 \mic\ panel, upside-down purple triangle). Our 250 \mic\ calibration is expected to under-predict the integrated L$_{TIR}$ for hot objects (Arp~220 has an IRAS 60-to-100 ratio $\sim$ 1)\footnote{IRAS flux densities taken from the NED database}, which could explain the difference. Assuming the 100 \mic\ flux is 130 Jy (from ISO-LWS), we derive a L$_{TIR}$ of 1.23$^{\pm0.6}$ $\times$ 10$^{12}$ \lsun\ for this object, thus $\sim$70$\%$ of value derived by \citet{Rangwala2011}. This highlights the good predictions from our 100 \mic\ monochromatic calibration, especially for unusual object such as Arp~220.

Finally, Pereira-Santaella et al. (in prep) present SPIRE-FTS \hersc\ observations of local active galaxies and provide SPIRE 250 \mic\ continuum flux densities for those sources. They also provide integrated IR luminosities taken from \citet{Sanders2003} and rescaled to the distance they adopt. 
For most of the galaxies of their sample, the SPIRE fluxes only include the nuclear far-IR emission while the IR luminosities are derived from the integrated IRAS fluxes. Comparisons should be safe for UGC05101 and NGC~7130. We consider the other 250 \mic\ fluxes as lower limits. We overplot these objects with red upward arrows in Figure~\ref{Comp_other} (250 \mic\ panel) and UGC05101 and NGC~7130  with red crosses. Our predictions using the 250 \mic\ calibration match the IR luminosities published in their study for these two objects within a factor of 2. 

\subsubsection{High redshift sources}

We combine these nearby results with a selection of z$\sim$1-3 objects taken from published catalogs. We derive distances from the redshifts provided in the catalogs, adopting an H$_o$ = 70 km~s$^{-1}$Mpc$^{-1}$, $\Omega$$_M$ = 0.3, $\Omega$$_{\Lambda}$ = 0.7 cosmology. In each catalog, we select objects observed at rest-frames close to 70, 100, 160 or 250 \mic, within a margin of $\pm$10 \mic, and compare the published integrated L$_{TIR}$ with those derived from our calibrations. For each sample, we quote the methods used to derive the published L$_{TIR}$ to help the reader assess the level of confidence of these modelled values and try to understand how much of the scatter in Fig.~\ref{Comp_other} could be driven by biases in our predictions or, on the contrary, linked with uncertainties in the published values we quote (derived with different methods and, for some of them, with limited data).

\vspace{4pt}
We first select z$\sim$2 dust obscured galaxies from the catalog of \citet{Melbourne2012} (Fig.~\ref{Comp_other}, green circles). Those galaxies are observed with \hersc\ at 250, 350, 500 \mic. They estimate the integrated L$_{TIR}$ by interpolating between the mid to far-IR flux densities and extrapolate the long wavelength tail contribution to the L$_{TIR}$ using a modified blackbody curve, assuming a dust emissivity index of 1.5. Even though they do not completely sample the submm slope of their high-redshift sources, they note that their estimated L$_{TIR}$ are relatively robust, with less than a 5$\%$ change in L$_{TIR}$ if the temperature of the modified blackbody they used varies by 25$\%$. In Fig.~\ref{Comp_other}, we observe that our 70, 100 and 160 \mic\ predictions are consistent with the integrated L$_{TIR}$ estimated by \citet{Melbourne2012} within a factor of 3. We remind the reader that we select objects observed at rest-frames close but not exactly equal to 70, 100, 160 or 250 \mic. This results in some cases in an under or over-prediction of the monochromatic flux used in the calibration and thus of the predicted L$_{TIR}$. Restricting our selection to a margin of $\pm$5$\%$ around our reference wavebands remove some of the objects showing high discrepancies. This does not, however, fully explain the vertical scatter we obtain. The galaxies for which our 100 \mic\ calibration under-predicts the L$_{TIR}$ are all classified as Mrk231-like objects (namely AGN-dominated ULIRGs) in \citet{Melbourne2012}, with temperature superior to 40K. The discrepancy between the modelled and our predicted L$_{TIR}$ can thus be attributed to the fact that these objects have SEDs that peak at much shorter wavelength than 100 \mic\ (Section 5.4). Stronger discrepancies are observed when using the 250 \mic\ calibration. For the three z$\sim$1 galaxies detected at 500 \mic\ (so rest-frame 250 \mic), namely J143052.8+342933, J143313.4+333510 and J143334.0+342518, our 250 \mic\ calibration predicts a L$_{TIR}$ higher by a factor of 5.5 to 8.7 compared to those derived in \citet{Melbourne2012}. Those galaxies are among the coldest objects of the sample (T$<$26K). Our 250 \mic\ calibration probably over-predicts the L$_{TIR}$ in those objects.  

\vspace{4pt}
We also add submillimeter galaxies (SMGs) from \citet{Magnelli2012} observed with \hersc\ at 160, 250, 350 and 500 \mic\ (magenta triangles). The published integrated L$_{TIR}$ are derived using a power-law temperature distribution model. We include the lensed-SMGs in our analysis. \hersc\ flux densities of these objects have been de-magnified using the magnification factors tabulated in the Table 11 of \citet{Magnelli2012}. Uncertainties on the modelled L$_{TIR}$ are provided for this sample and were added to Fig.~\ref{Comp_other}. We observe that the L$_{TIR}$ of submillimeter galaxies estimated in \citet{Magnelli2012} are systematically lower than our predictions, whatever the rest-frame used. Our monochromatic 70, 100 and 160 \mic\ calibrations predict a L$_{TIR}$ higher by a factor of 3 at most compared to those derived by \citet{Magnelli2012}, our 250 \mic\ calibration by up to an order of magnitude. For the top left panel (70\mic), deviations from our predictions seem to increase with luminosities. As reminded in \citet{Magnelli2012}, the SMG population is very heterogenous and biased towards cold dust temperatures compared to the entire infrared galaxy population. This can partly explain why the L$_{TIR}$ predictions using our 160 or 250 \mic\ calibrations are higher than the modelled values. If the SED profiles of SMGs were similar in the shape than those of local objects but simply shifted to shorter wavelengths, the inverse trend would be observed at 70 \mic, namely that our calibration would under-predict the L$_{TIR}$. This is not what is observed in Fig.~\ref{Comp_other}. We finally note that part of the TIR luminosity could be missing in the modelled L$_{TIR}$ values of \citet{Magnelli2012} due to the fact that their fits do not include a detailed modelling of the rest-frame [8 to 70 \mic] spectrum. Indeed, SMGs often show broad emission features from PAHs and starburst activity could dominate the L$_{TIR}$ in those objects \citep[e.g.][]{Menendez2009}.

\vspace{4pt}
We finally add WISE-selected hyper-luminous galaxies from \citet{Wu2012} observed with CSO/SHARC-II at 350 and 450 \mic\ and CSO/Bolocam at 1.1 mm (yellow diamonds) and a z=2.452 hyper-luminous galaxy from \citet{Eisenhardt2012} (WISE 1814+3412) observed with CSO/SHARC-II at 350 \mic\ (blue star, 100 \mic\ panel). The same SED fitting technique is applied in the two studies to derive integrated TIR luminosities: they use a single modified blackbody model combined with power-laws to connect the mid-IR to mm SED points and a modified blackbody component (with a dust emissivity index of 1.5) to fit the longer wavelengths. Their integrated L$_{TIR}$ can thus be considered as low limits for the total luminosities. The modelled L$_{TIR}$ values from \citet{Wu2012} and \citet{Eisenhardt2012} match with our 100 \mic\ calibration, here again within a factor of 3. \citet{Wu2012} note that their objects could host highly obscured AGNs heating their dust cocoon to very high temperatures. This could explain why our 100 \mic\ calibration under-predict the L$_{TIR}$. We also note that in the fourth panel of Fig.~\ref{Comp_other}, our predicted L$_{TIR}$ for the galaxy W0149+2350 \citep[from][]{Wu2012} is lower than their modelled value by a factor of 7.6 while our prediction using the 100 \mic\ calibration match their modelled value within 30$\%$. This galaxy has a redshift of z=3.228 and was observed with the Submillimeter Array (SMA) at 1.3 mm. The 1.1mm flux density quoted in \citet{Wu2012} is a conversion of the SMA measurement assuming an emissivity index $\beta$=1.5 and 2, then taking the average of the two values. Discrepancy could thus be linked with the uncertainty on the predicted 1.1mm flux for this object.

We conclude that while we observe a wide spread for objects with integrated L$_{TIR}$ $>$ 10$^{11}$ \lsun\ on the 250 \mic\ panel, predictions of the PACS monochromatic calibrations match modelled L$_{TIR}$ from resolved elements of nearby galaxies to global objects at further redshift over a surprisingly large luminosity range (from 10$^4$ to 10$^{14}$ \lsun). This reinforces the usefulness of PACS wavelengths as reliable monochromatic calibrators for the L$_{TIR}$ of nearby galaxies. This also means that reliable estimates of the L$_{TIR}$ of high-redshift objects can be obtained using the SPIRE filters and our 70, 100 and 160 \mic\ calibrations. For instance, SPIRE 250 \mic\ observations combined with our 100 \mic\ calibration coefficients could be used to study the peak of the star formation history at z=1.5. \\

\section{Combining mid to far-IR bands}

\citet{Dale_Helou_2002} derived a bolometric relation to estimate the integrated L$_{TIR}$ from a combination of MIPS filters. This relation is applicable to a wide range of galaxy luminosities. With \hersc\ observations, our wavelength coverage now goes longward of 160 \mic. This enables us to better sample the submm slope of the SED and reduce the uncertainties on integrated L$_{TIR}$ linked with the presence of cold dust not detected by previous MIPS 160 \mic\ observations. 
In this section, we thus aim to make the most of the good resolution of \hersc\ to perform a similar multi-wavelength empirical calibration on a resolved basis. We derive calibrations linking combined far-IR brightnesses (Section 6.1) or luminosities (Section 6.2) to the TIR surface brightnesses or luminosities and study their dependence and biases.

   \begin{figure}
   \centering
       \vspace{-30pt}
    \includegraphics[width=9cm]{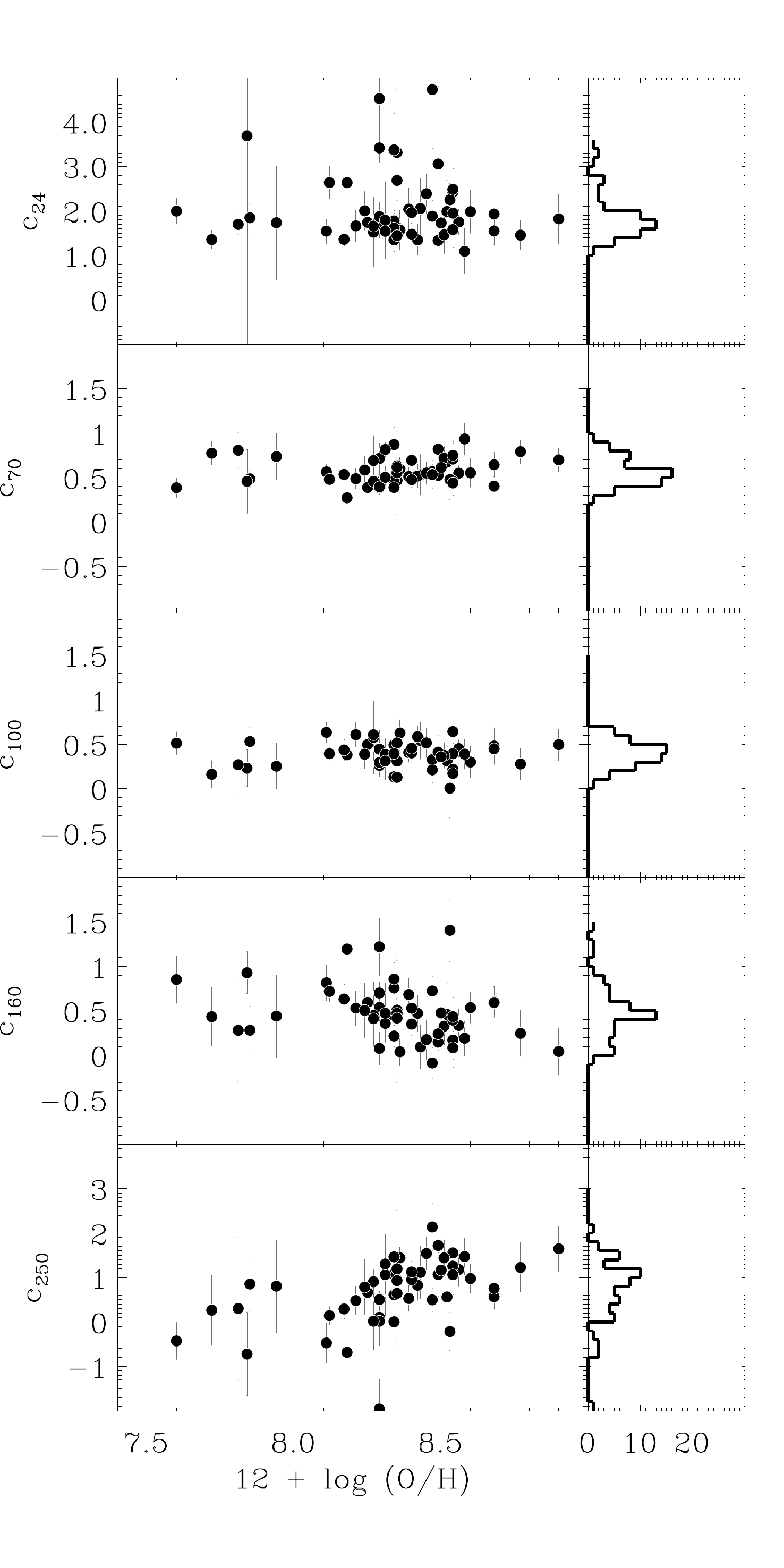} \\
    \vspace{-15pt}
    \caption{Calibration coefficients to derive S$_{TIR}$ from a combination of \spitz\ and \hersc\ brightnesses (24, 70, 100, 160 + 250 \mic) plotted as a function of metallicity expressed as 12+log(O/H). The relation is : S$_{TIR}$ = $\sum$ c$_i$~S$_i$ with S$_{TIR}$ and S$_i$ in W~kpc$^{-2}$. We join an histogram of the coefficients on the right hand side of each plot. }
    \label{Coeffs_combined_fits}
    \end{figure}

 \begin{table*}
\caption{\large Calibration coefficients to predict the TIR brightness/luminosity from combined \spitz\ and \hersc\ brightnesses/luminosities. }
\label{Coeff_total_Comb}
 \centering
  \begin{tabular}{cccccccccc}
\hline
\hline
&\\
{\large Surface brightnesses} & \\
&\\
\cline{1-8}
&c$_{24}$	   &	c$_{70}$     	&	c$_{100}$   	&	c$_{160}$    	&	c$_{250}$ & R$^2$ & CV(RMSE)\\
\cline{2-8}
&\\
{\large S$_{TIR}$= $\Sigma$ c$_i$~S$_i$}& 
3.925$\pm$ 0.284 & 1.551$\pm$ 0.059 & - & - & - &0.86&1.29\\
S$_{TIR}$, S$_{i}$ in W~kpc$^{-2}$ & 
2.421$\pm$ 0.086 & - & 1.410$\pm$ 0.014 & - &  - &0.99&0.29\\
(Eq.~\ref{eq4})& 3.854$\pm$ 0.088 &  - &  - & 1.373$\pm$ 0.015 &  - &0.95&0.79\\
& 5.179$\pm$ 0.132 & - &  - &  - & 3.196$\pm$ 0.059 &0.93&0.90\\
& - & 0.458$\pm$ 0.034 & 1.444$\pm$ 0.023 &  - &  -  &0.98&0.47\\
& - & 0.999$\pm$ 0.023 &  - & 1.226$\pm$ 0.017 & - &0.97&0.62\\
&- & 1.306$\pm$ 0.021 & - & - & 2.752$\pm$ 0.044 &0.98&0.53\\
& - &  - & 1.239$\pm$ 0.025 & 0.620$\pm$ 0.028 &  - &0.92&0.98\\
&- & - & 1.403$\pm$ 0.016 &  - & 1.242$\pm$ 0.048 &0.92&0.98\\
& - &  - &  - & 2.342$\pm$ 0.040 & -0.944$\pm$ 0.111 &0.74&1.73\\
& 2.162$\pm$ 0.113 & 0.185$\pm$ 0.035 & 1.319$\pm$ 0.016 & - & - &0.99&0.26\\
& 2.126$\pm$ 0.093 & 0.670$\pm$ 0.028 &- & 1.134$\pm$ 0.010 & - &0.99&0.37\\
& 2.317$\pm$ 0.114 & 0.922$\pm$ 0.028 & - & - & 2.525$\pm$ 0.030&0.99&0.29\\
& 2.708$\pm$ 0.071 &  - & 0.734$\pm$ 0.022 &  0.739$\pm$ 0.018 & - &0.97&0.55\\
& 2.561$\pm$ 0.072 & - & 0.993$\pm$ 0.017 &  - & 1.338$\pm$ 0.032 &0.98&0.53\\
& 3.826$\pm$ 0.089 & - & - & 1.460$\pm$ 0.032 &  -0.237$\pm$ 0.067&0.95&0.77\\
&  - & 0.789$\pm$ 0.032 & 0.387$\pm$ 0.029 & 0.960$\pm$ 0.020 & - &0.97&0.62\\
& - & 0.688$\pm$ 0.028 & 0.795$\pm$ 0.022 & - &  1.634$\pm$ 0.043&0.97&0.62\\
&  - & 1.018$\pm$ 0.021 & - & 1.068$\pm$ 0.035 &  0.402$\pm$ 0.097&0.97&0.63\\
& - & - & 1.363$\pm$ 0.031 &  0.097$\pm$ 0.065 &  1.090$\pm$ 0.110&0.91&0.99\\
& 2.051$\pm$ 0.089 & 0.521$\pm$ 0.030 & 0.294$\pm$ 0.019 & 0.934$\pm$ 0.014 & -  &0.99&0.38\\
& 1.983$\pm$ 0.084 & 0.427$\pm$ 0.026 & 0.708$\pm$ 0.017 & - & 1.561$\pm$ 0.030&0.99&0.38\\
& 2.119$\pm$ 0.090 & 0.688$\pm$ 0.025 &- & 0.995$\pm$ 0.027 & 0.354$\pm$ 0.068&0.99&0.38\\
& 2.643$\pm$ 0.069 &- & 0.836$\pm$ 0.024 & 0.357$\pm$ 0.042 & 0.791$\pm$ 0.072&0.97&0.57\\
& - & 0.767$\pm$ 0.032 & 0.503$\pm$ 0.038 & 0.558$\pm$ 0.059 & 0.814$\pm$ 0.111&0.96&0.64\\
& 2.013$\pm$ 0.081 & 0.508$\pm$ 0.029 & 0.393$\pm$ 0.025 & 0.599$\pm$ 0.042 & 0.680$\pm$ 0.078&0.99&0.40\\
&\\												
\hline
\hline					
&\\							
{\large Luminosities} & \\
&\\
\cline{1-8}
&c$_{24}$	   &	c$_{70}$     	&	c$_{100}$   	&	c$_{160}$    	&	c$_{250}$ & R$^2$ & CV(RMSE)\\
\cline{2-8}
&\\
{\large L$_{TIR}$= $\Sigma$ c$_i$~$\nu$L$_{\nu}$(i)}& 
3.980$\pm$ 0.283 & 1.553$\pm$ 0.058 & - & - & - &0.84&2.78\\
L$_{TIR}$, $\nu$L$_{\nu}$(i) in \lsun\ & 
2.453$\pm$ 0.085 & - & 1.407$\pm$ 0.013 & - &  - &0.99&0.68\\
(Eq.~\ref{eq5})& 3.901$\pm$ 0.090 &  - &  - & 1.365$\pm$ 0.015 &  - &0.90&2.12\\
 & 5.288$\pm$ 0.134 & - &  - &  - & 3.150$\pm$ 0.060 &0.88&2.41\\
&  - & 0.463$\pm$ 0.035 & 1.442$\pm$ 0.023 &  - &  -  &0.99&0.71\\
&  - & 1.010$\pm$ 0.023 &  - & 1.218$\pm$ 0.017 & - &0.98&0.94\\
& - & 1.325$\pm$ 0.020 & - & - & 2.717$\pm$ 0.042 &0.99&0.70\\
&  - &  - & 1.238$\pm$ 0.024 & 0.620$\pm$ 0.027 &  - &0.93&1.85\\
& - & - & 1.403$\pm$ 0.016 &  - & 1.242$\pm$ 0.048 &0.93&1.84\\
&  - &  - &  - & 2.370$\pm$ 0.039 & -1.029$\pm$ 0.108 &0.73&3.58\\
&  2.192$\pm$ 0.114 & 0.187$\pm$ 0.035 & 1.314$\pm$ 0.016 & - & - &0.99&0.56\\
&  2.133$\pm$ 0.095 & 0.681$\pm$ 0.028 &- & 1.125$\pm$ 0.010 & - &0.98&0.86\\
&  2.333$\pm$ 0.113 & 0.938$\pm$ 0.027 & - & - & 2.490$\pm$ 0.029&0.99&0.66\\
&  2.739$\pm$ 0.070 &  - & 0.732$\pm$ 0.021 &  0.736$\pm$ 0.017 & - &0.96&1.42\\
&  2.594$\pm$ 0.068 & - & 0.990$\pm$ 0.016 &  - & 1.334$\pm$ 0.031 &0.96&1.36\\
&  3.868$\pm$ 0.091 & - & - & 1.458$\pm$ 0.031 &  -0.252$\pm$ 0.065&0.91&2.08\\
&   - & 0.808$\pm$ 0.031 & 0.367$\pm$ 0.026 & 0.968$\pm$ 0.018 & - &0.98&0.95\\
&  - & 0.705$\pm$ 0.027 & 0.784$\pm$ 0.020 & - &  1.639$\pm$ 0.042&0.98&0.92\\
&   - & 1.032$\pm$ 0.020 & - & 1.051$\pm$ 0.033 &  0.423$\pm$ 0.092&0.98&0.95\\
&  - & - & 1.379$\pm$ 0.025 &  0.058$\pm$ 0.049 &  1.150$\pm$ 0.092&0.93&1.86\\
&  2.064$\pm$ 0.091 & 0.539$\pm$ 0.030 & 0.277$\pm$ 0.017 & 0.938$\pm$ 0.012 & -  &0.98&0.86\\
&  1.999$\pm$ 0.083 & 0.443$\pm$ 0.025 & 0.696$\pm$ 0.014 & - & 1.563$\pm$ 0.028&0.99&0.84\\
&  2.127$\pm$ 0.092 & 0.702$\pm$ 0.024 &- & 0.974$\pm$ 0.024 & 0.382$\pm$ 0.063&0.98&0.87\\
&  2.667$\pm$ 0.067 &- & 0.848$\pm$ 0.019 & 0.319$\pm$ 0.031 & 0.847$\pm$ 0.060&0.96&1.44\\
& - & 0.783$\pm$ 0.030 & 0.497$\pm$ 0.033 & 0.540$\pm$ 0.051 & 0.852$\pm$ 0.103&0.98&0.97\\
&  2.023$\pm$ 0.082 & 0.523$\pm$ 0.028 & 0.390$\pm$ 0.021 & 0.577$\pm$ 0.036 & 0.721$\pm$ 0.070&0.98&0.89\\
&&&&&&\\
\cline{2-8}
&&&&&&\\
\end{tabular}
\end{table*} 

\subsection{Calibration from combined brightnesses}

Using a combination of resolved brightnesses at MIPS 24 \mic, PACS 70, 100 and 160 \mic\ and SPIRE 250 \mic, we calculate calibration coefficients c$_i$ for each galaxy, such as:
\begin{eqnarray}
S_{TIR} = \sum c_i~ S_i 
\label{eq4}
\end{eqnarray}

\noindent where S$_{TIR}$ refers to the TIR surface brightness and S$_i$ the brightness in a given \spitz\ or \hersc\ band {\it i}. Here and for the rest of Section 6, we derive calibrations in linear space. We use a multiple linear regression fit (function {\it mregress}, a variant from the IDL function {\it regress} by Ph. Prugniel, 2008) combined with our ``jack-knife" technique applied on the resolved elements of our galaxies to conservatively estimate the calibration coefficients and their uncertainties. We list the coefficients obtained for individual galaxies in the Appendix in Table~\ref{Coeff_Indiv_combined}. We indicate the calibration coefficients obtained for the whole gathering the resolved ISM elements of the whole sample in Table~\ref{Coeff_total_Comb}. 

To quantify the scatter between the modelled and the predicted brightnesses for each combination, we also provide indicators of the goodness-of-fit in the last two columns: the coefficient of determination R$^2$ and the coefficient of variation of the root-mean-square error CV(RMSE). R$^2$ ranges between 0 and 1 and indicates the proportion of variability of the resolved TIR brightnesses accounted for by our calibration. For instance, R$^2$=0.90 means that our calibration accounts for 90$\%$ of the total variation of our TIR brightnesses. CV(RMSE) is the standard deviation (measuring the differences between the TIR brightnesses predicted by our calibrations and the TIR brightnesses we obtained using the [DL07] modelling) normalised to the mean values of our resolved TIR brightnesses. Therefore, the lower the CV(RMSE), the better. These two quantities are defined as:
\begin{equation}
R^2=1-\frac{\Sigma~(M_i-P_i)^2}{\Sigma(M_i-{\overline M_i})^2}
\end{equation}

\begin{equation}
CV(RMSE)=\frac{RMSE}{\overline M_i}=\frac{1}{\overline M_i}\sqrt{\frac{\Sigma~(M_i-P_i)^2}{n}}
\end{equation}

\noindent with P$_i$ the predicted surface brightnesses, M$_i$ the resolved TIR surface brightnesses modelled using [DL07], $\overline{M_i}$ the mean of the modelled brightnesses and n the number of ISM elements. \\

We plot the coefficients c$_i$  derived using Eq.~\ref{eq4} as a function of metallicity in Fig.~\ref{Coeffs_combined_fits}. Calibration coefficients weighting the PACS 70 and 100 \mic\ brightnesses are similar though the sample as suggested by the peaked distributions of the histograms. We nevertheless observe a larger distribution in the 160 and 250 \mic\ coefficients. No strong trend is observed with metallicity for the 24, 70, 100 and 160 \mic\ coefficients. However, the c$_{250}$ coefficients are quite low for objects with 12+log(O/H) $<$ 8.2 (median=-0.01), probably linked with he fact that submm emission has a smaller contribution to the total IR budget in low-metallicity galaxies compared to more metal-rich objects. \\

\subsection{Calibration from combined luminosities}

Similarly to Section 5.3, we gather the resolved elements of the complete sample and derive calibration coefficients c$_i$ similar to Eq.~\ref{eq4} but linking the L$_{TIR}$ to different combinations of \spitz\ and \hersc\ luminosities. Eq.~\ref{eq4} can thus be re-written as:
\begin{eqnarray}
L_{TIR} = \sum c_i~\nu L_{\nu}(i) 
\label{eq5}
\end{eqnarray}

\noindent where L$_{TIR}$ now refers to the TIR luminosity and $\nu$L$_{\nu}$(i) the resolved luminosities in a given \spitz\ or \hersc\ band {\it i}. L$_{TIR}$ and the different $\nu$L$_{\nu}$(i) are in \lsun. We list the calibration coefficients obtained for various combinations of \spitz\ and \hersc\ bands in Table~\ref{Coeff_total_Comb} with respective R$^2$ and CV(RMSE) coefficients.  \\

   \begin{figure*}
   \centering
   \hspace{-10pt}
    \includegraphics[width=18cm]{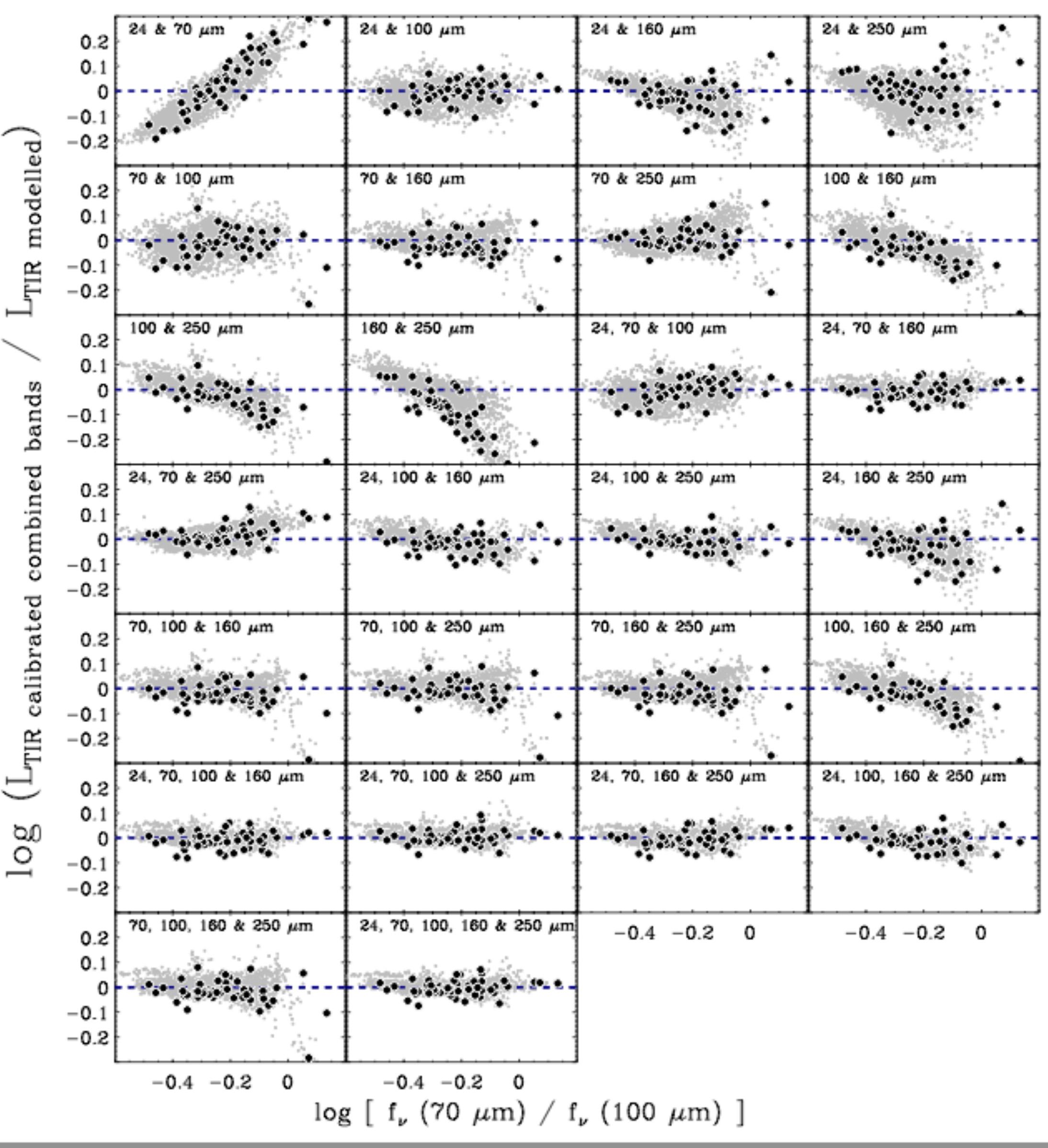} \\
    \caption{Comparison between the resolved (grey points) and integrated (black circles) L$_{TIR}$ obtained with a combination of monochromatic fluxes (among 24, 70, 100, 160 and 250 \mic) and the properly modelled L$_{TIR}$ as a function of the f$_{\nu}$(70\mic)/f$_{\nu}$(100\mic) far-IR colour. The calibration relations are of the form L$_{TIR}$= $\Sigma$ c$_i$~$\nu$L$_{\nu}$(i) with L$_{TIR}$ and $\nu$L$_{\nu}$(i) in \lsun. We indicate the 2, 3, 4 or 5 bands used to obtain the calibrated TIR luminosities in each panel. We report the various calibration coefficients in Table~\ref{Coeff_total_Comb}. We sort and average the resolved elements (in grey) 15-by-15 for clarity. Note that the Y-axis range is different from that of Fig.~\ref{Comp_mono}. }
    \label{Comp_combined}
    \end{figure*}

   \subsection{Validity of the calibrations for KINGFISH galaxies}
   
Following the same scheme as Section 5, we test the ability of our combined calibrations (derived using all the resolved elements of KINGFISH galaxies) to predict resolved and integrated L$_{TIR}$ and analyse their intrinsic biases. We thus compare the resolved and integrated L$_{TIR}$ of the KINGFISH sample calibrated from various combinations of the 24, 70, 100, 160 or 250 \mic\ luminosities versus the L$_{TIR}$ derived with a proper SED modelling. For integrated values, we use the integrated flux densities of \citet{Dale2007,Dale2012} to predict integrated L$_{TIR}$ and compare them to the modelled integrated L$_{TIR~S500}$ (Table~\ref{Global_TIR}). Figure~\ref{Comp_combined} illustrates these comparisons as a function of the 70/100 colour. Results are identical for integrated or resolved L$_{TIR}$.

We do observe a correlation between the goodness of some calibrations ( ``24+70", ``160+250", ``100+160+250" for instance) and the 70/100 colour. Nevertheless, our combined calibrations lead in most cases to better estimates of the L$_{TIR}$ than monochromatic calibrations. Combined calibrations are especially much more reliable than our 70, 160 or 250 \mic\ monochromatic calibrations for galaxies whose 70-to-100 flux density ratios are below 0.4 or above 0.8. 
We note that combining MIPS 24, PACS 70 and PACS 160 \mic\ data alone, we obtain $c_{24}$~=~2.126, c$_{70}$~=~0.670 and c$_{160}$~=~1.134. We remind the reader that the coefficients obtained by \citet{Dale_Helou_2002} and calibrated from \spitz\ 24, 70 and 160 \mic\ fluxes are 1.559, 0.7686 and 1.347 for c$_{24}$, c$_{70}$ and c$_{160}$ \mic\ respectively. Global L$_{TIR}$ derived using the \citet{Dale_Helou_2002} calibration differ from our modelled L$_{TIR~S500}$ by $\sim$26$\%$ at most (for IC~2574) with a median of the differences of $\sim$6$\%$. The integrated values obtained with the ``24+70+160" \mic\ calibration presented in this paper differ from the modelled integrated L$_{TIR}$ by $\sim$ 23$\%$ at most (for NGC~1512), with a median of the differences of $\sim$4$\%$. Both predictions are thus very close at global scale (integrated luminosities).

The calibration performed using the complete coverage of the FIR / submm emission (24, 70, 100, 160 and 250 \mic\ data) leads to a reliable approximation of the modelled L$_{TIR}$ (low CV(RMSE) value) : using this calibration, all the estimated luminosities reside within 19$\%$ of the modelled L$_{TIR}$, with a median of the differences of 3.5$\%$. Using 4 of these wavelengths leads to similarly good results. Any L$_{TIR}$ predicted from a fewer number of fluxes should contain the 100 \mic\
flux, or a combination of 70+160 to lead to L$_{TIR}$ predictions reliable within 25$\%$. We note that calibrations including 24 \mic\ data lead to better estimates of the L$_{TIR}$ for galaxies showing high 70/100 colour than calibrations that do not include this wavelength. We thus advice using the 24 \mic\ flux (if available) in the L$_{TIR}$ predictions for these environments.

\section{Conclusions}

\begin{itemize}

\item We investigate how SPIRE wavelengths influence the determination of the L$_{TIR}$ and conclude that using data up to 250 \mic\ leads to L$_{TIR}$ values that are in very good agreement with that obtained with a complete SED modelling of the dust thermal emission (within 10$\%$ for most of our resolved elements).  \\

\item The [70-160] band contains 30 to 50$\%$ of the IR emission. We observe an overall shift in the SED to shorter wavelengths with decreasing metallicity. Indeed, the [24-70 \mic] fraction increases for warmer sources (often found in low-metallicity objects) while the [160-1100 \mic] fraction accounts for only a few $\%$ for low-metallicity galaxies (to up to 25$\%$ of the total infrared luminosity budget for metal-rich environments). The [3-24 \mic] fraction accounts for $\sim$20$\%$ of the L$_{TIR}$, with a significant scatter from one environment to another. \\

\item We study the correlation between TIR and monochromatic \spitz\ and \hersc\ surface brightnesses/luminosities and derive calibration coefficients to quantify these correlations. For most of the galaxies of our sample, the three PACS bands can be used as reliable monochromatic estimators of L$_{TIR}$, with slopes on average $<$1.0, $\sim$1.0 and $>$1.0 for 70, 100 and 160 \mic\ respectively. We also observe a strong correlation between the SPIRE 250 \mic\ and L$_{TIR}$, although with more scatter than the PACS relations. We estimate calibration coefficients for waveband beyond 250 \mic\ in Appendix A. \\

\item  We conclude that the 100 \mic\ band is the best band to use as a monochromatic estimator (scatter of 0.05 dex) of L$_{TIR}$. \\

\item We show that the calibrations at 70, 100 and 160 \mic\ reproduce modelled L$_{TIR}$ over a very large luminosity range, from nearby galaxies to galaxies at z$\sim$1-3. L$_{TIR}$ values are reproduced with larger uncertainties from 250 \mic\ fluxes. We nevertheless caution the use of our 70, 160 and 250 \mic\ calibration for strong star-bursting environments and for objects showing cold dust temperatures.\\

\item We finally derive calibration coefficients to derive TIR surface brightnesses/luminosities from a combination of \spitz\ and \hersc\ surface brightnesses/fluxes. These calibrations lead to better estimates of L$_{TIR}$ than monochromatic calibrations and show much smaller biases.  We update the widely used L$_{TIR}$ calibration of \citet{Dale_Helou_2002} using \hersc/PACS 70 and 160 \mic\ data in lieu of \spitz/MIPS data at the same wavelengths. The two calibrations lead to similar estimates (with similar uncertainties) for integrated luminosities. As expected, the calibration using the complete sampling of the FIR/submm emission (data at 24, 70, 100, 160 and 250 \mic) leads to a reliable estimation of the L$_{TIR}$ but using 4 of those wavelengths leads to similarly satisfying predictions. We note that including 24 \mic\ data in the calibration is essential to properly estimate L$_{TIR}$ in strongly star-forming environments (high 70/100 colour).

\end{itemize}


\section*{Acknowledgments}
We would like to thank the referee for his/her careful reading of the paper and a helpful report. PACS has been developed by MPE (Germany); UVIE (Austria); KU Leuven, CSL, IMEC (Belgium); CEA, LAM (France); MPIA (Germany); INAF-IFSI/OAA/OAP/OAT, LENS, SISSA (Italy); IAC (Spain). This development has been supported by BMVIT (Austria), ESA-PRODEX (Belgium), CEA/CNES (France), DLR (Germany), ASI/INAF (Italy), and CICYT/MCYT (Spain). 
SPIRE has been developed by a consortium of institutes led by Cardiff Univ. (UK) and including: Univ. Lethbridge (Canada); NAOC (China); CEA, LAM (France); IFSI, Univ. Padua (Italy);IAC (Spain); Stockholm Observatory (Sweden); Imperial College London, RAL, UCL-MSSL, UKATC, Univ. Sussex (UK); and Caltech, JPL, NHSC, Univ. Colorado (USA). This development has been supported by national funding agencies: CSA (Canada); NAOC (China); CEA, CNES, CNRS (France); ASI (Italy); MCINN (Spain); SNSB (Sweden); STFC, UKSA (UK); and NASA (USA). This research has also made use of the NASA/IPAC Extragalactic Database (NED) which is operated by the Jet Propulsion Laboratory, California Institute of Technology, under contract with the National Aeronautics and Space Administration.


\bibliographystyle{apj}
\bibliography{/Users/maudgalametz/Documents/Work/Papers/mybiblio.bib}

 \appendix
 
 \section{TIR calibration beyond 250 \mic} 

\subsection{350 and 500 \mic}

To estimate similar monochromatic calibrations for the 350 and 500 \mic\ wavebands, we use the L$_{TIR}$ maps obtained at SPIRE 500 resolution. The FWHM of the PSF for SPIRE 500 \mic\ is $\sim$36\arcsec\ and the pixel size of our L$_{TIR}$ maps is 14\arcsec. We refer to Section 3.1 for explanations of the methodology.
We gather the resolved elements of the complete sample to derive calibrations coefficients linking the resolved TIR luminosities with the 350 or 500 \mic\ fluxes. We remind Eq.~\ref{eq3} here: 

\begin{equation}
log~L_{TIR} = a_i~log ~\nu L_{\nu}(i) + b_i 
\end{equation}

where L$_{TIR}$ refers to the TIR luminosity and $\nu$ L$_{\nu}$(i) the flux at 350 or 500 \mic. Both L$_{TIR}$ and $\nu$ L$_{\nu}$(i) are in \lsun.
We refer to Section 5 for details on the regression technique. \\

We obtain the following calibration coefficients :

\begin {itemize}
\item ( a$_{350}$ , b$_{350}$ ) = ( 1.106 $\pm$ 0.011 , 0.661 $\pm$ 0.064 ) 
\item ( a$_{500}$ , b$_{500}$ ) = ( 1.160 $\pm$ 0.012 , 1.008 $\pm$ 0.062 ) 
\end{itemize}

The 350 and 500 \mic\ relations with the TIR luminosities are thus over-linear relations. As for the 250 \mic\ band (Fig.~\ref{LTIRvsSB_separatewave}), we observe a large spread of L$_{TIR}$ values for a given 350 or 500 \mic\ flux as well as a strong correlation of the goodness of our 350 and 500 \mic\ monochromatic calibrations with the 70-to-100 flux density ratio. We thus caution the use of these calibrations for extreme (star-forming or very cold) environments.

\subsection{850 and 1000 \mic}

From our resolved SED modelling performed at the resolution of SPIRE 500 \mic, we also extrapolate maps of the KINGFISH galaxies at 850 and 1000 \mic, wavebands that are frequently observed from ground-based telescopes (SCUBA-2, LABOCA, MAMBO etc.). We gather the resolved elements of the complete sample and estimate calibration coefficients to link the resolved TIR luminosities to the extrapolated 850 or 1000 \mic\ resolved fluxes, both in \lsun\ (Eq.~\ref{eq3} ). \\

We obtain the following calibration coefficients :

\begin {itemize}
\item ( a$_{850}$ , b$_{850}$ ) = ( 1.150 $\pm$ 0.013 , 2.161 $\pm$ 0.057 ) 
\item ( a$_{1000}$ , b$_{1000}$ ) = ( 1.152 $\pm$ 0.014 , 2.533 $\pm$ 0.054 ) 
\end{itemize}

\vspace{10pt}

We remind the reader that the extrapolated 850 and 1000 \mic\ fluxes used in this calibration are coming from pure thermal dust emission. Any non-dust contamination contributing to observations at those wavelengths  (free-free or synchrotron emission, molecular line contamination etc) has to be removed if the calibration is used. The slopes of the 500, 850 and 1000 \mic\ calibrations are very close because these observations sample the submm slope of the SEDs where fluxes are evolving in a similar way. The 850 and 1000 maps are moreover directly extrapolated from the SED model performed using data up to 500 \mic. Our results are thus consistent with a scaling of the fluxes (L$_{850}$/L$_{1000}$$\sim$constant) from one relation to the other, translated in log space by a simple shift of the intercept. Small differences are nevertheless due to the fact that the submm slope varies from one galaxy to another.

  \section{TIR surface brightness versus Spitzer/Herschel brightnesses}

   \begin{figure*}
   \centering
    \includegraphics[width=18cm]{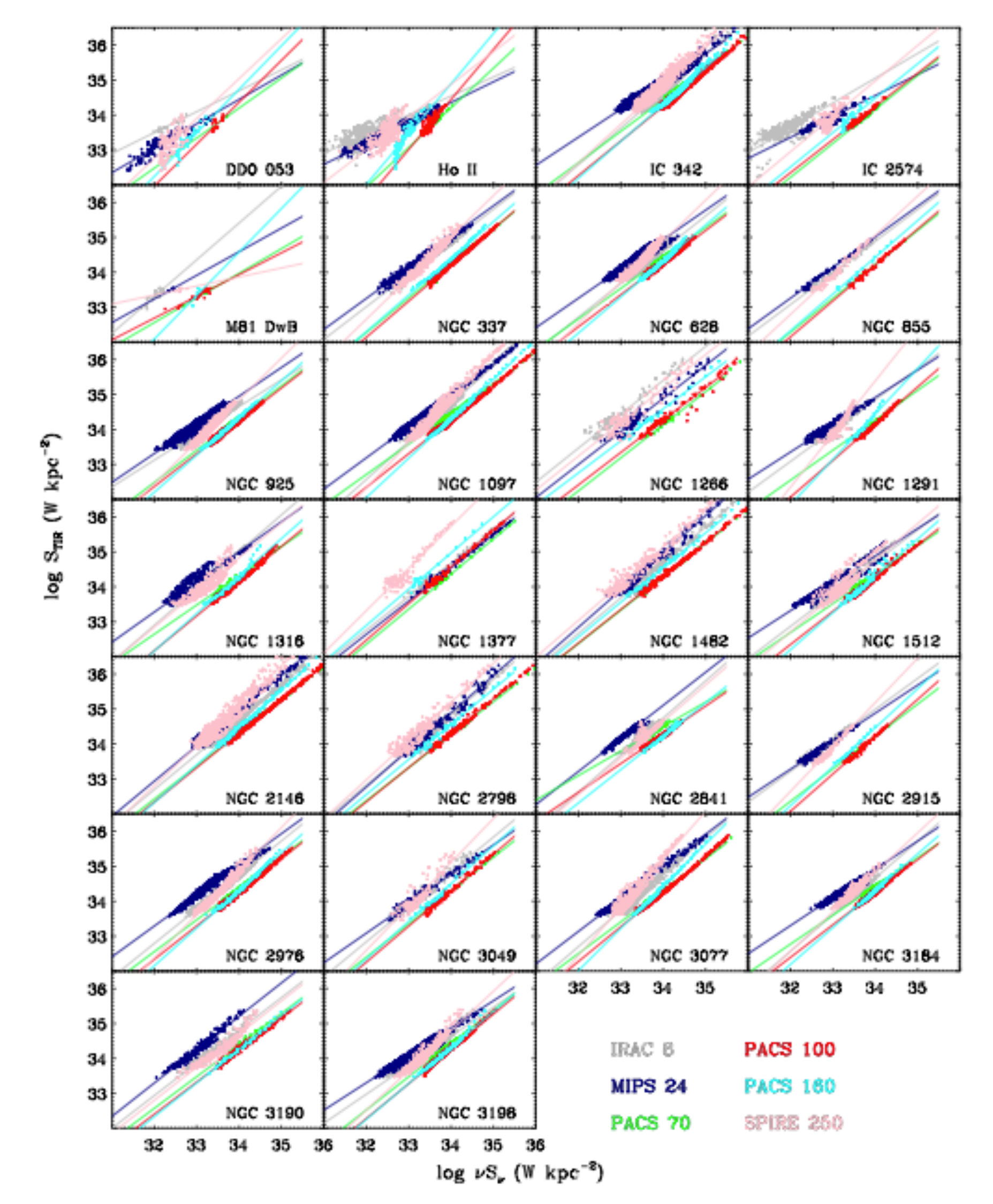}\\
    \caption{TIR surface brightness as a function of the different \spitz\ and \hersc\ bands for the galaxies of the KINGFISH sample. We overlay the regressions estimated for each band. For each galaxy, we sort and average resolved element 5-by-5 for clarity. We report the coefficients of the fits in Table~\ref{Coeff_Indiv_mono}.}
    \label{LTIRvsSB}
    \end{figure*}
      \addtocounter {figure}{-1}    
   \begin{figure*}
   \centering
    \includegraphics[width=18cm]{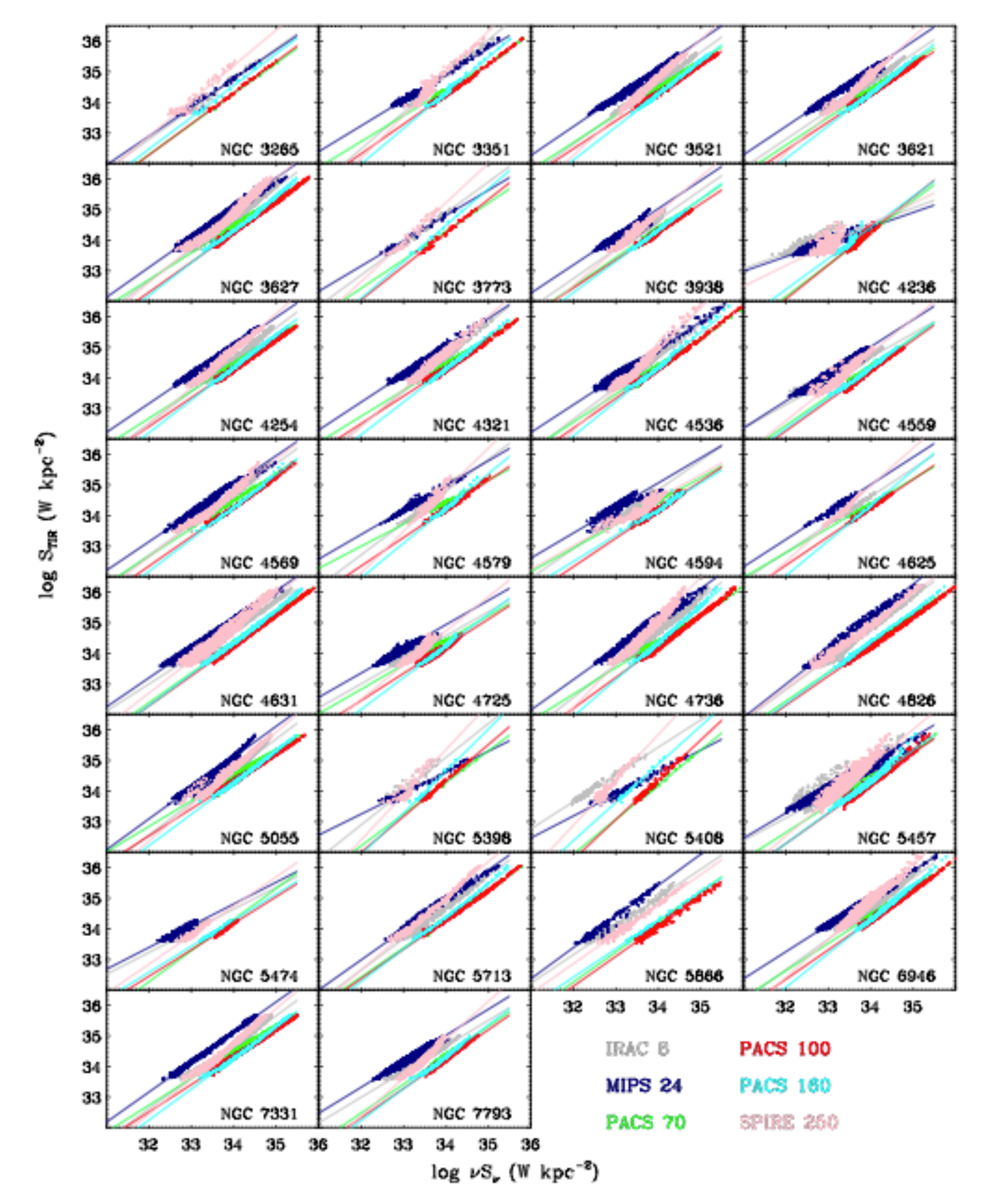}\\
        \caption{continued.}
    \end{figure*}

\newpage

\begin{minipage}{19cm}
\section{Calibration coefficients for individual galaxies}
\end{minipage}
\nopagebreak

\begin{table*}
\caption{\large Calibration coefficients derived for individual galaxies to convert the monochromatic brightnesses in various \spitz\ or \hersc\ bands (8, 24, 70, 100, 160 or 250 \mic) into TIR brightness.}
\label{Coeff_Indiv_mono}
\centering
  \begin{tabular}{c|cc|cc|cc|}
\hline
\hline
Name	&a$_{8}$&b$_{8}$&a$_{24}$&b$_{24}$&a$_{70}$&b$_{70}$\\	
\hline
DDO053&
     0.630$\pm$     0.043&    13.339$\pm$     1.390&     0.771$\pm$     0.028&
     8.350$\pm$     0.900&     1.031$\pm$     0.219&    -0.779$\pm$     7.364\\
HolmbergII&
     0.718$\pm$     0.028&    10.492$\pm$     0.895&     0.623$\pm$     0.011&
    13.238$\pm$     0.352&     1.161$\pm$     0.019&    -5.243$\pm$     0.638\\
IC342&
     1.079$\pm$     0.004&    -2.024$\pm$     0.131&     0.825$\pm$     0.003&
     6.926$\pm$     0.096&     0.916$\pm$     0.003&     3.293$\pm$     0.106\\
IC2574&
     0.613$\pm$     0.031&    13.926$\pm$     0.999&     0.653$\pm$     0.051&
    12.433$\pm$     1.661&     0.899$\pm$     0.053&     3.542$\pm$     1.801\\
M81Dw&
     1.210$\pm$     0.120&    -5.409$\pm$     3.827&     0.705$\pm$     0.094&
    10.675$\pm$     3.050&     0.724$\pm$     0.029&     9.385$\pm$     0.966\\
NGC0337&
     0.945$\pm$     0.010&     2.756$\pm$     0.343&     0.896$\pm$     0.008&
     4.563$\pm$     0.278&     0.998$\pm$     0.009&     0.334$\pm$     0.301\\
NGC0628&
     0.992$\pm$     0.006&     0.917$\pm$     0.205&     0.881$\pm$     0.007&
     5.000$\pm$     0.247&     0.904$\pm$     0.007&     3.672$\pm$     0.229\\
NGC0855&
     0.914$\pm$     0.012&     3.836$\pm$     0.386&     0.881$\pm$     0.008&
     5.069$\pm$     0.258&     0.972$\pm$     0.006&     1.177$\pm$     0.198\\
NGC0925&
     0.812$\pm$     0.008&     7.101$\pm$     0.279&     0.867$\pm$     0.008&
     5.517$\pm$     0.255&     0.943$\pm$     0.006&     2.273$\pm$     0.198\\
NGC1097&
     1.103$\pm$     0.005&    -2.749$\pm$     0.184&     0.904$\pm$     0.003&
     4.259$\pm$     0.111&     0.875$\pm$     0.004&     4.702$\pm$     0.136\\
NGC1266&
     1.021$\pm$     0.027&     0.746$\pm$     0.898&     0.977$\pm$     0.023&
     1.699$\pm$     0.781&     0.965$\pm$     0.015&     1.329$\pm$     0.526\\
NGC1291&
     0.949$\pm$     0.020&     2.655$\pm$     0.671&     0.817$\pm$     0.017&
     7.186$\pm$     0.557&     0.864$\pm$     0.009&     4.895$\pm$     0.318\\
NGC1316&
     1.076$\pm$     0.015&    -1.573$\pm$     0.483&     0.965$\pm$     0.024&
     2.318$\pm$     0.788&     0.849$\pm$     0.009&     5.435$\pm$     0.297\\
NGC1377&
     0.963$\pm$     0.008&     1.938$\pm$     0.270&     0.957$\pm$     0.005&
     2.018$\pm$     0.183&     1.008$\pm$     0.005&     0.075$\pm$     0.174\\
NGC1482&
     1.036$\pm$     0.011&    -0.516$\pm$     0.372&     1.081$\pm$     0.012&
    -1.776$\pm$     0.399&     0.961$\pm$     0.004&     1.619$\pm$     0.139\\
NGC1512&
     0.997$\pm$     0.019&     0.887$\pm$     0.637&     0.821$\pm$     0.017&
     7.018$\pm$     0.565&     0.865$\pm$     0.008&     4.961$\pm$     0.284\\
NGC2146&
     1.009$\pm$     0.006&     0.394$\pm$     0.195&     1.019$\pm$     0.005&
     0.318$\pm$     0.181&     0.952$\pm$     0.003&     1.943$\pm$     0.116\\
NGC2798&
     1.033$\pm$     0.013&    -0.278$\pm$     0.438&     1.094$\pm$     0.018&
    -2.271$\pm$     0.609&     0.948$\pm$     0.005&     2.021$\pm$     0.188\\
NGC2841&
     1.090$\pm$     0.010&    -2.312$\pm$     0.352&     0.963$\pm$     0.008&
     2.371$\pm$     0.281&     0.793$\pm$     0.023&     7.540$\pm$     0.778\\
NGC2915&
     0.916$\pm$     0.021&     3.883$\pm$     0.695&     0.819$\pm$     0.015&
     7.058$\pm$     0.486&     0.931$\pm$     0.009&     2.560$\pm$     0.297\\
NGC2976&
     0.993$\pm$     0.008&     0.998$\pm$     0.269&     0.895$\pm$     0.008&
     4.627$\pm$     0.262&     0.916$\pm$     0.005&     3.227$\pm$     0.167\\
NGC3049&
     1.036$\pm$     0.023&    -0.350$\pm$     0.778&     0.861$\pm$     0.013&
     5.508$\pm$     0.432&     0.966$\pm$     0.012&     1.451$\pm$     0.421\\
NGC3077&
     1.059$\pm$     0.009&    -1.224$\pm$     0.309&     0.934$\pm$     0.007&
     3.237$\pm$     0.242&     0.912$\pm$     0.005&     3.314$\pm$     0.175\\
NGC3184&
     1.080$\pm$     0.015&    -2.055$\pm$     0.491&     0.835$\pm$     0.008&
     6.564$\pm$     0.265&     0.855$\pm$     0.007&     5.355$\pm$     0.249\\
NGC3190&
     0.961$\pm$     0.018&     2.153$\pm$     0.593&     0.974$\pm$     0.014&
     2.109$\pm$     0.447&     0.947$\pm$     0.021&     2.183$\pm$     0.729\\
NGC3198&
     0.854$\pm$     0.012&     5.604$\pm$     0.387&     0.806$\pm$     0.007&
     7.485$\pm$     0.237&     0.975$\pm$     0.010&     1.288$\pm$     0.328\\
NGC3265&
     1.008$\pm$     0.016&     0.520$\pm$     0.528&     0.951$\pm$     0.010&
     2.445$\pm$     0.334&     0.998$\pm$     0.013&     0.342$\pm$     0.443\\
NGC3351&
     1.215$\pm$     0.014&    -6.456$\pm$     0.457&     0.853$\pm$     0.006&
     5.927$\pm$     0.197&     0.852$\pm$     0.007&     5.425$\pm$     0.254\\
NGC3521&
     1.005$\pm$     0.002&     0.461$\pm$     0.076&     0.947$\pm$     0.002&
     2.889$\pm$     0.072&     0.895$\pm$     0.002&     4.039$\pm$     0.072\\
NGC3621&
     0.940$\pm$     0.004&     2.687$\pm$     0.134&     0.939$\pm$     0.004&
     3.152$\pm$     0.121&     0.916$\pm$     0.004&     3.257$\pm$     0.132\\
NGC3627&
     1.028$\pm$     0.005&    -0.236$\pm$     0.170&     0.984$\pm$     0.005&
     1.618$\pm$     0.157&     0.885$\pm$     0.004&     4.372$\pm$     0.143\\
NGC3773&
     1.024$\pm$     0.017&     0.089$\pm$     0.553&     0.835$\pm$     0.013&
     6.425$\pm$     0.439&     0.937$\pm$     0.015&     2.441$\pm$     0.508\\
NGC3938&
     0.979$\pm$     0.007&     1.394$\pm$     0.219&     0.947$\pm$     0.008&
     2.866$\pm$     0.250&     0.962$\pm$     0.006&     1.685$\pm$     0.193\\
NGC4236&
     0.730$\pm$     0.024&    10.056$\pm$     0.779&     0.541$\pm$     0.012&
    16.085$\pm$     0.402&     1.119$\pm$     0.027&    -3.799$\pm$     0.917\\
NGC4254&
     1.052$\pm$     0.005&    -1.162$\pm$     0.169&     0.948$\pm$     0.003&
     2.818$\pm$     0.110&     0.893$\pm$     0.003&     4.085$\pm$     0.117\\
NGC4321&
     1.129$\pm$     0.004&    -3.704$\pm$     0.143&     0.923$\pm$     0.004&
     3.663$\pm$     0.142&     0.862$\pm$     0.003&     5.149$\pm$     0.113\\
NGC4536&
     1.055$\pm$     0.009&    -1.129$\pm$     0.289&     0.923$\pm$     0.007&
     3.553$\pm$     0.224&     0.924$\pm$     0.005&     2.963$\pm$     0.176\\
NGC4559&
     0.797$\pm$     0.007&     7.604$\pm$     0.227&     0.910$\pm$     0.007&
     4.109$\pm$     0.214&     0.982$\pm$     0.008&     0.971$\pm$     0.272\\
NGC4569&
     1.010$\pm$     0.010&     0.373$\pm$     0.328&     0.951$\pm$     0.011&
     2.692$\pm$     0.362&     0.911$\pm$     0.007&     3.501$\pm$     0.244\\
NGC4579&
     1.103$\pm$     0.021&    -2.681$\pm$     0.689&     0.827$\pm$     0.011&
     6.888$\pm$     0.376&     0.747$\pm$     0.008&     9.054$\pm$     0.264\\
NGC4594&
     0.983$\pm$     0.009&     1.419$\pm$     0.303&     0.927$\pm$     0.028&
     3.644$\pm$     0.922&     0.888$\pm$     0.034&     4.304$\pm$     1.146\\
NGC4625&
     0.925$\pm$     0.015&     3.172$\pm$     0.494&     0.919$\pm$     0.016&
     3.789$\pm$     0.538&     0.815$\pm$     0.015&     6.698$\pm$     0.500\\
NGC4631&
     0.911$\pm$     0.004&     3.795$\pm$     0.129&     0.957$\pm$     0.003&
     2.568$\pm$     0.086&     0.980$\pm$     0.002&     1.017$\pm$     0.075\\
NGC4725&
     0.853$\pm$     0.010&     5.618$\pm$     0.348&     0.820$\pm$     0.008&
     7.100$\pm$     0.257&     0.849$\pm$     0.013&     5.577$\pm$     0.453\\
NGC4736&
     1.090$\pm$     0.004&    -2.326$\pm$     0.145&     0.998$\pm$     0.004&
     1.188$\pm$     0.141&     0.869$\pm$     0.002&     4.835$\pm$     0.082\\
NGC4826&
     1.023$\pm$     0.005&     0.056$\pm$     0.176&     1.099$\pm$     0.006&
    -2.197$\pm$     0.215&     0.946$\pm$     0.003&     2.140$\pm$     0.092\\
NGC5055&
     1.060$\pm$     0.007&    -1.395$\pm$     0.222&     1.019$\pm$     0.004&
     0.450$\pm$     0.141&     0.849$\pm$     0.005&     5.662$\pm$     0.158\\
NGC5398&
     1.214$\pm$     0.046&    -6.073$\pm$     1.528&     0.697$\pm$     0.012&
    10.932$\pm$     0.386&     1.045$\pm$     0.017&    -1.237$\pm$     0.569\\
NGC5408&
     0.911$\pm$     0.011&     4.485$\pm$     0.366&     0.736$\pm$     0.013&
     9.604$\pm$     0.449&     1.142$\pm$     0.016&    -4.624$\pm$     0.539\\
NGC5457&
     0.804$\pm$     0.006&     7.296$\pm$     0.185&     0.852$\pm$     0.003&
     5.977$\pm$     0.113&     0.929$\pm$     0.004&     2.846$\pm$     0.136\\
NGC5474&
     0.767$\pm$     0.027&     8.637$\pm$     0.896&     0.794$\pm$     0.022&
     7.926$\pm$     0.728&     1.091$\pm$     0.085&    -2.778$\pm$     2.871\\
NGC5713&
     1.033$\pm$     0.007&    -0.427$\pm$     0.245&     0.992$\pm$     0.008&
     1.236$\pm$     0.267&     0.957$\pm$     0.005&     1.789$\pm$     0.176\\
NGC5866&
     0.949$\pm$     0.013&     2.752$\pm$     0.430&     1.031$\pm$     0.009&
     0.371$\pm$     0.310&     0.993$\pm$     0.013&     0.457$\pm$     0.442\\
NGC6946&
     1.011$\pm$     0.003&     0.285$\pm$     0.104&     0.890$\pm$     0.002&
     4.754$\pm$     0.053&     0.903$\pm$     0.002&     3.728$\pm$     0.063\\
NGC7331&
     1.032$\pm$     0.007&    -0.428$\pm$     0.225&     0.985$\pm$     0.002&
     1.625$\pm$     0.066&     0.883$\pm$     0.004&     4.441$\pm$     0.154\\
NGC7793&
     0.851$\pm$     0.004&     5.742$\pm$     0.120&     0.869$\pm$     0.004&
     5.502$\pm$     0.143&     0.962$\pm$     0.005&     1.671$\pm$     0.158\\
\hline
\end{tabular}
\begin{list}{}{}
\item NB: The relation is : log~S$_{TIR}$= a$_i$~log~S$_i$ + b$_i$ with S$_{TIR}$ and S$_i$ in W~kpc$^{-2}$. 
\end{list}
\end{table*} 

      \addtocounter {table}{-1}
   
\begin{table*}
\caption{\large continued}
\centering
  \begin{tabular}{c|cc|cc|cc|}
\hline
\hline
Name	&a$_{100}$&b$_{100}$&a$_{160}$&b$_{160}$&a$_{250}$&b$_{250}$\\	
\hline    
DDO053&
     1.655$\pm$     0.148&   -21.647$\pm$     4.964&     1.749$\pm$     0.190&
   -24.095$\pm$     6.249&     1.741$\pm$     0.118&   -23.183$\pm$     3.820\\
HolmbergII&
     1.615$\pm$     0.048&   -20.360$\pm$     1.608&     1.809$\pm$     0.044&
   -26.203$\pm$     1.438&     1.382$\pm$     0.085&   -11.428$\pm$     2.761\\
IC342&
     1.001$\pm$     0.003&     0.207$\pm$     0.093&     1.083$\pm$     0.005&
    -2.542$\pm$     0.156&     1.156$\pm$     0.006&    -4.463$\pm$     0.203\\
IC2574&
     0.893$\pm$     0.027&     3.851$\pm$     0.906&     1.235$\pm$     0.067&
    -7.464$\pm$     2.227&     1.407$\pm$     0.120&   -12.589$\pm$     3.963\\
M81Dw&
     0.663$\pm$     0.060&    11.434$\pm$     1.971&     1.866$\pm$     0.266&
   -28.235$\pm$     8.779&     0.545$\pm$     0.214&    15.739$\pm$     6.977\\
NGC0337&
     1.038$\pm$     0.005&    -1.079$\pm$     0.160&     1.058$\pm$     0.008&
    -1.580$\pm$     0.274&     1.212$\pm$     0.015&    -6.177$\pm$     0.498\\
NGC0628&
     0.929$\pm$     0.004&     2.688$\pm$     0.128&     1.052$\pm$     0.005&
    -1.440$\pm$     0.183&     1.212$\pm$     0.008&    -6.297$\pm$     0.272\\
NGC0855&
     1.037$\pm$     0.016&    -1.054$\pm$     0.548&     1.066$\pm$     0.020&
    -1.817$\pm$     0.679&     1.146$\pm$     0.022&    -3.888$\pm$     0.721\\
NGC0925&
     0.953$\pm$     0.004&     1.862$\pm$     0.120&     1.077$\pm$     0.004&
    -2.278$\pm$     0.148&     1.238$\pm$     0.008&    -7.217$\pm$     0.280\\
NGC1097&
     0.966$\pm$     0.004&     1.410$\pm$     0.131&     1.137$\pm$     0.004&
    -4.363$\pm$     0.142&     1.310$\pm$     0.008&    -9.630$\pm$     0.267\\
NGC1266&
     0.946$\pm$     0.015&     2.116$\pm$     0.512&     0.952$\pm$     0.022&
     2.217$\pm$     0.732&     1.044$\pm$     0.029&    -0.247$\pm$     0.985\\
NGC1291&
     1.070$\pm$     0.019&    -2.212$\pm$     0.658&     1.357$\pm$     0.039&
   -11.666$\pm$     1.325&     1.657$\pm$     0.062&   -20.849$\pm$     2.059\\
NGC1316&
     1.006$\pm$     0.012&    -0.036$\pm$     0.392&     1.118$\pm$     0.026&
    -3.702$\pm$     0.881&     1.107$\pm$     0.038&    -2.719$\pm$     1.252\\
NGC1377&
     1.053$\pm$     0.007&    -1.233$\pm$     0.238&     1.074$\pm$     0.019&
    -1.516$\pm$     0.657&     1.269$\pm$     0.022&    -7.336$\pm$     0.745\\
NGC1482&
     0.967$\pm$     0.006&     1.450$\pm$     0.195&     1.001$\pm$     0.007&
     0.504$\pm$     0.252&     1.075$\pm$     0.014&    -1.422$\pm$     0.456\\
NGC1512&
     0.963$\pm$     0.008&     1.502$\pm$     0.261&     1.131$\pm$     0.019&
    -4.117$\pm$     0.648&     1.215$\pm$     0.031&    -6.398$\pm$     1.015\\
NGC2146&
     0.986$\pm$     0.003&     0.792$\pm$     0.098&     1.083$\pm$     0.005&
    -2.359$\pm$     0.162&     1.186$\pm$     0.012&    -5.207$\pm$     0.419\\
NGC2798&
     0.985$\pm$     0.005&     0.805$\pm$     0.160&     1.028$\pm$     0.010&
    -0.373$\pm$     0.351&     1.154$\pm$     0.017&    -4.021$\pm$     0.562\\
NGC2841&
     0.834$\pm$     0.006&     5.909$\pm$     0.188&     0.978$\pm$     0.007&
     0.958$\pm$     0.246&     1.130$\pm$     0.012&    -3.745$\pm$     0.402\\
NGC2915&
     1.086$\pm$     0.016&    -2.708$\pm$     0.546&     1.136$\pm$     0.026&
    -4.158$\pm$     0.858&     1.293$\pm$     0.024&    -8.743$\pm$     0.800\\
NGC2976&
     0.966$\pm$     0.005&     1.419$\pm$     0.177&     1.092$\pm$     0.008&
    -2.821$\pm$     0.274&     1.265$\pm$     0.011&    -8.138$\pm$     0.374\\
NGC3049&
     1.043$\pm$     0.013&    -1.150$\pm$     0.443&     1.109$\pm$     0.020&
    -3.198$\pm$     0.681&     1.397$\pm$     0.053&   -12.287$\pm$     1.760\\
NGC3077&
     0.998$\pm$     0.005&     0.324$\pm$     0.185&     1.219$\pm$     0.007&
    -7.009$\pm$     0.226&     1.433$\pm$     0.015&   -13.503$\pm$     0.491\\
NGC3184&
     0.946$\pm$     0.005&     2.105$\pm$     0.161&     1.079$\pm$     0.008&
    -2.373$\pm$     0.268&     1.189$\pm$     0.014&    -5.570$\pm$     0.454\\
NGC3190&
     0.928$\pm$     0.008&     2.698$\pm$     0.257&     0.979$\pm$     0.012&
     0.969$\pm$     0.419&     0.994$\pm$     0.025&     0.926$\pm$     0.836\\
NGC3198&
     1.006$\pm$     0.006&     0.082$\pm$     0.211&     1.049$\pm$     0.011&
    -1.334$\pm$     0.375&     1.182$\pm$     0.018&    -5.383$\pm$     0.595\\
NGC3265&
     1.023$\pm$     0.009&    -0.480$\pm$     0.295&     1.061$\pm$     0.020&
    -1.516$\pm$     0.690&     1.212$\pm$     0.033&    -5.970$\pm$     1.110\\
NGC3351&
     0.975$\pm$     0.005&     1.097$\pm$     0.177&     1.181$\pm$     0.006&
    -5.829$\pm$     0.216&     1.433$\pm$     0.017&   -13.756$\pm$     0.585\\
NGC3521&
     0.914$\pm$     0.002&     3.233$\pm$     0.056&     1.080$\pm$     0.002&
    -2.435$\pm$     0.054&     1.266$\pm$     0.003&    -8.261$\pm$     0.118\\
NGC3621&
     0.925$\pm$     0.002&     2.816$\pm$     0.071&     1.078$\pm$     0.004&
    -2.351$\pm$     0.122&     1.280$\pm$     0.005&    -8.662$\pm$     0.182\\
NGC3627&
     0.985$\pm$     0.002&     0.751$\pm$     0.073&     1.074$\pm$     0.004&
    -2.208$\pm$     0.135&     1.158$\pm$     0.007&    -4.479$\pm$     0.251\\
NGC3773&
     1.104$\pm$     0.020&    -3.288$\pm$     0.675&     1.212$\pm$     0.031&
    -6.712$\pm$     1.057&     1.326$\pm$     0.043&    -9.831$\pm$     1.418\\
NGC3938&
     0.921$\pm$     0.003&     2.944$\pm$     0.114&     1.029$\pm$     0.004&
    -0.657$\pm$     0.137&     1.214$\pm$     0.008&    -6.363$\pm$     0.264\\
NGC4236&
     1.252$\pm$     0.040&    -8.214$\pm$     1.340&     1.205$\pm$     0.039&
    -6.446$\pm$     1.312&     1.172$\pm$     0.044&    -4.842$\pm$     1.436\\
NGC4254&
     0.946$\pm$     0.002&     2.134$\pm$     0.085&     1.070$\pm$     0.003&
    -2.051$\pm$     0.104&     1.218$\pm$     0.005&    -6.520$\pm$     0.173\\
NGC4321&
     0.905$\pm$     0.003&     3.517$\pm$     0.118&     1.081$\pm$     0.003&
    -2.467$\pm$     0.108&     1.226$\pm$     0.007&    -6.855$\pm$     0.241\\
NGC4536&
     0.977$\pm$     0.005&     1.099$\pm$     0.171&     1.145$\pm$     0.008&
    -4.557$\pm$     0.283&     1.368$\pm$     0.014&   -11.546$\pm$     0.456\\
NGC4559&
     0.995$\pm$     0.003&     0.418$\pm$     0.109&     1.005$\pm$     0.005&
     0.153$\pm$     0.158&     1.157$\pm$     0.011&    -4.496$\pm$     0.368\\
NGC4569&
     0.967$\pm$     0.005&     1.398$\pm$     0.165&     1.059$\pm$     0.009&
    -1.738$\pm$     0.323&     1.090$\pm$     0.015&    -2.264$\pm$     0.505\\
NGC4579&
     0.947$\pm$     0.010&     2.023$\pm$     0.325&     1.174$\pm$     0.022&
    -5.672$\pm$     0.757&     1.482$\pm$     0.038&   -15.509$\pm$     1.264\\
NGC4594&
     0.948$\pm$     0.010&     1.970$\pm$     0.335&     0.898$\pm$     0.017&
     3.687$\pm$     0.564&     0.849$\pm$     0.020&     5.719$\pm$     0.671\\
NGC4625&
     0.945$\pm$     0.012&     2.123$\pm$     0.416&     1.157$\pm$     0.012&
    -5.011$\pm$     0.404&     1.263$\pm$     0.025&    -7.953$\pm$     0.844\\
NGC4631&
     0.986$\pm$     0.002&     0.712$\pm$     0.053&     1.036$\pm$     0.003&
    -0.889$\pm$     0.088&     1.102$\pm$     0.005&    -2.627$\pm$     0.183\\
NGC4725&
     0.911$\pm$     0.006&     3.261$\pm$     0.206&     1.057$\pm$     0.008&
    -1.706$\pm$     0.268&     1.176$\pm$     0.015&    -5.211$\pm$     0.506\\
NGC4736&
     1.013$\pm$     0.002&    -0.235$\pm$     0.072&     1.206$\pm$     0.003&
    -6.660$\pm$     0.116&     1.436$\pm$     0.007&   -13.737$\pm$     0.225\\
NGC4826&
     0.961$\pm$     0.002&     1.553$\pm$     0.086&     1.010$\pm$     0.004&
     0.020$\pm$     0.125&     1.038$\pm$     0.005&    -0.346$\pm$     0.158\\
NGC5055&
     0.891$\pm$     0.002&     4.024$\pm$     0.053&     1.033$\pm$     0.003&
    -0.848$\pm$     0.108&     1.206$\pm$     0.007&    -6.271$\pm$     0.241\\
NGC5398&
     1.182$\pm$     0.008&    -5.840$\pm$     0.281&     1.407$\pm$     0.029&
   -13.154$\pm$     0.977&     1.690$\pm$     0.087&   -21.910$\pm$     2.893\\
NGC5408&
     1.311$\pm$     0.018&   -10.214$\pm$     0.605&     1.288$\pm$     0.025&
    -8.880$\pm$     0.833&     1.555$\pm$     0.033&   -16.951$\pm$     1.076\\
NGC5457&
     0.994$\pm$     0.002&     0.443$\pm$     0.084&     1.044$\pm$     0.004&
    -1.173$\pm$     0.146&     1.216$\pm$     0.006&    -6.463$\pm$     0.194\\
NGC5474&
     0.903$\pm$     0.019&     3.477$\pm$     0.635&     0.884$\pm$     0.014&
     4.251$\pm$     0.477&     1.006$\pm$     0.023&     0.601$\pm$     0.755\\
NGC5713&
     0.965$\pm$     0.005&     1.494$\pm$     0.163&     1.066$\pm$     0.006&
    -1.846$\pm$     0.211&     1.187$\pm$     0.009&    -5.344$\pm$     0.300\\
NGC5866&
     0.985$\pm$     0.009&     0.587$\pm$     0.313&     0.939$\pm$     0.008&
     2.340$\pm$     0.282&     0.979$\pm$     0.008&     1.527$\pm$     0.275\\
NGC6946&
     0.963$\pm$     0.002&     1.523$\pm$     0.059&     1.116$\pm$     0.002&
    -3.664$\pm$     0.074&     1.259$\pm$     0.004&    -7.913$\pm$     0.124\\
NGC7331&
     0.900$\pm$     0.002&     3.711$\pm$     0.073&     1.018$\pm$     0.003&
    -0.353$\pm$     0.114&     1.160$\pm$     0.005&    -4.689$\pm$     0.176\\
NGC7793&
     0.966$\pm$     0.003&     1.384$\pm$     0.098&     1.055$\pm$     0.005&
    -1.541$\pm$     0.183&     1.284$\pm$     0.008&    -8.769$\pm$     0.257\\
\hline
\end{tabular}
\end{table*}


 \begin{table*}
\caption{\large Calibration coefficients derived for individual galaxies to predict the TIR brightness from a combination of the 24 to 250 \mic\ brightnesses.}
\label{Coeff_Indiv_combined}
 \centering
  \begin{tabular}{cccccc}
\hline
\hline
Name	&	c$_{24}$	    &	c$_{70}$	     &	c$_{100}$   &	c$_{160}$    &	c$_{250}$ \\
\hline
DDO053&
 1.998$\pm$ 0.287& 0.386$\pm$ 0.109&
 0.512$\pm$ 0.125& 0.852$\pm$ 0.264&
 -0.427$\pm$  0.412\\
HolmbergII&
 1.356$\pm$ 0.205& 0.775$\pm$ 0.131&
 0.162$\pm$ 0.151& 0.434$\pm$ 0.328&
 0.265$\pm$  0.786\\
IC342&
 1.338$\pm$ 0.104& 0.820$\pm$ 0.035&
 0.390$\pm$ 0.042& 0.148$\pm$ 0.059&
 1.721$\pm$  0.122\\
IC2574&
 1.847$\pm$ 0.321& 0.486$\pm$ 0.094&
 0.531$\pm$ 0.161& 0.282$\pm$ 0.270&
 0.854$\pm$  0.604\\
M81Dw&
 3.691$\pm$ 5.133& 0.458$\pm$ 0.357&
 0.231$\pm$ 0.207& 0.929$\pm$ 0.235&
 -0.722$\pm$  0.936\\
NGC0337&
 2.638$\pm$ 0.519& 0.273$\pm$ 0.104&
 0.382$\pm$ 0.191& 1.196$\pm$ 0.261&
 -0.683$\pm$  0.432\\
NGC0628&
 1.634$\pm$ 0.219& 0.468$\pm$ 0.068&
 0.515$\pm$ 0.074& 0.508$\pm$ 0.124&
 0.950$\pm$  0.235\\
NGC0855&
 4.530$\pm$ 1.160& 0.448$\pm$ 0.111&
 0.262$\pm$ 0.121& 0.538$\pm$ 0.188&
 0.104$\pm$  0.493\\
NGC0925&
 1.744$\pm$ 0.272& 0.389$\pm$ 0.079&
 0.497$\pm$ 0.087& 0.595$\pm$ 0.135&
 0.666$\pm$  0.210\\
NGC1097&
 1.882$\pm$ 0.350& 0.568$\pm$ 0.081&
 0.327$\pm$ 0.115& 0.724$\pm$ 0.160&
 0.498$\pm$  0.263\\
NGC1266&
 3.419$\pm$ 0.335& 0.388$\pm$ 0.083&
 0.295$\pm$ 0.100& 0.077$\pm$ 0.181&
 0.018$\pm$  0.551\\
NGC1291&
 1.988$\pm$ 0.700& 0.674$\pm$ 0.172&
 0.311$\pm$ 0.163& 0.459$\pm$ 0.347&
 0.563$\pm$  0.889\\
NGC1316&
 1.457$\pm$ 0.344& 0.792$\pm$ 0.133&
 0.278$\pm$ 0.176& 0.247$\pm$ 0.263&
 1.225$\pm$  0.560\\
NGC1377&
 1.630$\pm$ 0.227& 0.715$\pm$ 0.172&
 0.440$\pm$ 0.212& 1.221$\pm$ 0.324&
 -1.959$\pm$  0.660\\
NGC1482&
 1.547$\pm$ 0.274& 0.565$\pm$ 0.091&
 0.634$\pm$ 0.110& 0.816$\pm$ 0.204&
 -0.472$\pm$  0.441\\
NGC1512&
 1.753$\pm$ 0.282& 0.550$\pm$ 0.087&
 0.451$\pm$ 0.097& 0.336$\pm$ 0.178&
 1.180$\pm$  0.380\\
NGC2146&
 1.550$\pm$ 0.304& 0.645$\pm$ 0.137&
 0.480$\pm$ 0.203& 0.602$\pm$ 0.171&
 0.569$\pm$  0.280\\
NGC2798&
 1.349$\pm$ 0.257& 0.872$\pm$ 0.193&
 0.134$\pm$ 0.320& 0.757$\pm$ 0.272&
 0.611$\pm$  0.515\\
NGC2841&
 2.428$\pm$ 1.076& 0.461$\pm$ 0.168&
 0.394$\pm$ 0.184& 0.171$\pm$ 0.307&
 1.554$\pm$  0.499\\
NGC2915&
 1.736$\pm$ 1.277& 0.737$\pm$ 0.264&
 0.253$\pm$ 0.257& 0.441$\pm$ 0.465&
 0.806$\pm$  1.040\\
NGC2976&
 1.570$\pm$ 0.443& 0.595$\pm$ 0.134&
 0.627$\pm$ 0.146& 0.040$\pm$ 0.152&
 1.443$\pm$  0.237\\
NGC3049&
 2.253$\pm$ 0.363& 0.478$\pm$ 0.225&
 0.005$\pm$ 0.333& 1.406$\pm$ 0.352&
 -0.216$\pm$  0.428\\
NGC3077&
 1.823$\pm$ 0.567& 0.700$\pm$ 0.134&
 0.496$\pm$ 0.180& 0.043$\pm$ 0.267&
 1.647$\pm$  0.508\\
NGC3184&
 1.464$\pm$ 0.418& 0.718$\pm$ 0.132&
 0.375$\pm$ 0.142& 0.327$\pm$ 0.207&
 1.442$\pm$  0.401\\
NGC3190&
 3.057$\pm$ 1.748& 0.525$\pm$ 0.140&
 0.411$\pm$ 0.186& 0.241$\pm$ 0.185&
 1.066$\pm$  0.493\\
NGC3198&
 1.777$\pm$ 0.232& 0.485$\pm$ 0.116&
 0.431$\pm$ 0.164& 0.443$\pm$ 0.171&
 1.074$\pm$  0.220\\
NGC3265&
 1.523$\pm$ 0.781& 0.692$\pm$ 0.278&
 0.576$\pm$ 0.407& 0.450$\pm$ 0.374&
 0.018$\pm$  0.646\\
NGC3351&
 1.984$\pm$ 0.480& 0.553$\pm$ 0.158&
 0.300$\pm$ 0.176& 0.536$\pm$ 0.173&
 0.976$\pm$  0.326\\
NGC3521&
 2.042$\pm$ 0.475& 0.512$\pm$ 0.097&
 0.407$\pm$ 0.105& 0.684$\pm$ 0.182&
 0.529$\pm$  0.287\\
NGC3621&
 1.660$\pm$ 0.326& 0.458$\pm$ 0.089&
 0.607$\pm$ 0.086& 0.412$\pm$ 0.148&
 0.904$\pm$  0.264\\
NGC3627&
 1.615$\pm$ 0.334& 0.476$\pm$ 0.105&
 0.494$\pm$ 0.118& 0.858$\pm$ 0.182&
 0.004$\pm$  0.379\\
NGC3773&
 2.055$\pm$ 0.673& 0.532$\pm$ 0.224&
 0.542$\pm$ 0.206& 0.094$\pm$ 0.241&
 1.116$\pm$  0.583\\
NGC3938&
 1.350$\pm$ 0.345& 0.518$\pm$ 0.132&
 0.585$\pm$ 0.118& 0.472$\pm$ 0.207&
 0.826$\pm$  0.317\\
NGC4236&
 1.362$\pm$ 0.136& 0.535$\pm$ 0.074&
 0.436$\pm$ 0.118& 0.633$\pm$ 0.162&
 0.292$\pm$  0.220\\
NGC4254&
 2.389$\pm$ 0.444& 0.551$\pm$ 0.124&
 0.515$\pm$ 0.159& 0.176$\pm$ 0.220&
 1.542$\pm$  0.368\\
NGC4321&
 1.730$\pm$ 0.458& 0.614$\pm$ 0.114&
 0.359$\pm$ 0.126& 0.477$\pm$ 0.163&
 1.167$\pm$  0.307\\
NGC4536&
 1.665$\pm$ 0.354& 0.488$\pm$ 0.113&
 0.608$\pm$ 0.137& 0.531$\pm$ 0.196&
 0.482$\pm$  0.322\\
NGC4559&
 1.873$\pm$ 0.316& 0.396$\pm$ 0.074&
 0.446$\pm$ 0.101& 0.701$\pm$ 0.117&
 0.502$\pm$  0.187\\
NGC4569&
 1.093$\pm$ 0.510& 0.935$\pm$ 0.184&
 0.389$\pm$ 0.171& 0.192$\pm$ 0.196&
 1.471$\pm$  0.414\\
NGC4579&
 1.952$\pm$ 0.627& 0.709$\pm$ 0.197&
 0.218$\pm$ 0.219& 0.396$\pm$ 0.255&
 1.256$\pm$  0.377\\
NGC4594&
 1.582$\pm$ 0.401& 0.750$\pm$ 0.134&
 0.172$\pm$ 0.187& 0.434$\pm$ 0.163&
 1.055$\pm$  0.159\\
NGC4625&
 3.317$\pm$ 1.414& 0.555$\pm$ 0.466&
 0.311$\pm$ 0.541& 0.413$\pm$ 0.715&
 0.930$\pm$  1.580\\
NGC4631&
 2.640$\pm$ 0.358& 0.481$\pm$ 0.060&
 0.394$\pm$ 0.070& 0.719$\pm$ 0.117&
 0.142$\pm$  0.200\\
NGC4725&
 2.688$\pm$ 0.452& 0.633$\pm$ 0.123&
 0.128$\pm$ 0.109& 0.473$\pm$ 0.170&
 1.195$\pm$  0.275\\
NGC4736&
 1.543$\pm$ 0.508& 0.817$\pm$ 0.079&
 0.386$\pm$ 0.102& 0.359$\pm$ 0.169&
 1.064$\pm$  0.362\\
NGC4826&
 2.487$\pm$ 0.465& 0.440$\pm$ 0.139&
 0.643$\pm$ 0.131& 0.085$\pm$ 0.130&
 1.064$\pm$  0.454\\
NGC5055&
 1.964$\pm$ 0.359& 0.477$\pm$ 0.085&
 0.403$\pm$ 0.103& 0.530$\pm$ 0.157&
 0.950$\pm$  0.231\\
NGC5398&
 1.446$\pm$ 0.351& 0.616$\pm$ 0.250&
 0.515$\pm$ 0.348& 0.419$\pm$ 0.648&
 0.644$\pm$  1.041\\
NGC5408&
 1.699$\pm$ 0.231& 0.808$\pm$ 0.198&
 0.270$\pm$ 0.365& 0.281$\pm$ 0.577&
 0.303$\pm$  1.607\\
NGC5457&
 1.932$\pm$ 0.085& 0.404$\pm$ 0.027&
 0.449$\pm$ 0.043& 0.594$\pm$ 0.054&
 0.755$\pm$  0.087\\
NGC5474&
 1.793$\pm$ 0.861& 0.503$\pm$ 0.125&
 0.312$\pm$ 0.212& 0.473$\pm$ 0.323&
 1.305$\pm$  0.668\\
NGC5713&
 2.005$\pm$ 0.428& 0.585$\pm$ 0.154&
 0.388$\pm$ 0.158& 0.505$\pm$ 0.302&
 0.785$\pm$  0.615\\
NGC5866&
 4.732$\pm$ 1.320& 0.534$\pm$ 0.159&
 0.212$\pm$ 0.144& -0.086$\pm$ 0.173&
 2.137$\pm$  0.525\\
NGC6946&
 1.478$\pm$ 0.245& 0.695$\pm$ 0.053&
 0.456$\pm$ 0.070& 0.351$\pm$ 0.126&
 1.124$\pm$  0.238\\
NGC7331&
 3.376$\pm$ 0.823& 0.389$\pm$ 0.079&
 0.396$\pm$ 0.089& 0.217$\pm$ 0.118&
 1.465$\pm$  0.219\\
NGC7793&
 1.601$\pm$ 0.375& 0.537$\pm$ 0.074&
 0.467$\pm$ 0.093& 0.676$\pm$ 0.132&
 0.351$\pm$  0.203\\
\hline
\end{tabular}
\begin{list}{}{}
\item NB: The relation is : S$_{TIR}$= $\Sigma$ c$_i$~S$_i$ with S$_{TIR}$ and S$_i$ in W~kpc$^{-2}$.
\end{list}
\end{table*}

\clearpage
\newpage
\section{L$_{TIR}$ maps}

   
\begin{figure*}
    \centering
    \begin{tabular}{ p{5.8cm}p{5.8cm}p{5.8cm} }
    {\large DDO 053}~(3.61 Mpc, Im) &
    {\large Holmberg II}~(3.05 Mpc, Im) &
    {\large IC 342}~(3.28 Mpc, SABcd) \\    
    \includegraphics[width=5.8cm]{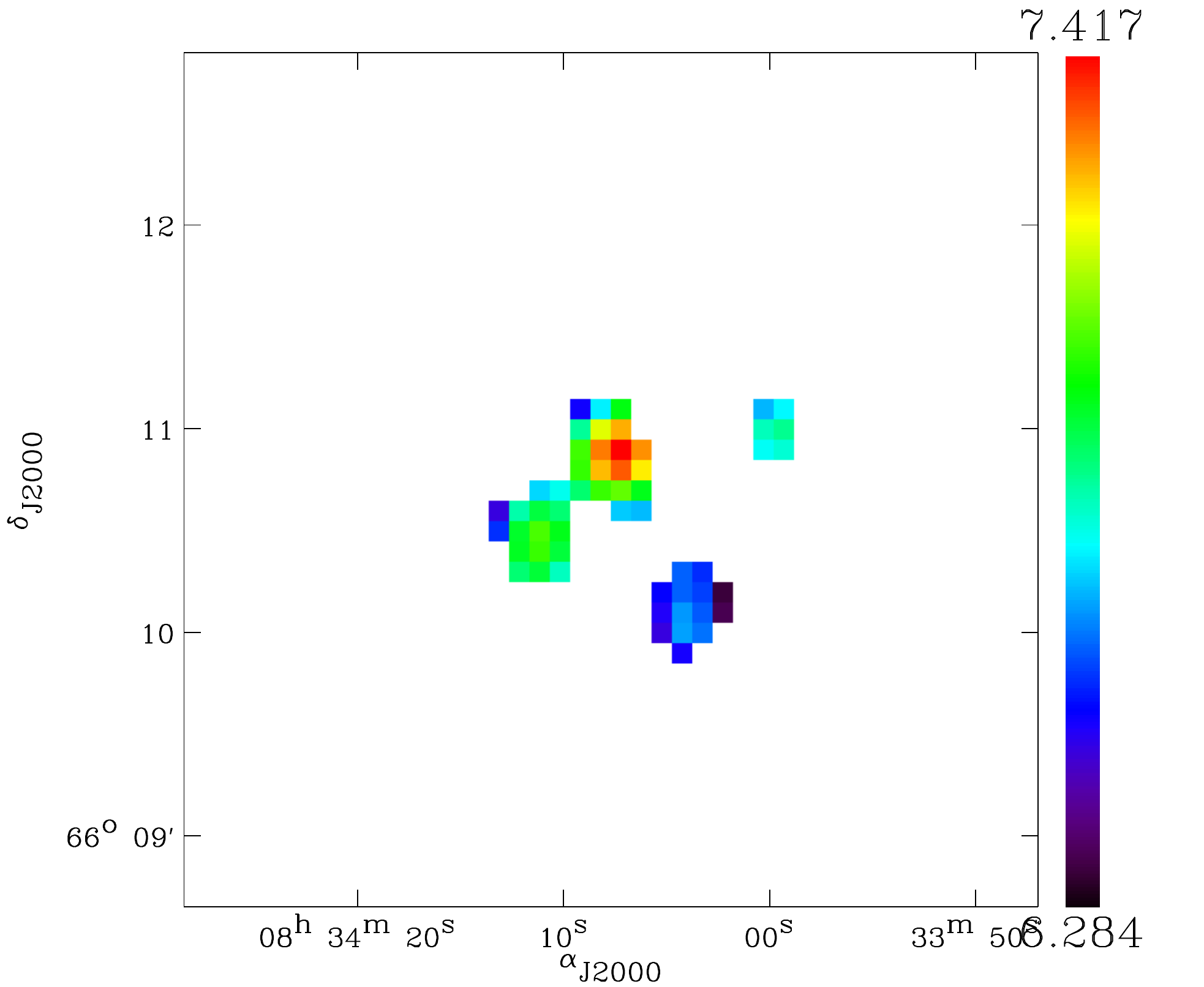} &
    \includegraphics[width=5.8cm]{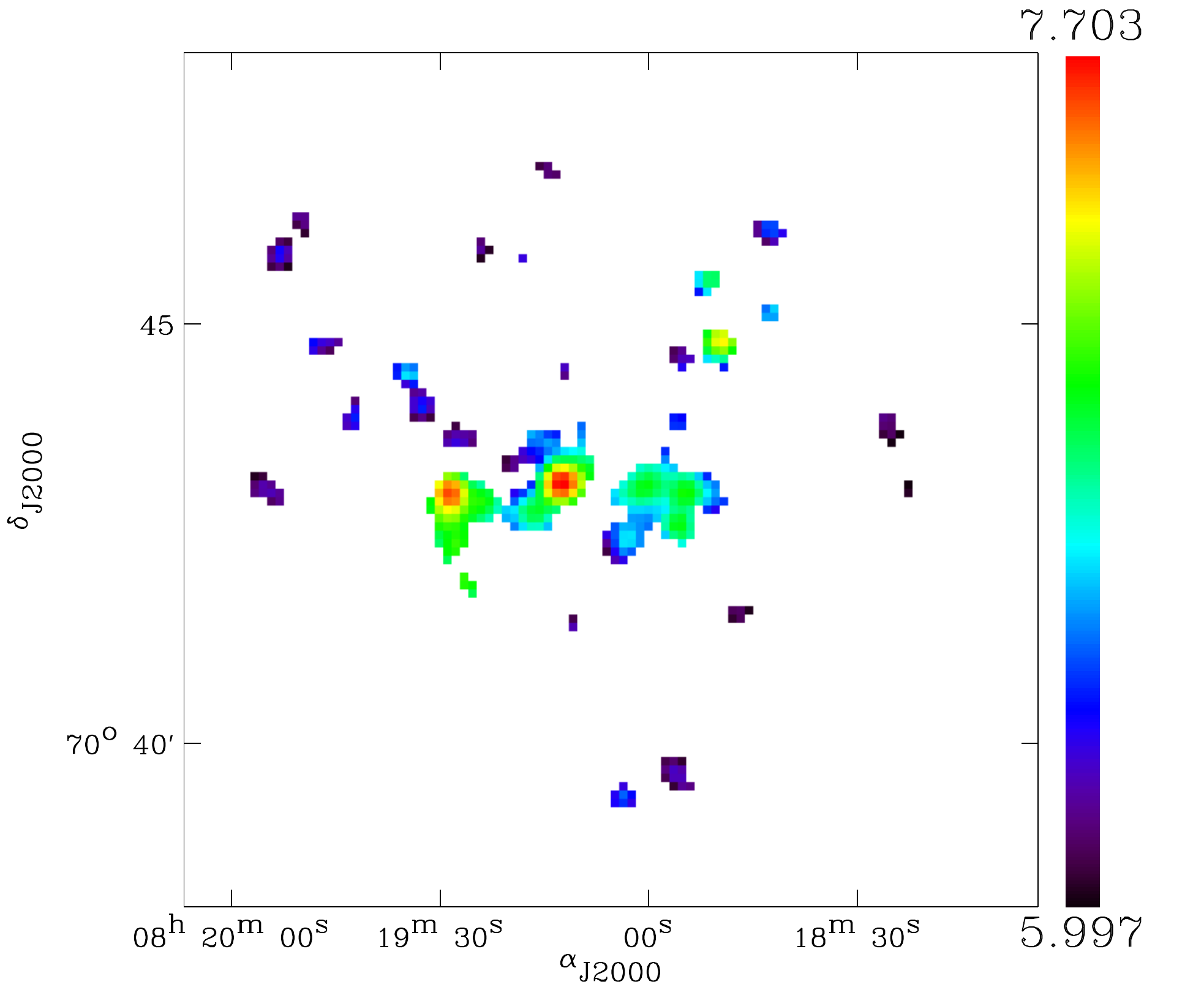}  &
    \includegraphics[width=5.8cm]{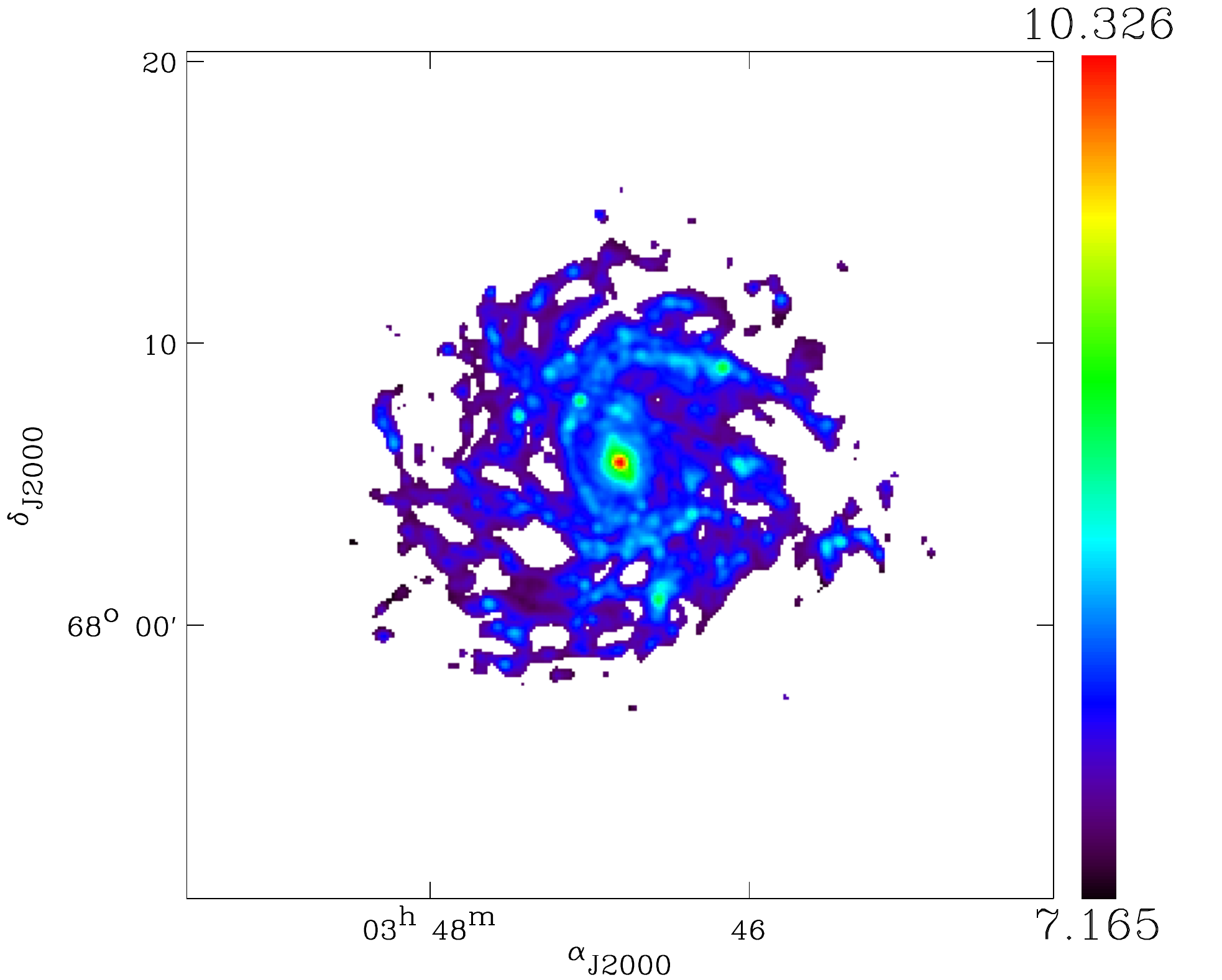} \\
    &&\\    
    {\large IC 2574}~(3.79 Mpc, SABm) &
    {\large M81 Dw B}~(3.6 Mpc, Im) &
    {\large NGC 337}~(19.3 Mpc, SBd) \\
    \includegraphics[width=5.8cm]{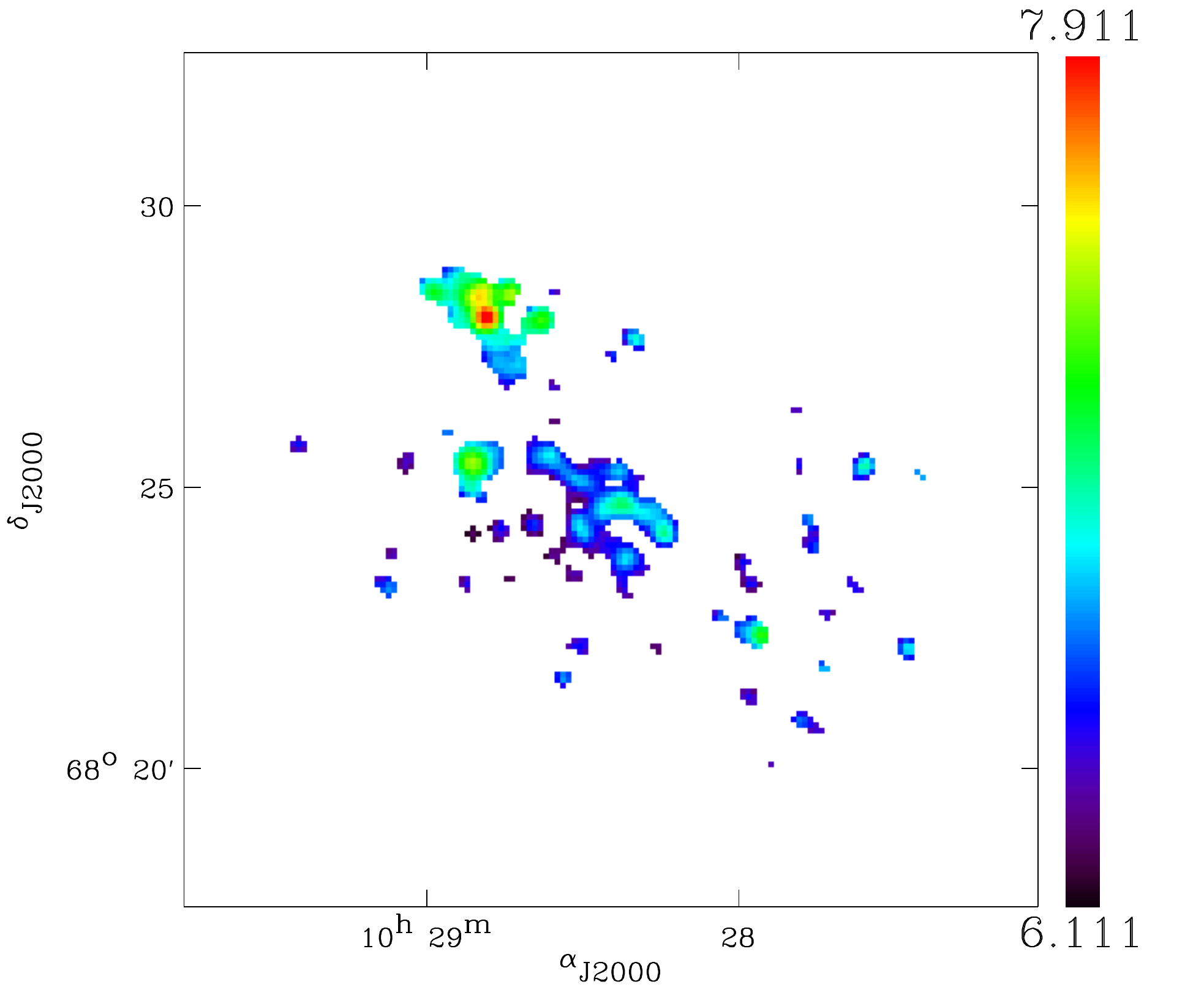} &
    \includegraphics[width=5.8cm]{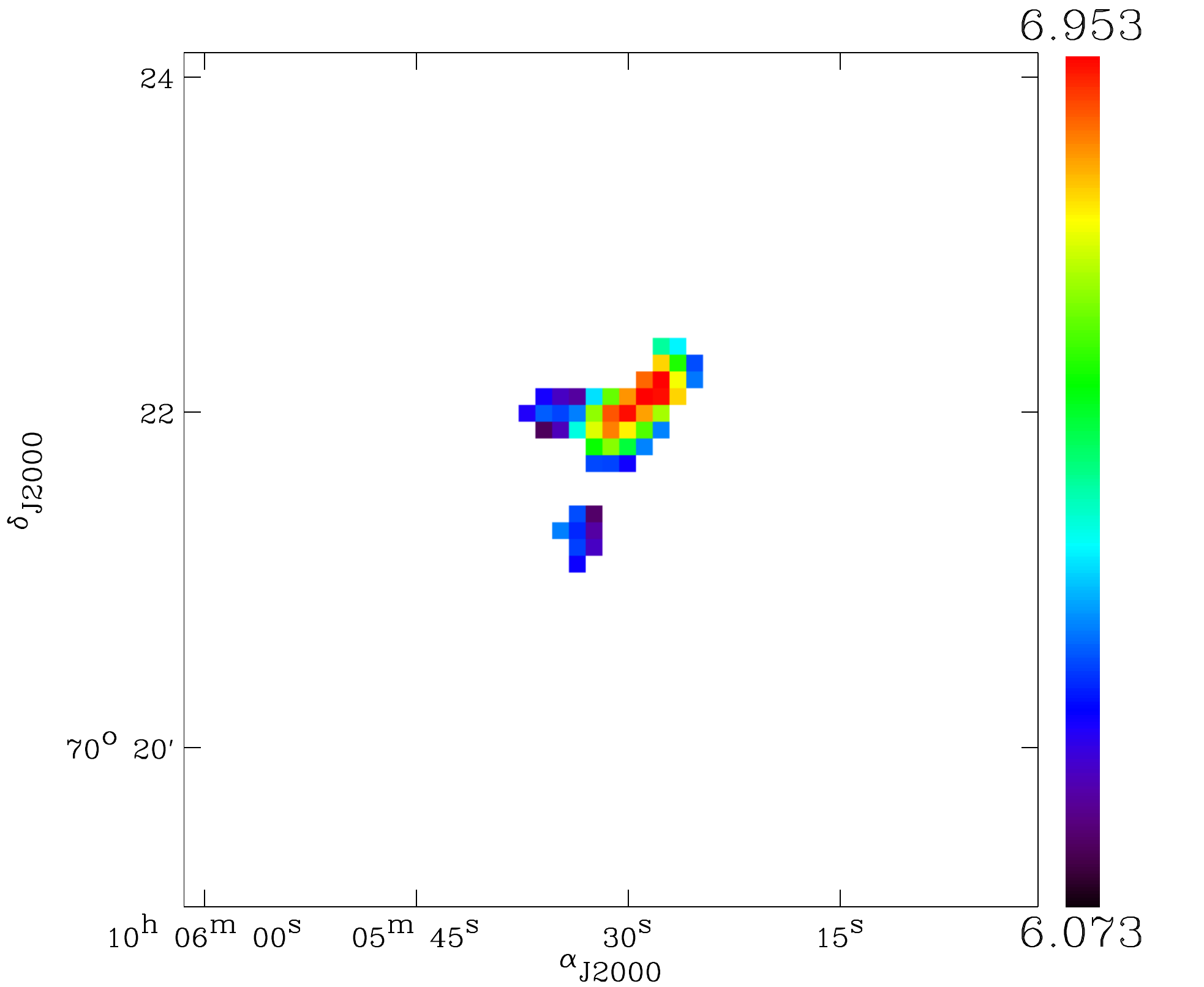} &
    \includegraphics[width=5.8cm]{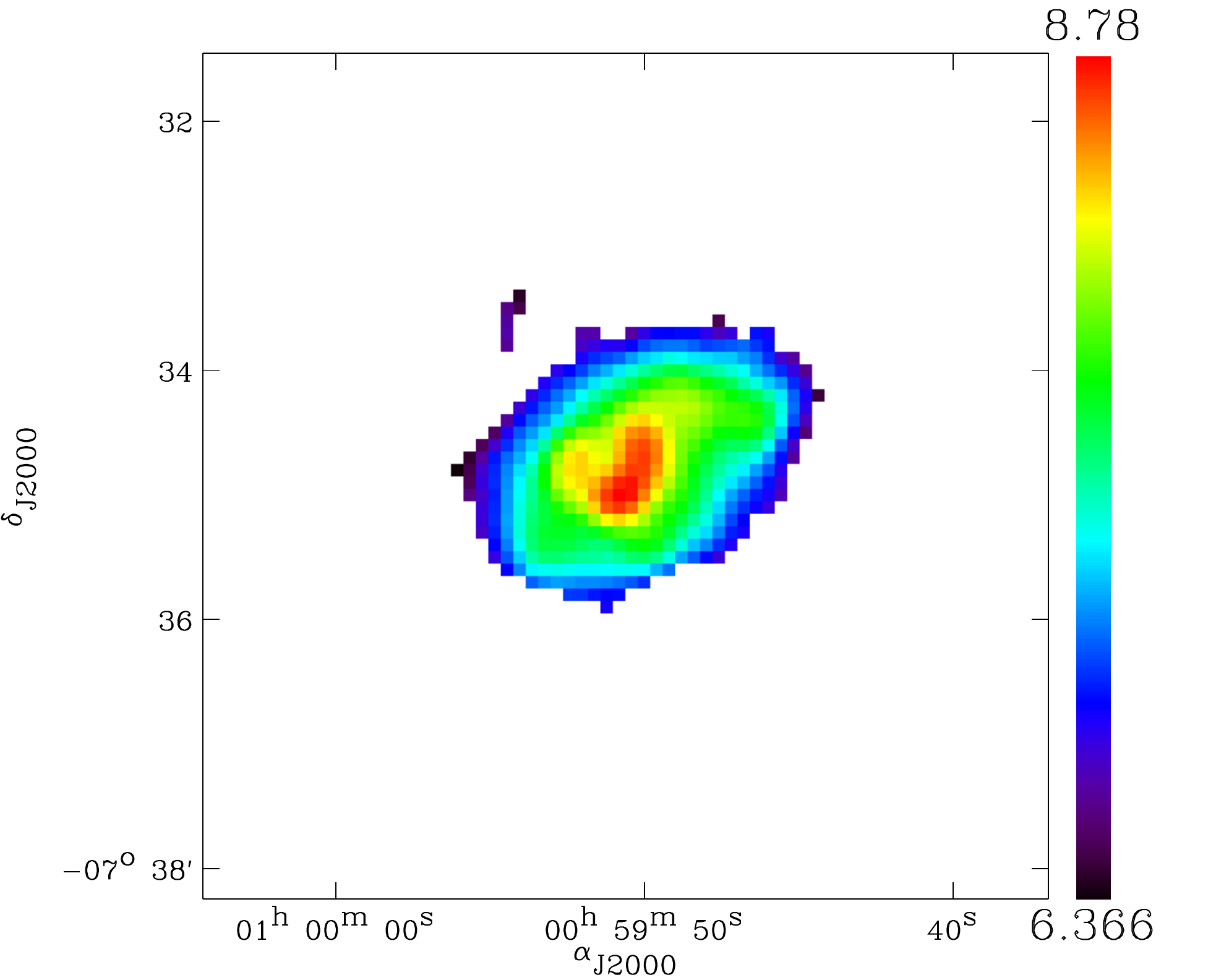} \\   
       &&\\        
    {\large NGC 628}~(7.2 Mpc, SAc) &
     {\large NGC 855}~(9.73 Mpc, E) &
    {\large NGC 925}~(9.12 Mpc, SABd) \\
     \includegraphics[width=5.8cm]{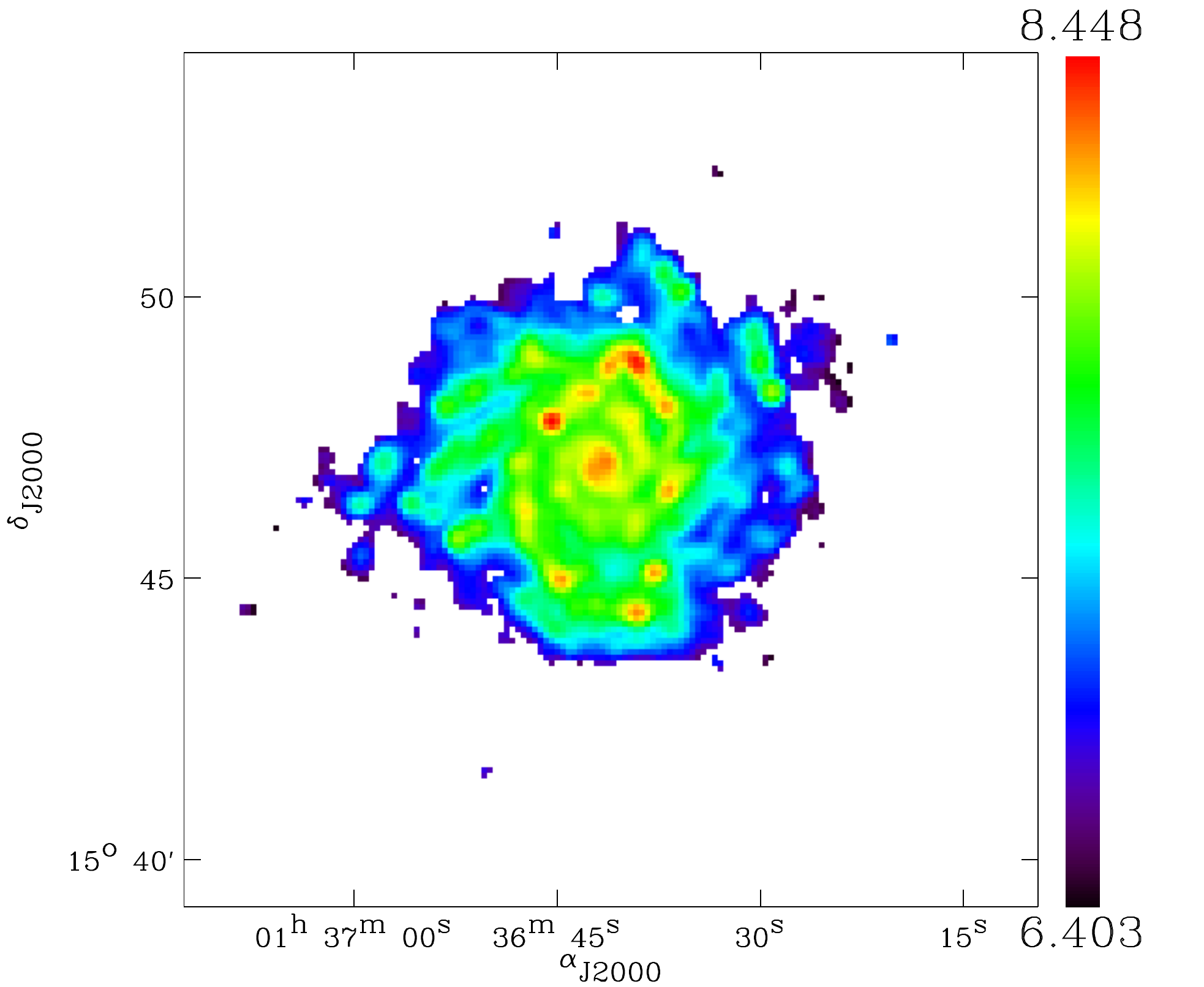} &
     \includegraphics[width=5.8cm]{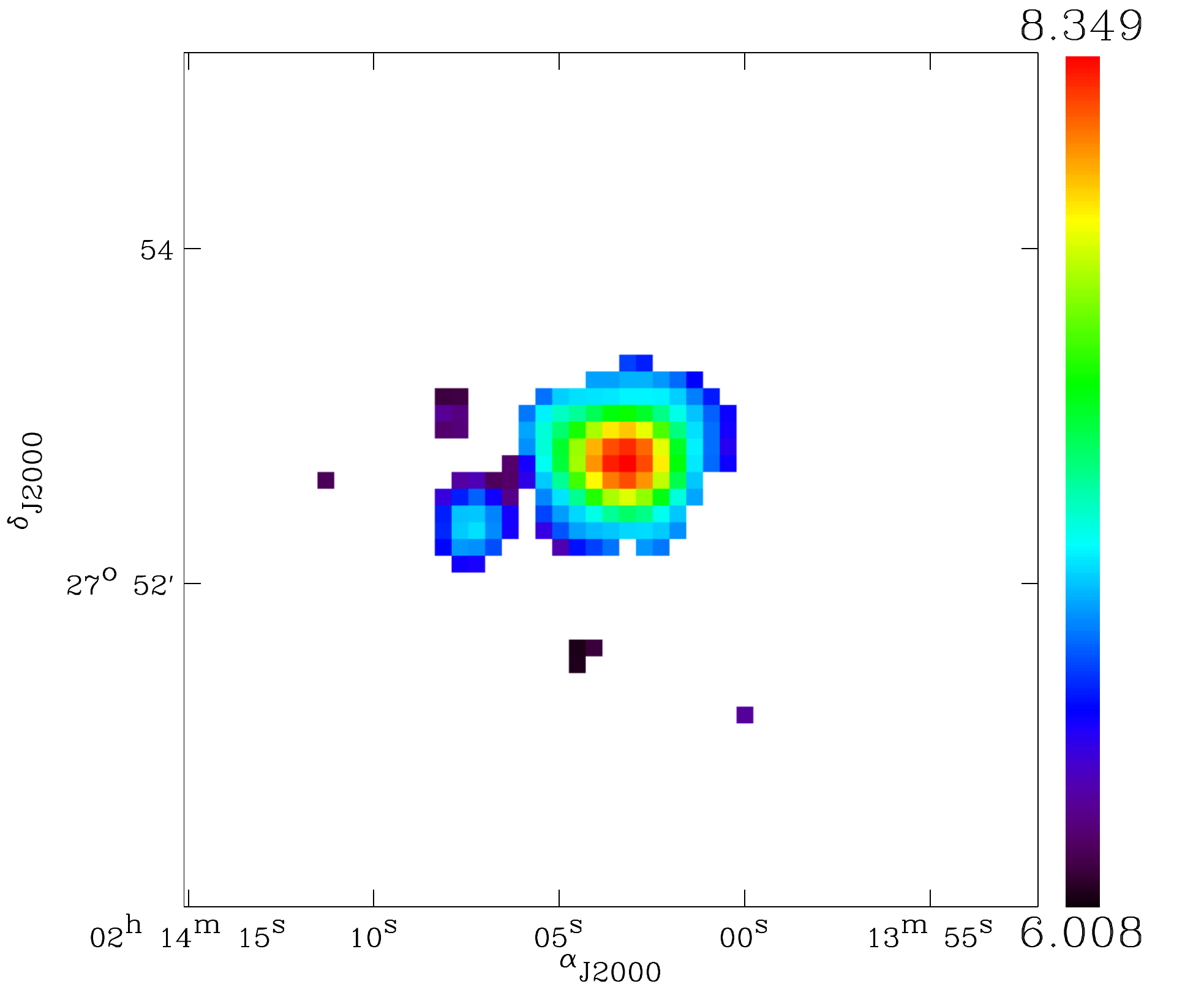} &     
    \includegraphics[width=5.8cm]{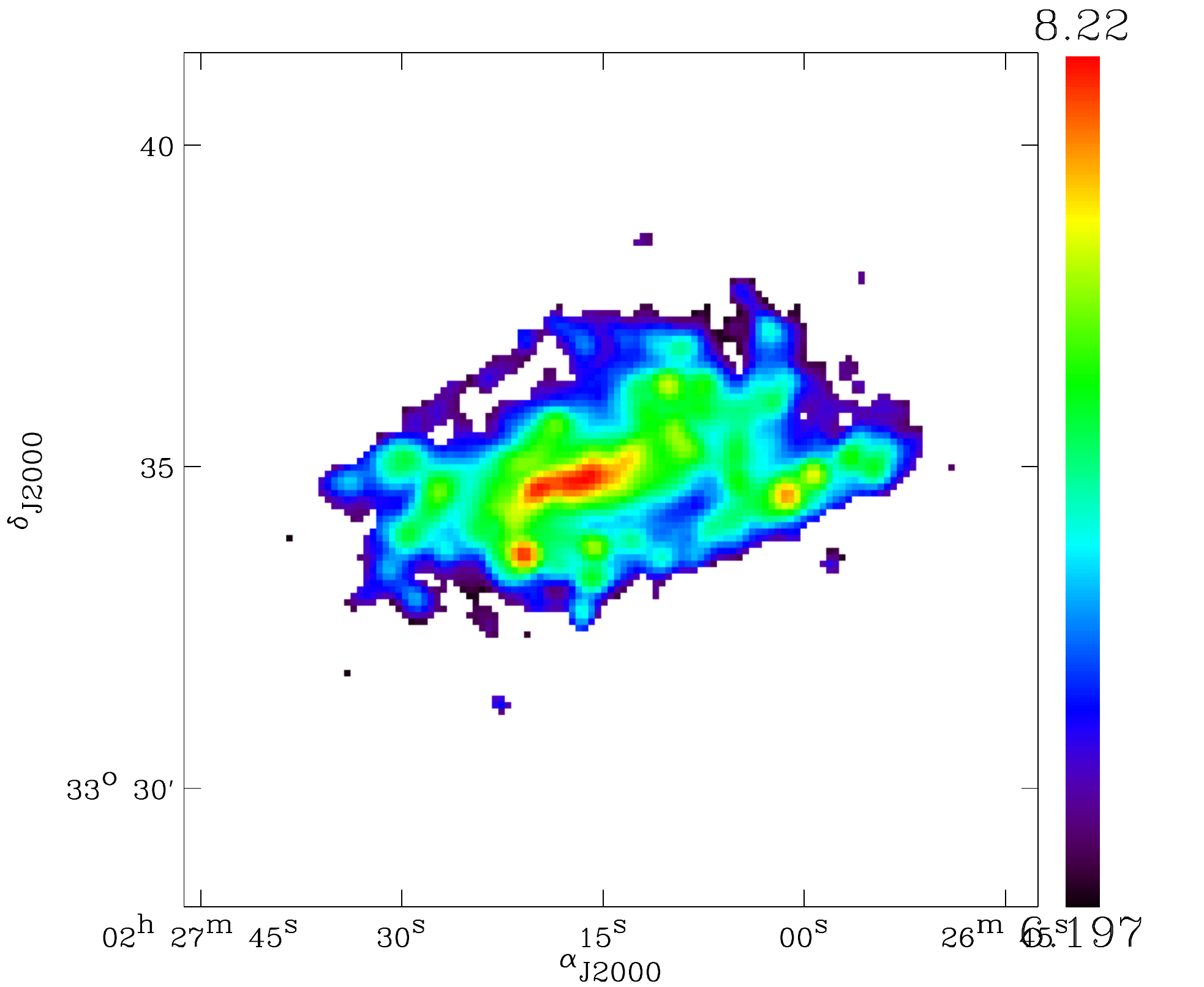} \\
    &&\\
    {\large NGC 1097}~(14.2 Mpc, SBb) &
     {\large NGC 1266}~(30.6 Mpc, SB0) &
    {\large NGC 1291}~(10.4 Mpc, SBa) \\
        \includegraphics[width=5.8cm]{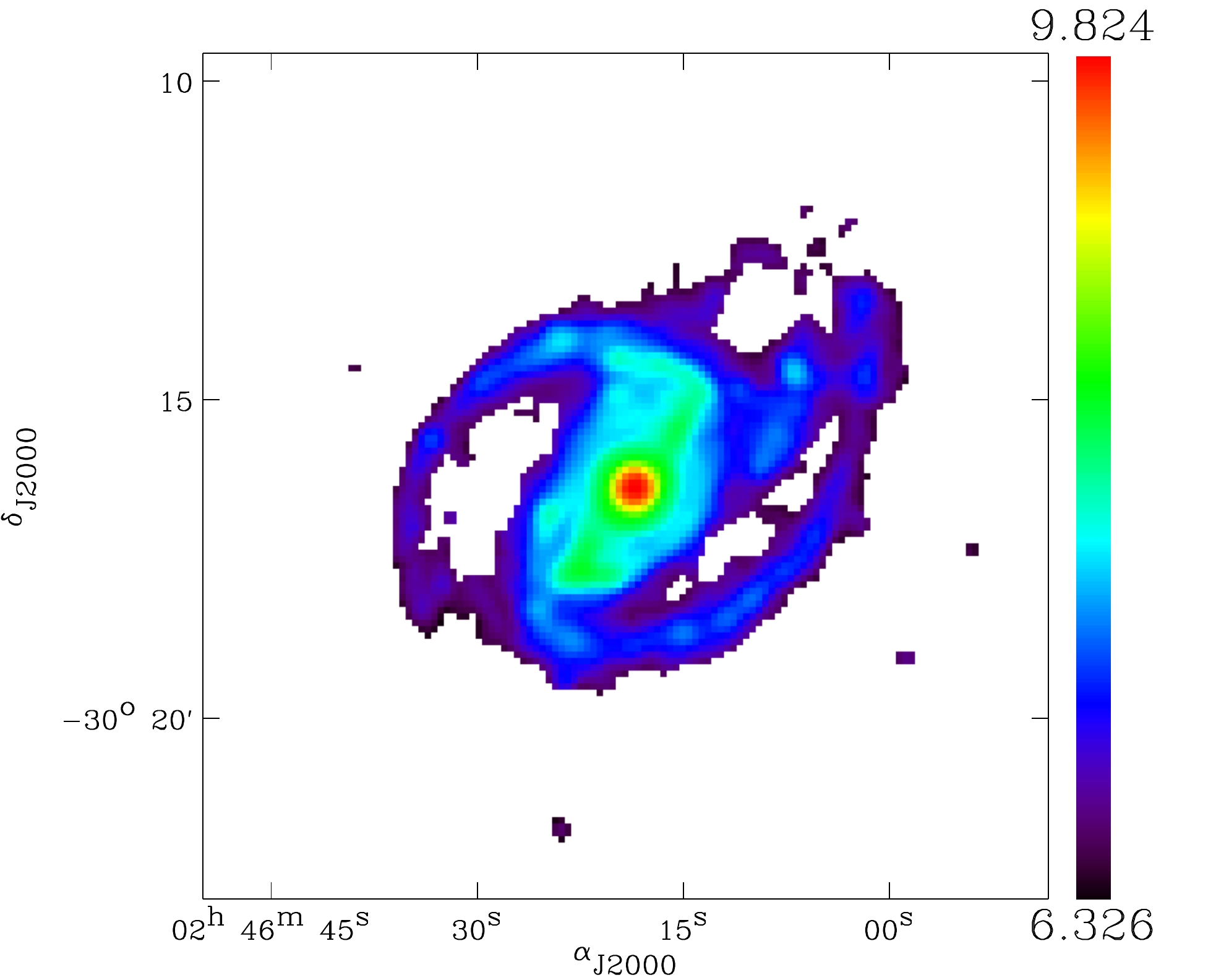} &
            \includegraphics[width=5.8cm]{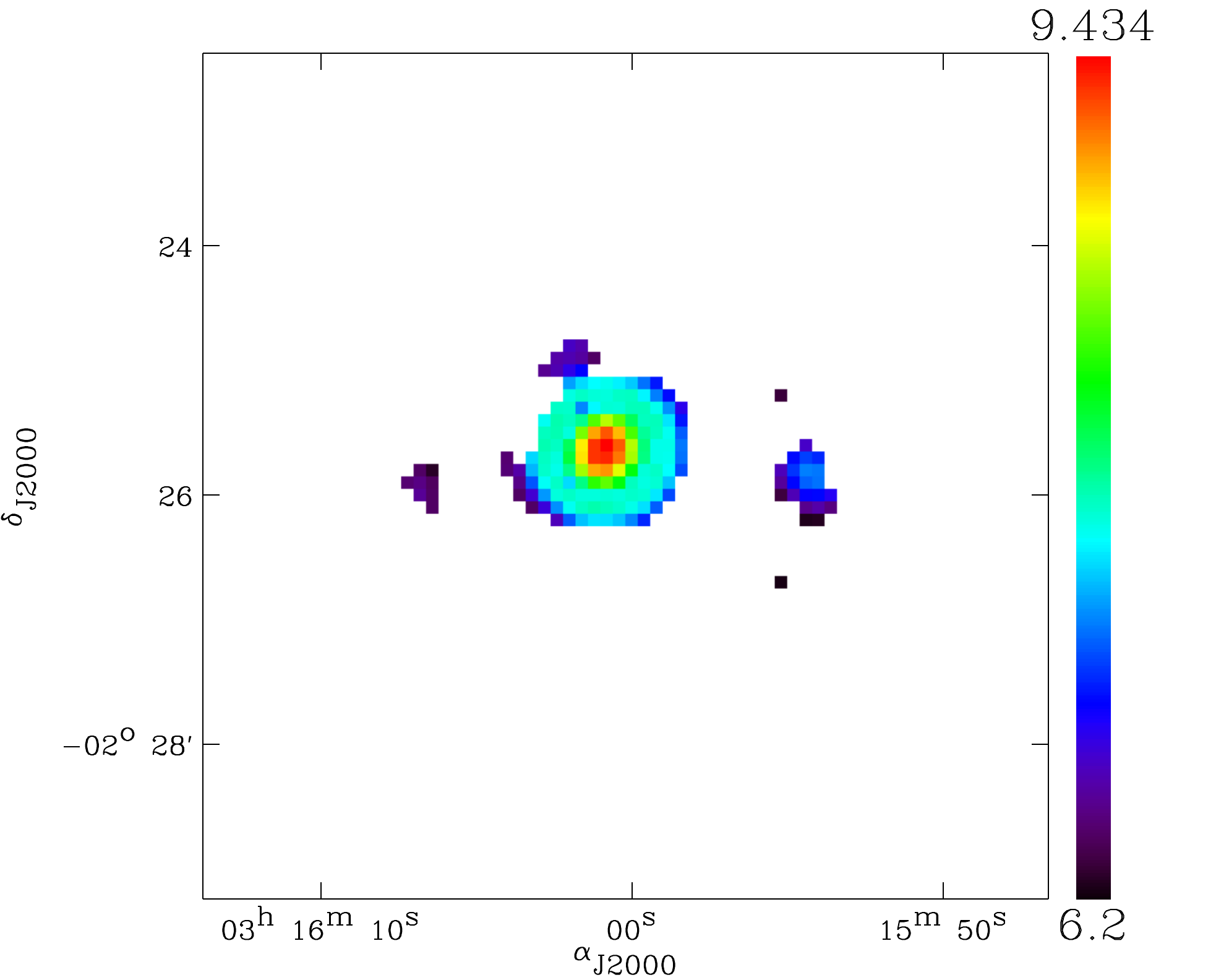} &          
    \includegraphics[width=5.8cm]{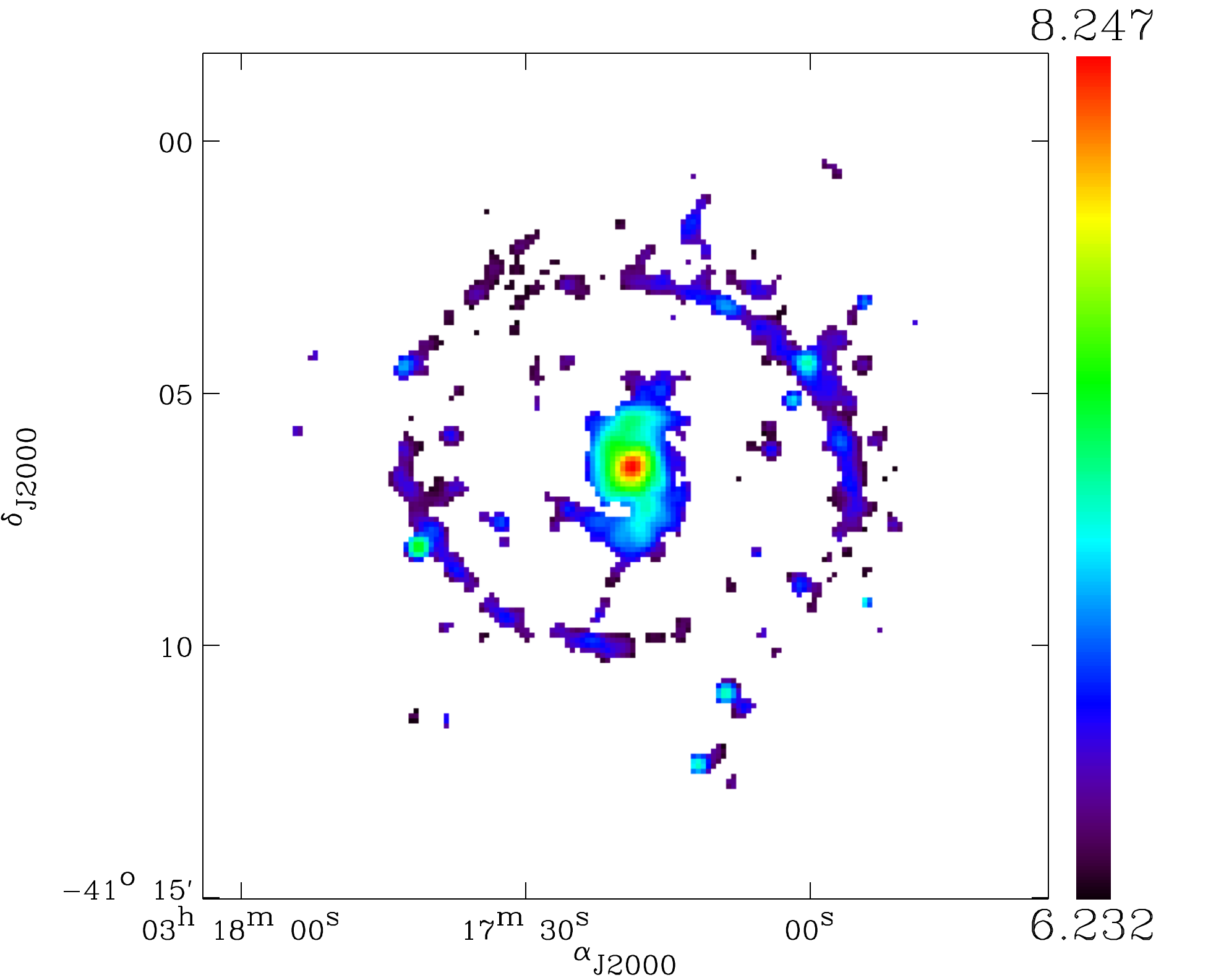} \\
    	\end{tabular}
		\caption{TIR brightnesses maps of the Kingfish sample in \lsun~kpc$^{-2}$ (log scale, resolution of SPIRE 250 \mic)}
	\label{LTIR_Maps}
\end{figure*}

\newpage
\addtocounter {figure}{-1}

\begin{figure*}
    \centering
    \begin{tabular}{ p{5.8cm}p{5.8cm}p{5.8cm} }    
    {\large NGC 1316}~(21.0 Mpc, SAB0) &
    {\large NGC 1377}~(24.6 Mpc, S0) &
       {\large NGC 1482}~(22.6 Mpc, SA0) \\
    \includegraphics[width=5.8cm]{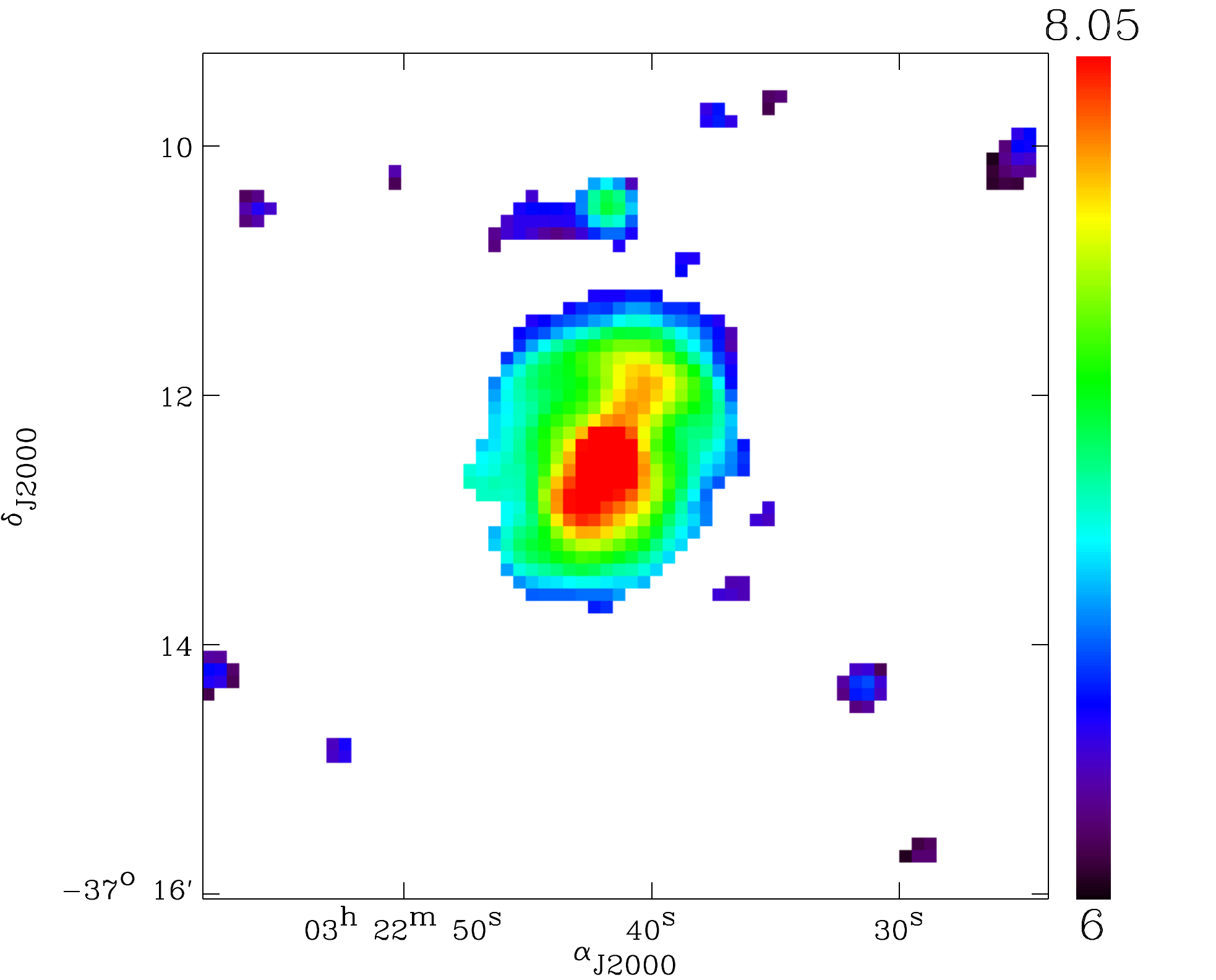}&
    \includegraphics[width=5.8cm]{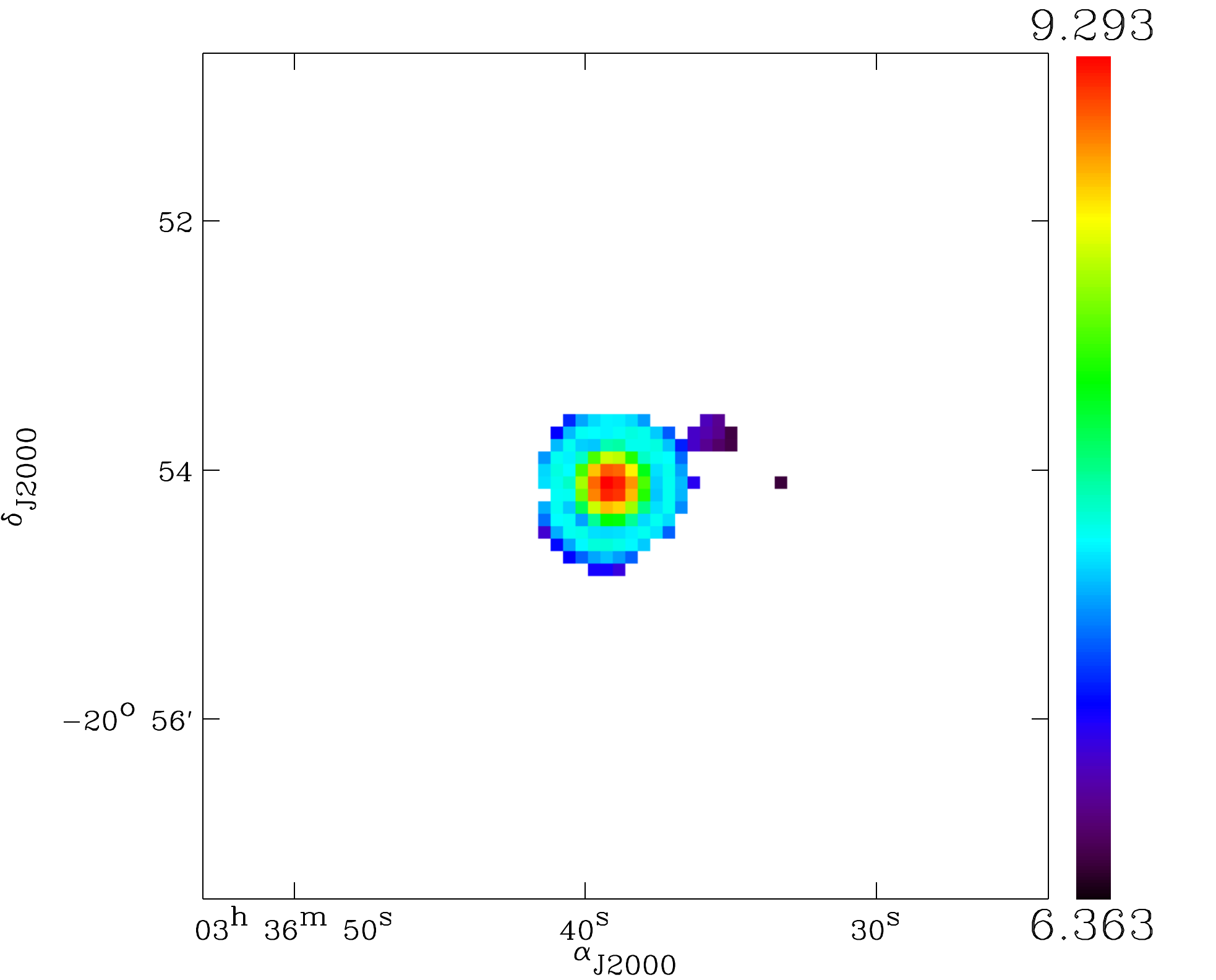} &    
    \includegraphics[width=5.8cm]{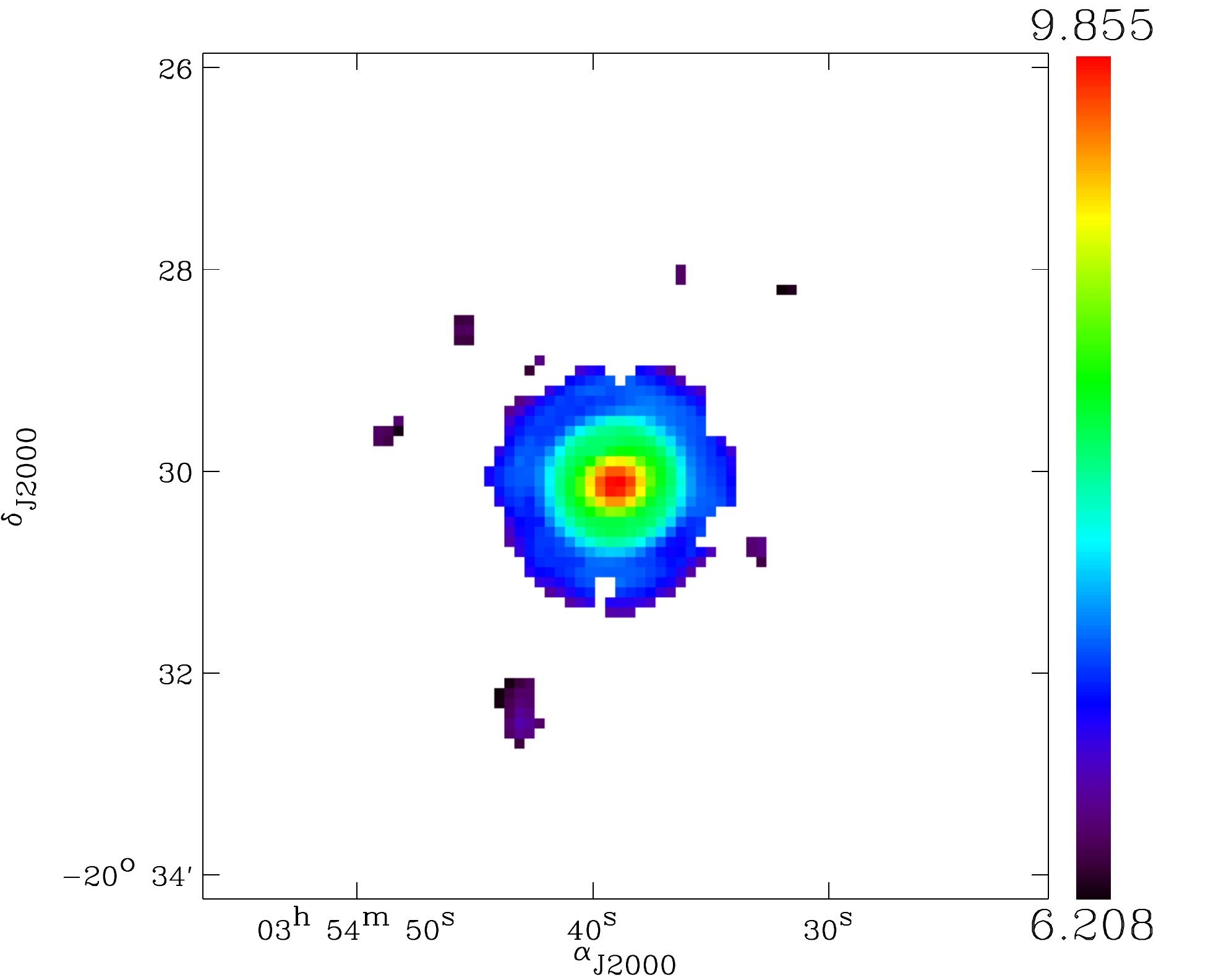} \\
    &&\\    
        {\large NGC 1512}~(11.6 Mpc, SBab) &
    {\large NGC 2146}~(17.2 Mpc, SBab) &
    {\large NGC 2798}~(25.8 Mpc, SBa) \\
    \includegraphics[width=5.8cm]{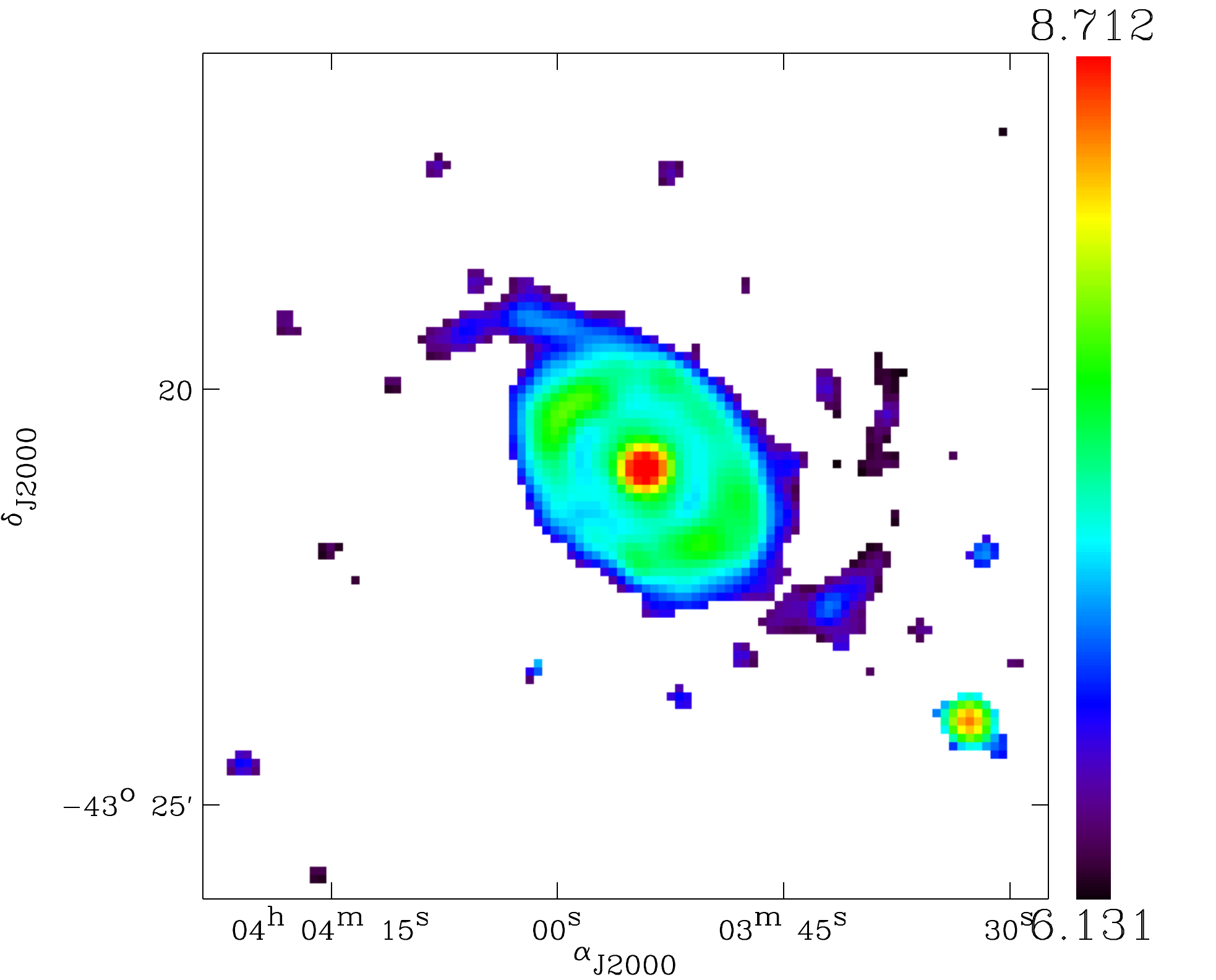} &
    \includegraphics[width=5.8cm]{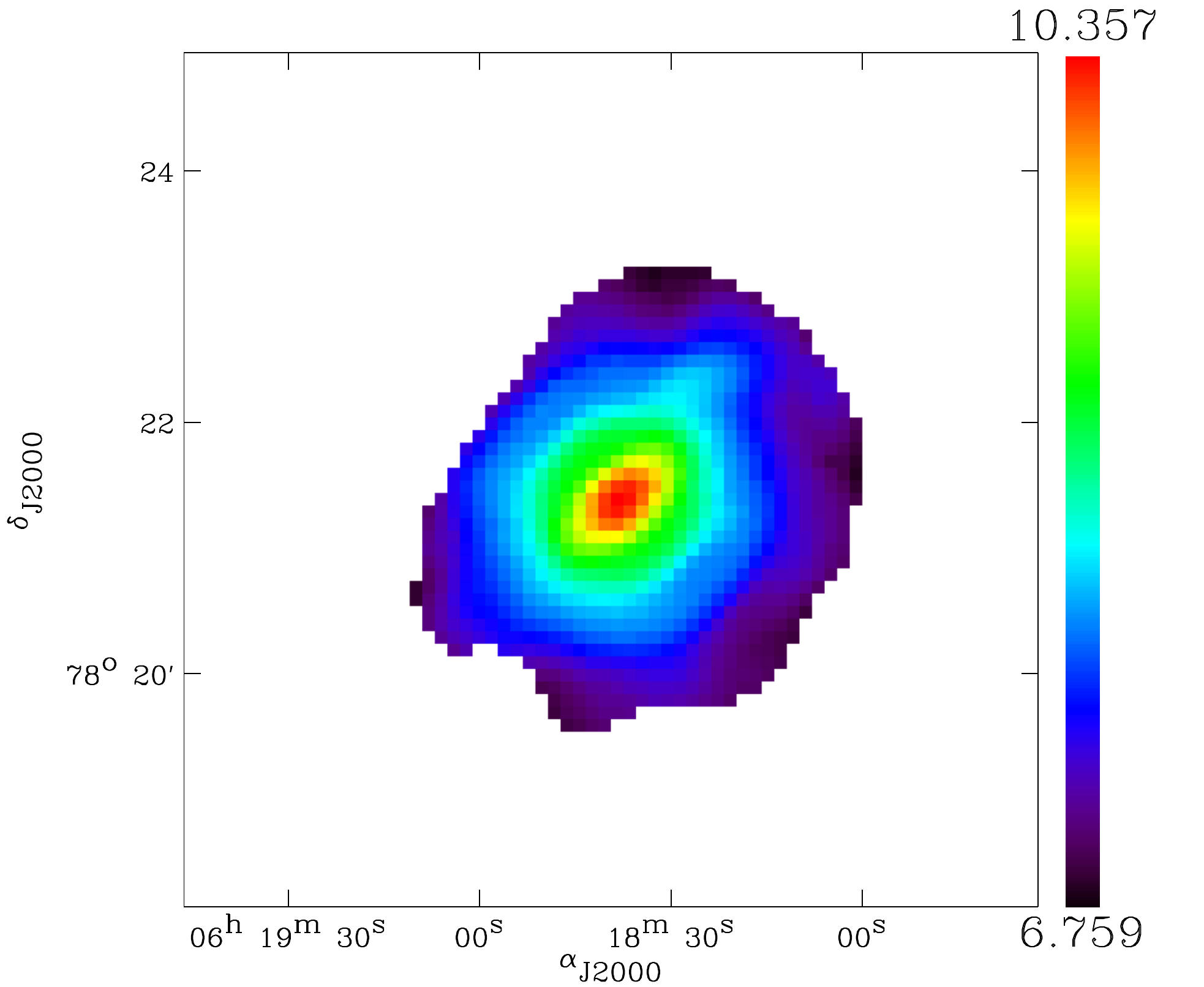} &
    \includegraphics[width=5.8cm]{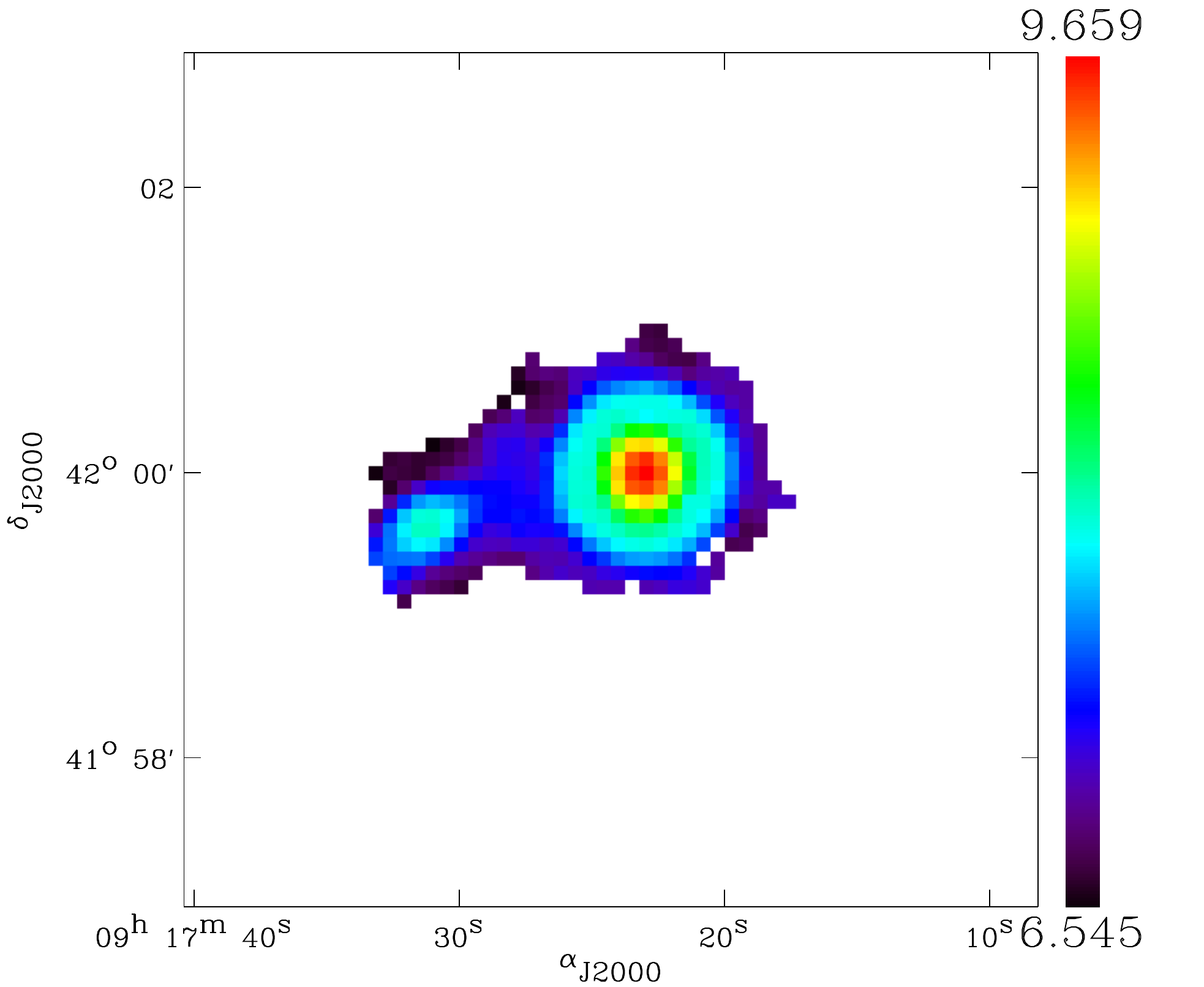} \\
    &&\\    
    {\large NGC 2841}~(14.1 Mpc, SAb)&
       {\large NGC 2915}~(3.78 Mpc, I0) &
    {\large NGC 2976}~(3.55 Mpc, SAc) \\
    \includegraphics[width=5.8cm]{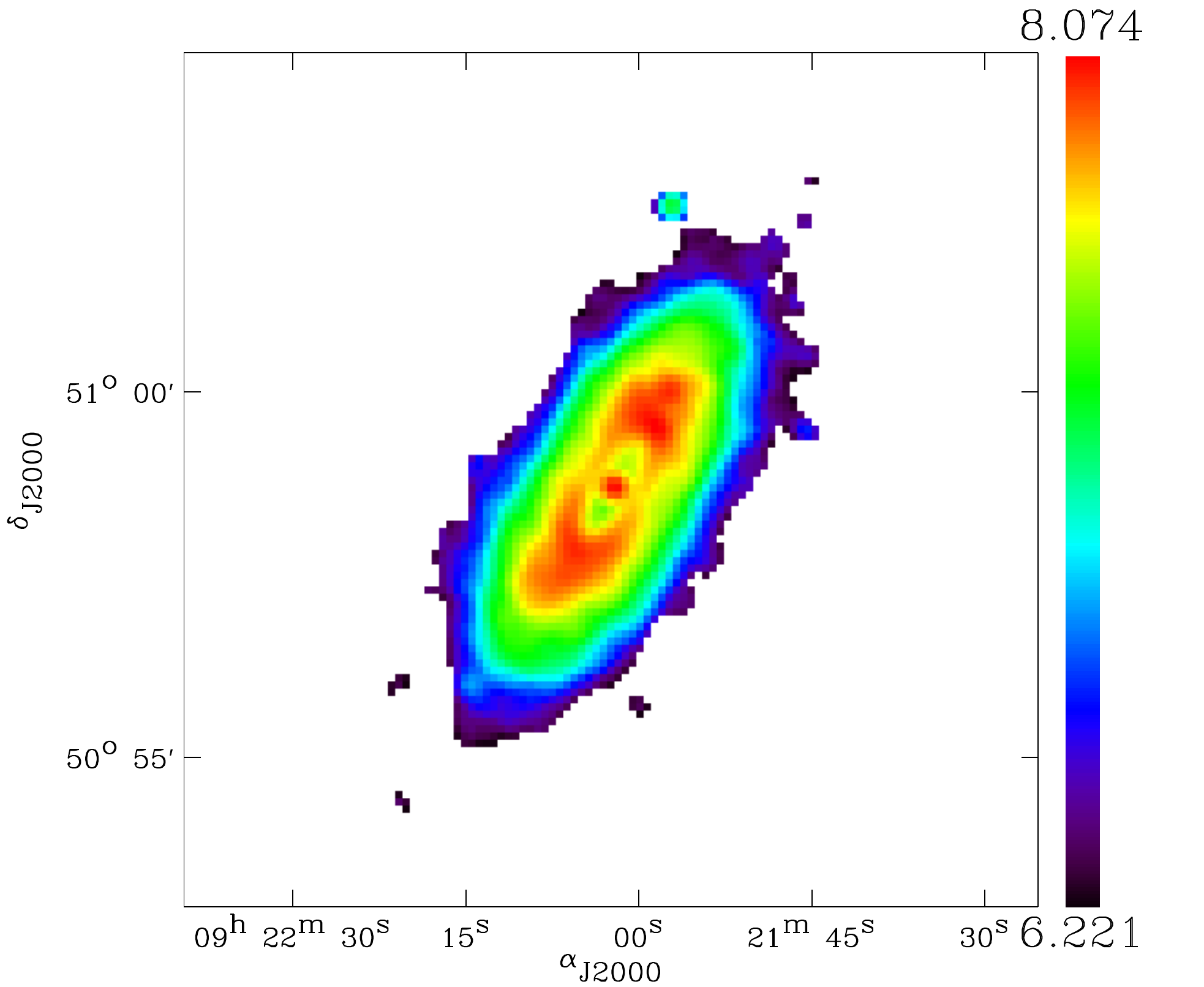} &
    \includegraphics[width=5.8cm]{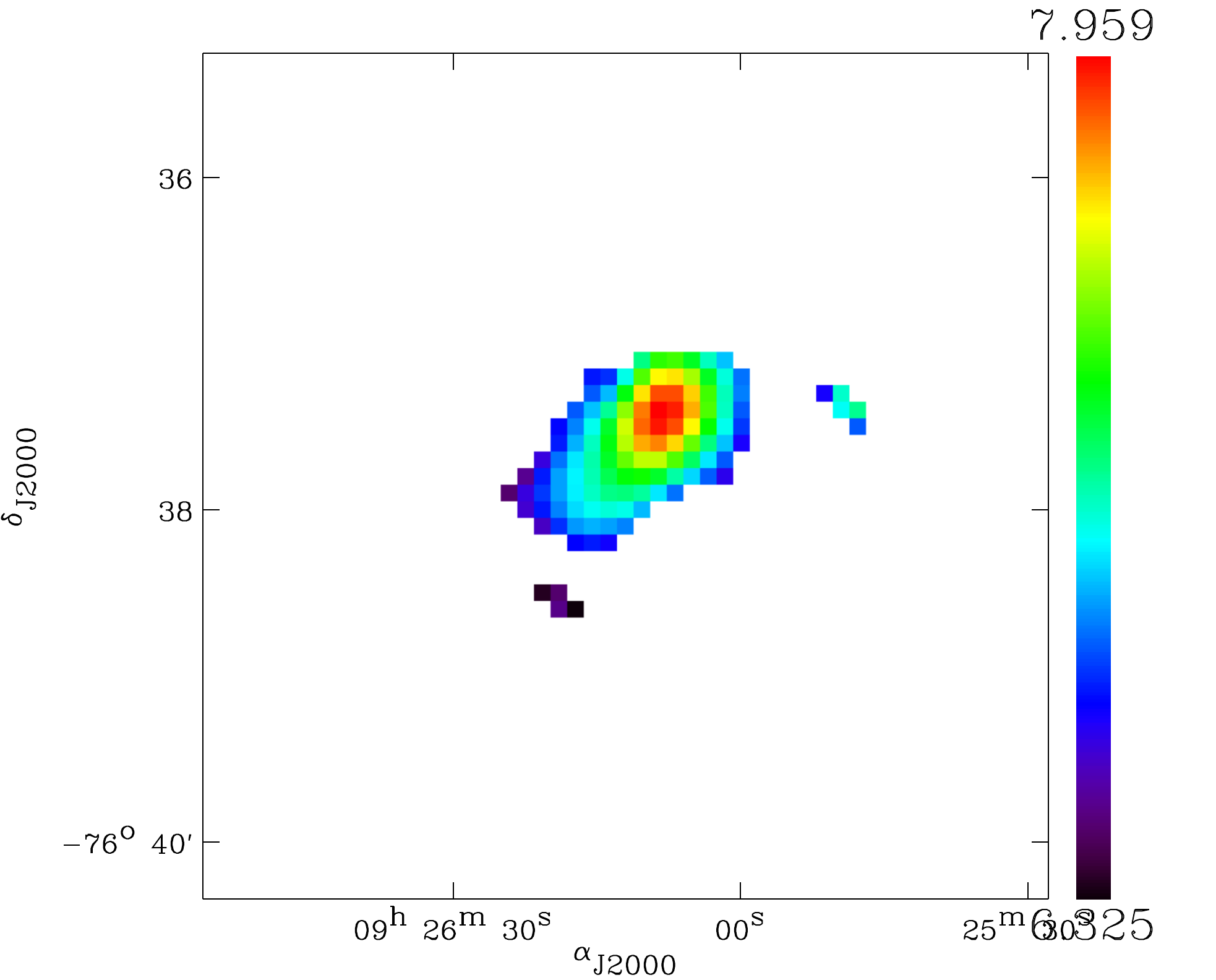} &
    \includegraphics[width=5.8cm]{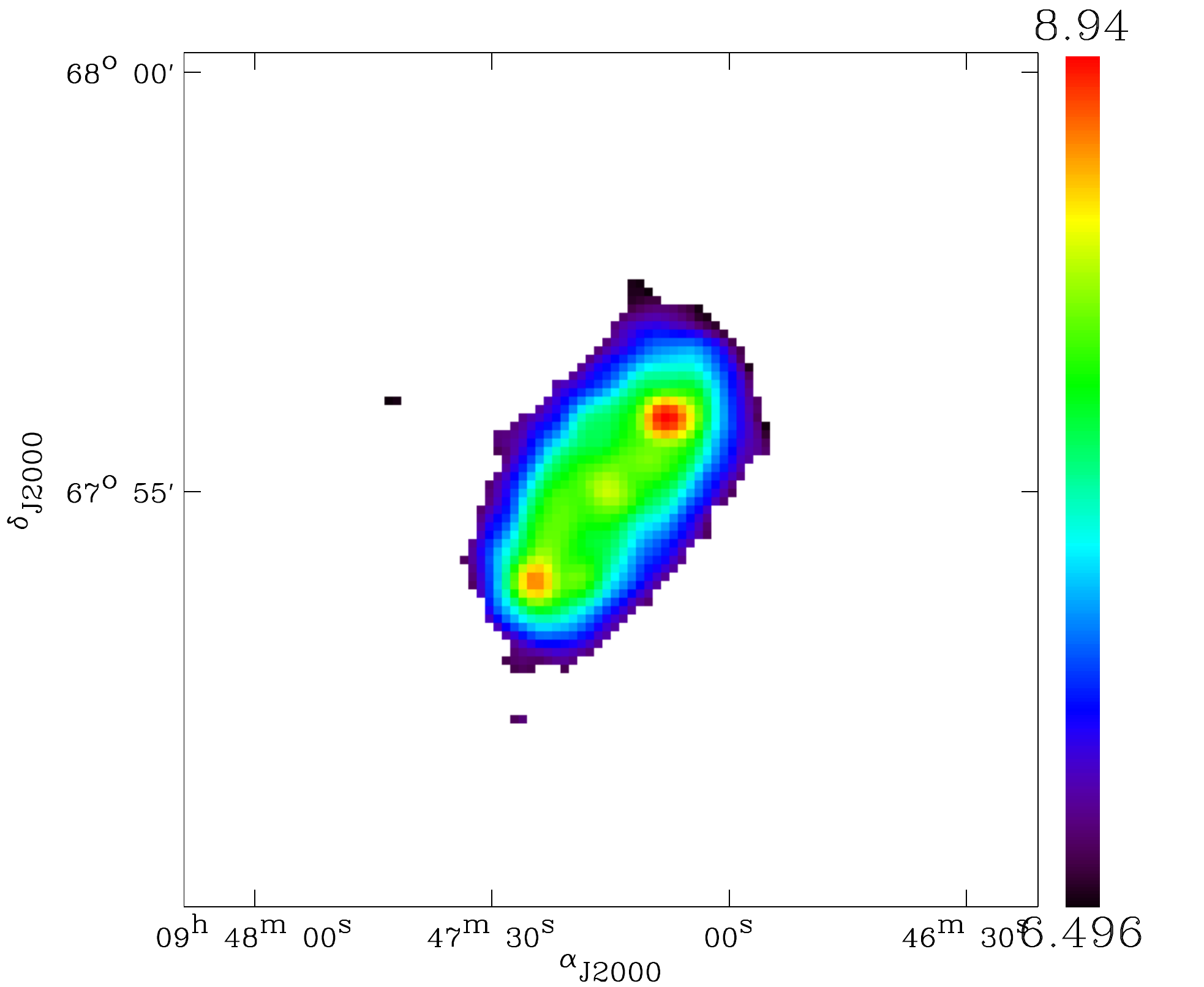} \\
    &&\\       
    {\large NGC 3049}~(19.2 Mpc, SBab) &           
    {\large NGC 3077}~(3.83 Mpc, I0pec) &
    {\large NGC 3184}~(11.7 Mpc, SABcd) \\
    \includegraphics[width=5.8cm]{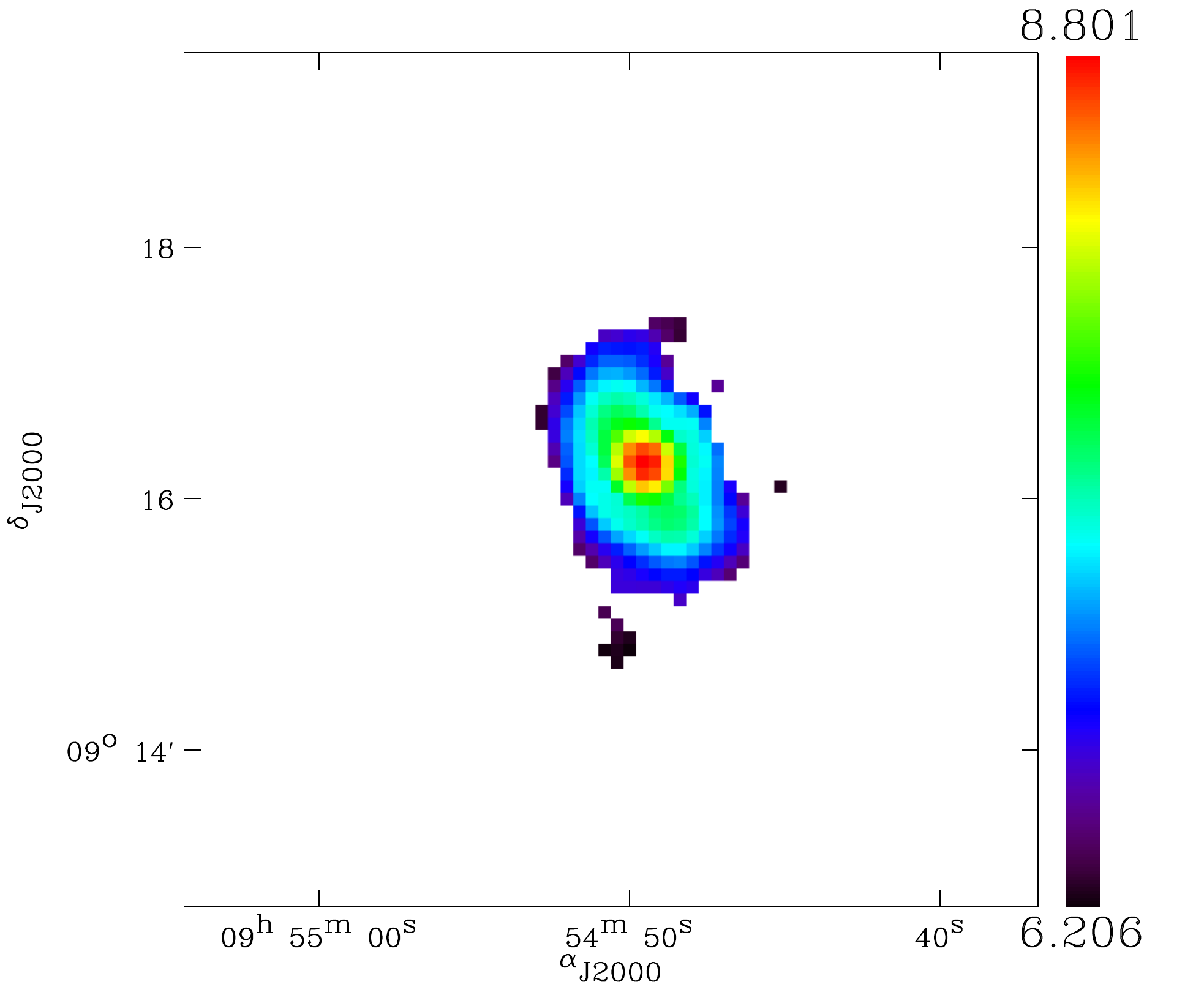} &
   \includegraphics[width=5.8cm]{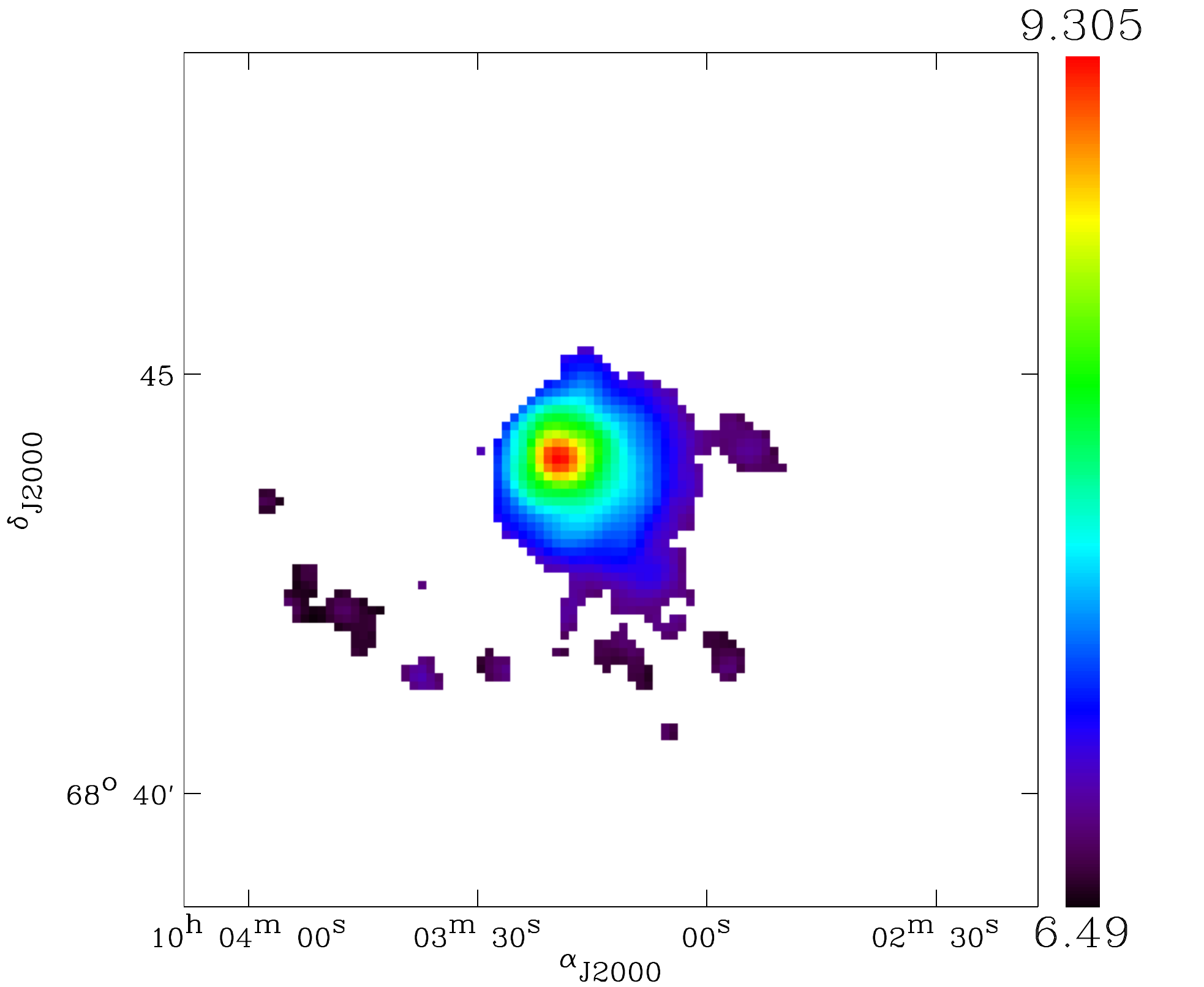} &
     \includegraphics[width=5.8cm]{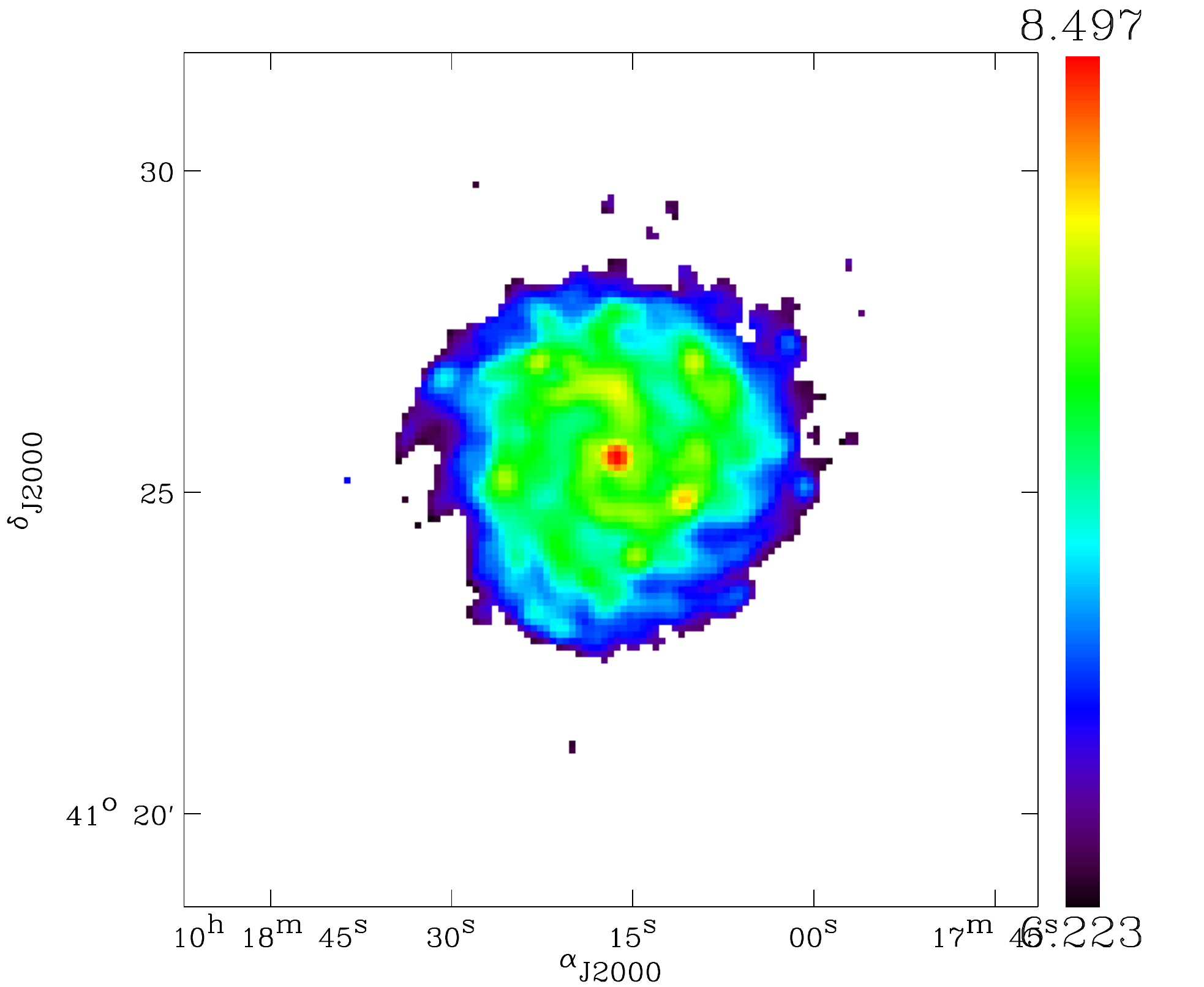} \\
    	\end{tabular}
		\caption{continued}
\end{figure*}

\newpage
\addtocounter {figure}{-1}

\begin{figure*}
    \centering
    \begin{tabular}{ p{5.8cm}p{5.8cm}p{5.8cm} }    
    {\large NGC 3190}~(19.3 Mpc, SAap) &
    {\large NGC 3198}~(14.1 Mpc, SBc) &
        {\large NGC 3265}~(19.6 Mpc, E) \\
      \includegraphics[width=5.8cm]{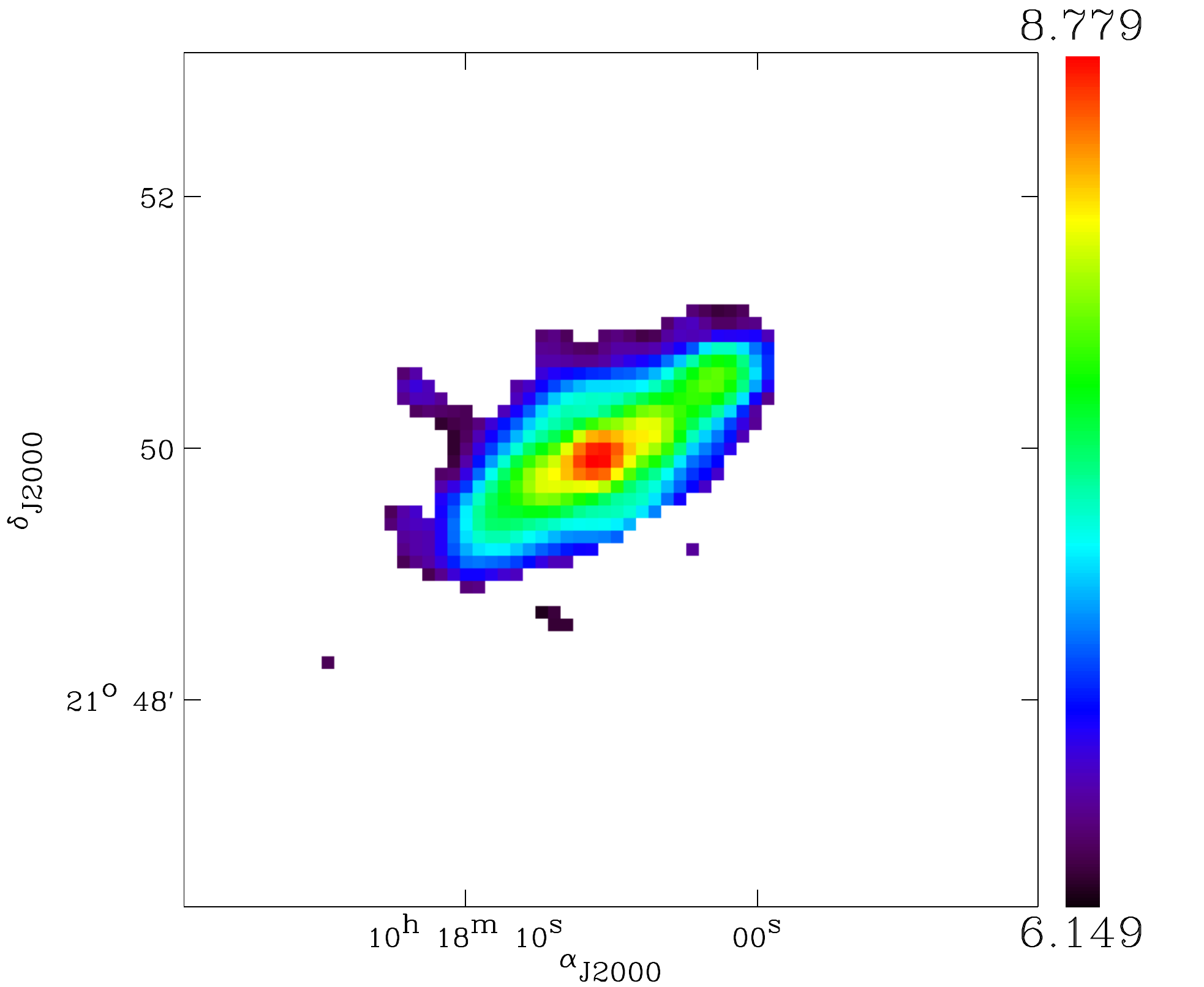}&   
    \includegraphics[width=5.8cm]{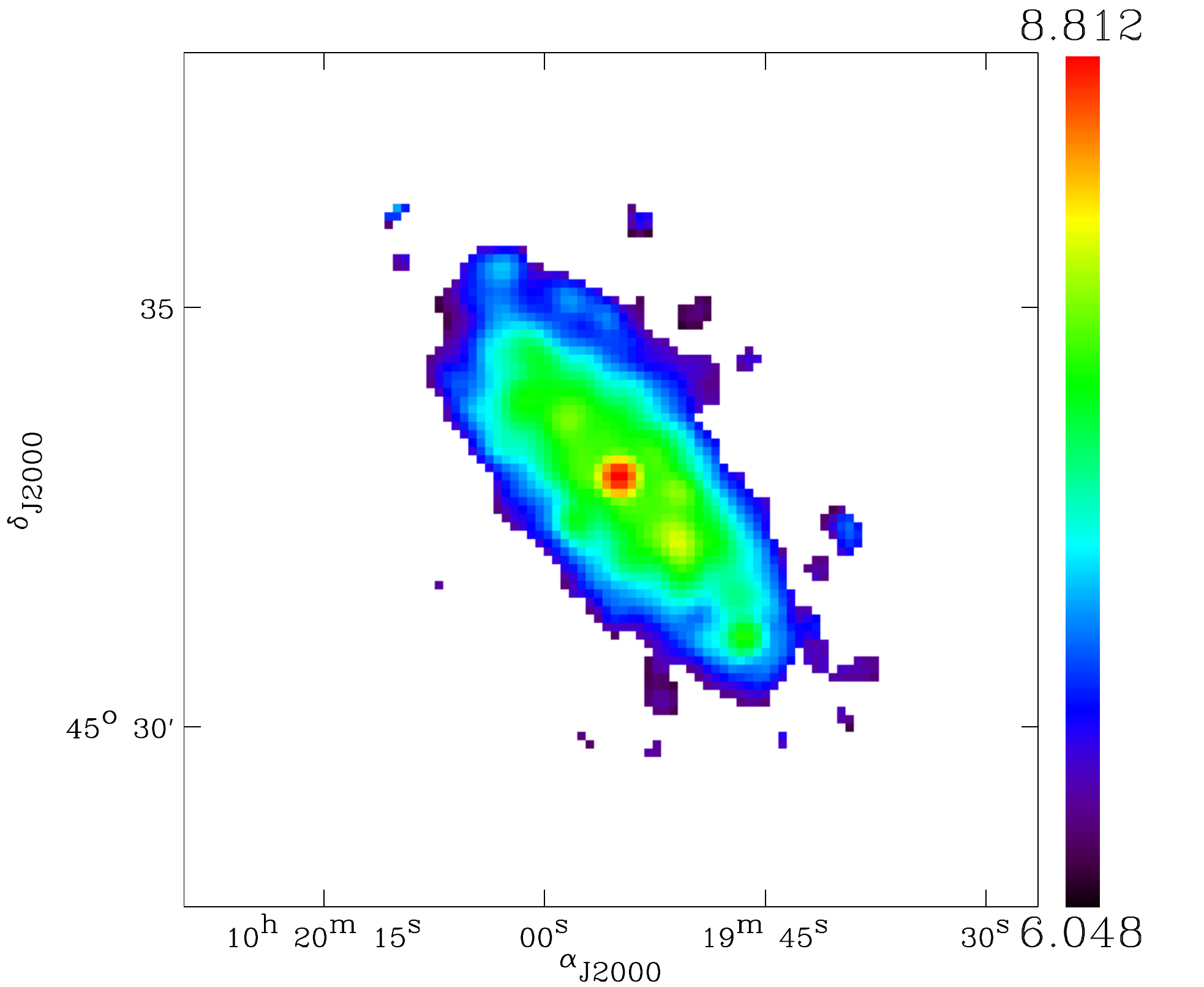} &
    \includegraphics[width=5.8cm]{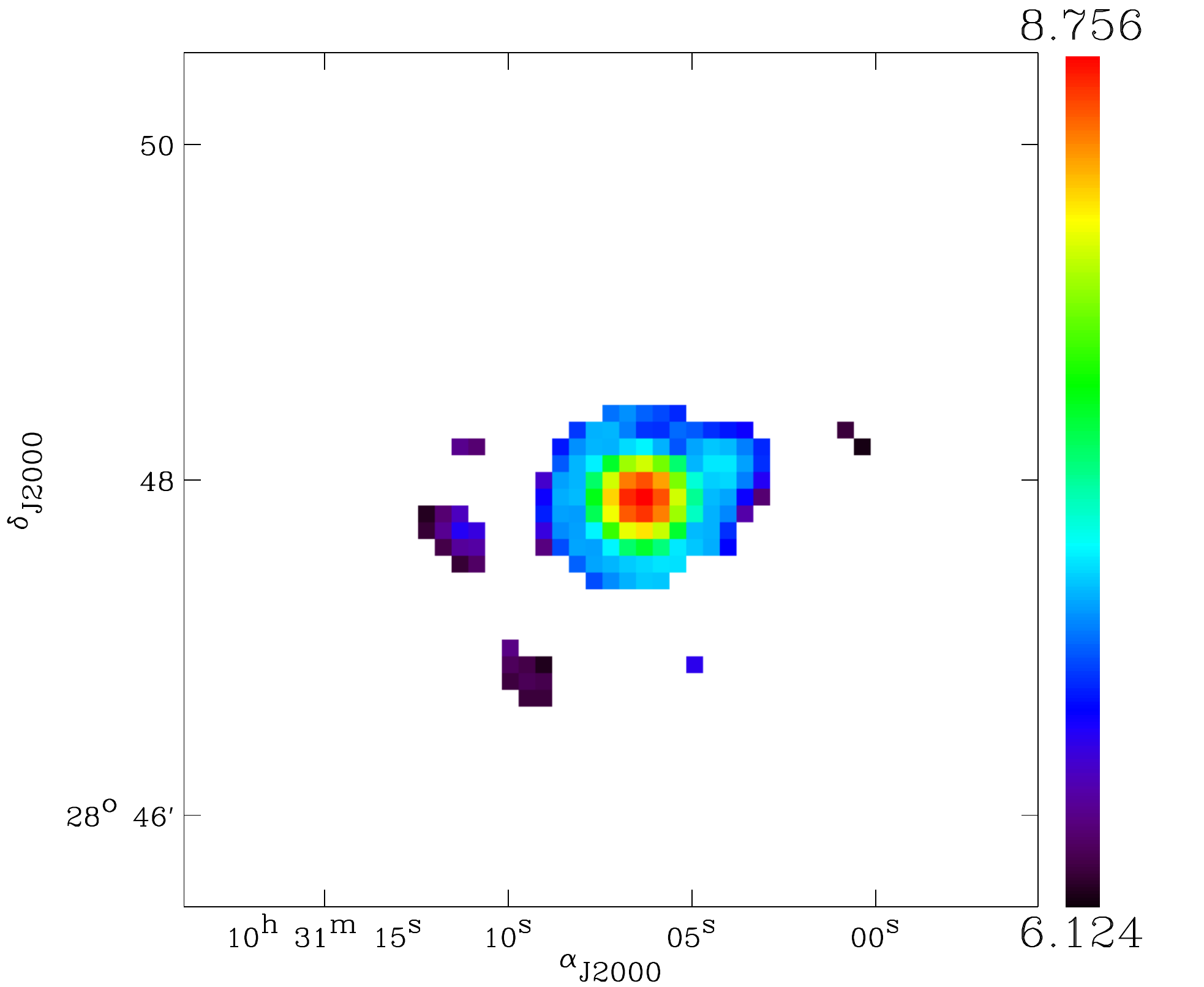} \\
        &&\\
    {\large NGC 3351}~(9.33 Mpc, SBb) &
    {\large NGC 3521}~(11.2 Mpc, SABbc) &
    {\large NGC 3621}~(6.55 Mpc, SAd) \\
    \includegraphics[width=5.8cm]{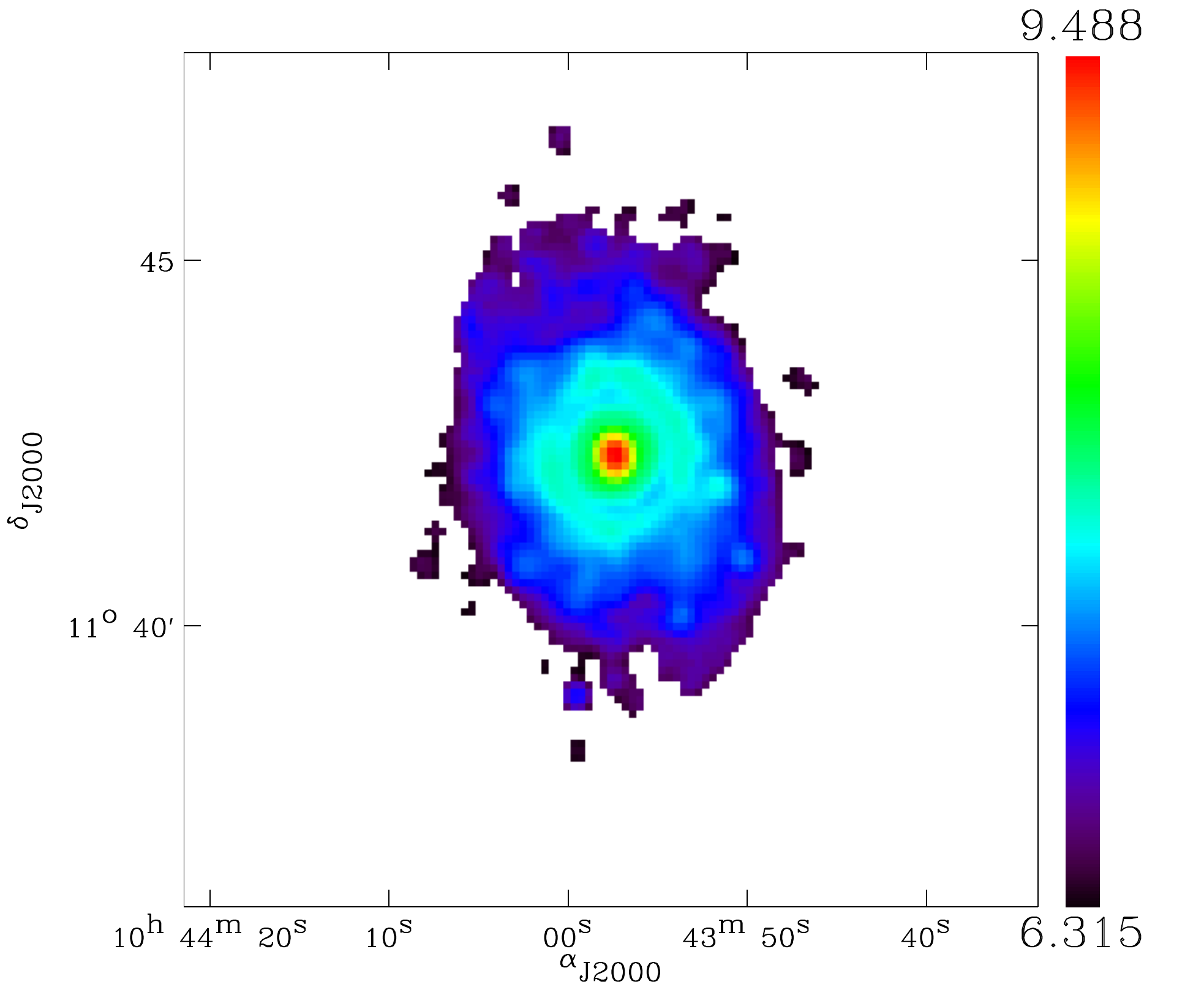} &   
    \includegraphics[width=5.8cm]{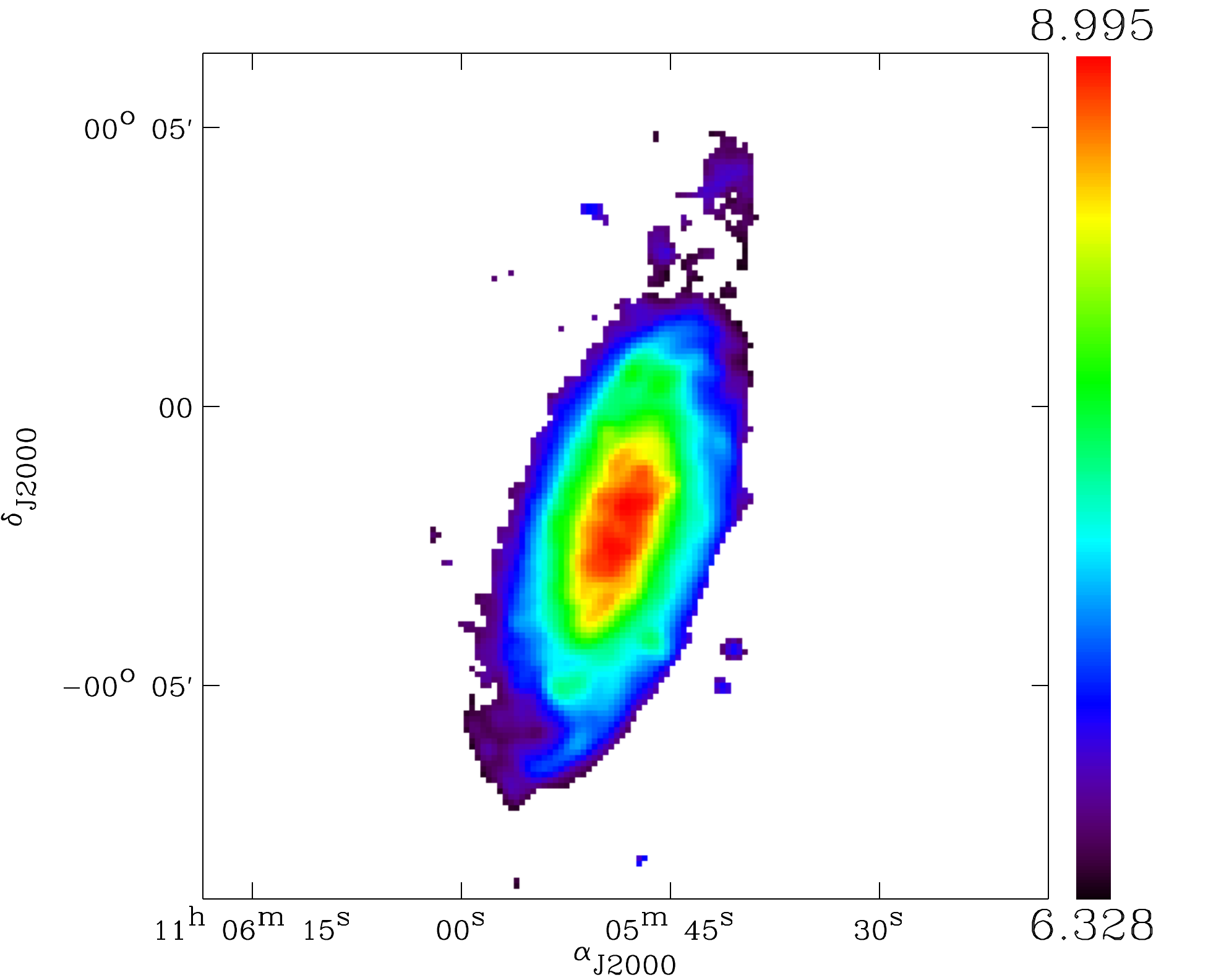} &
    \includegraphics[width=5.8cm]{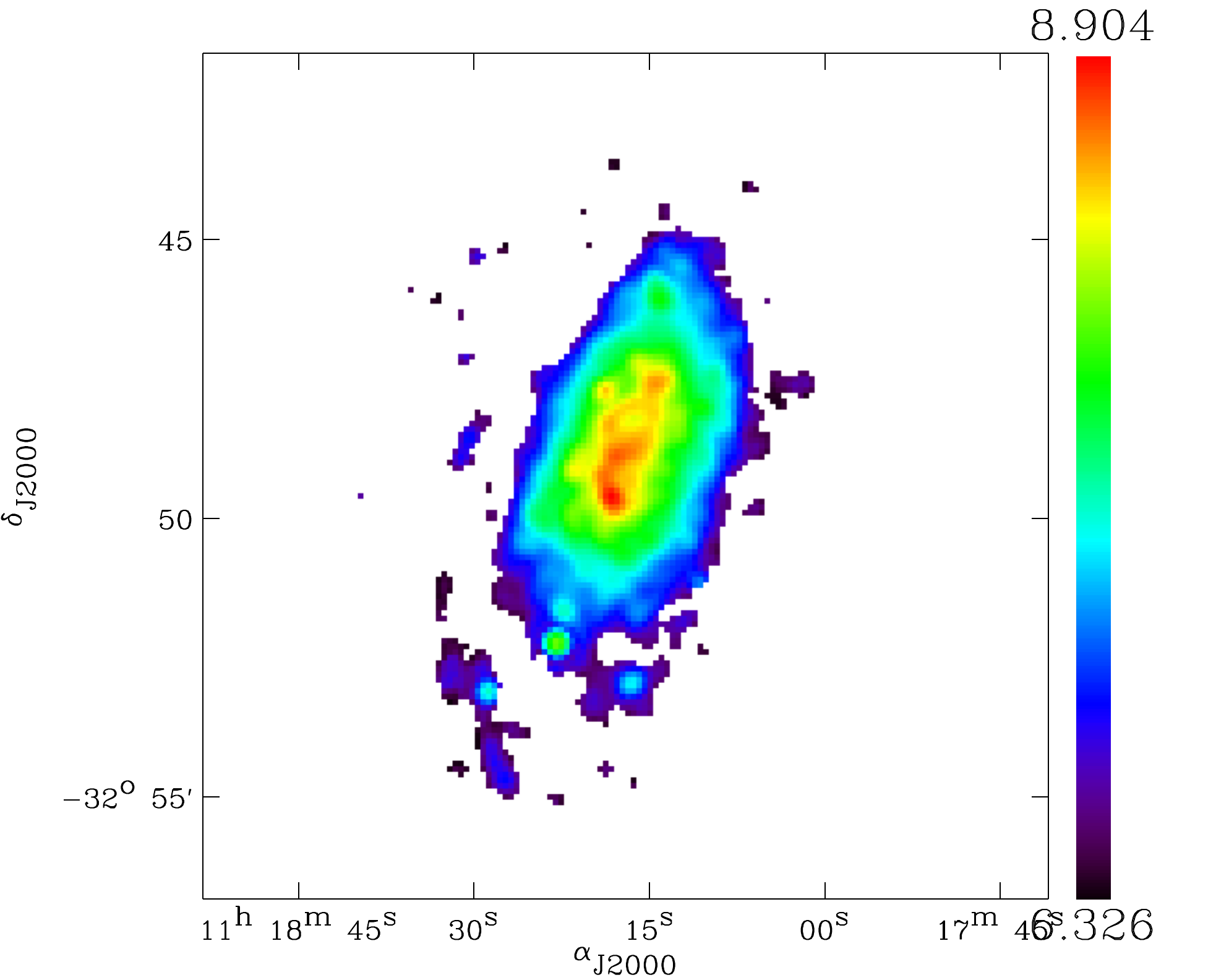} \\
        &&\\
    {\large NGC 3627}~(9.38 Mpc, SABb) &
    {\large NGC 3773}~(12.4 Mpc, SA0) &
    {\large NGC 3938}~(17.9 Mpc, SAc) \\
    \includegraphics[width=5.8cm]{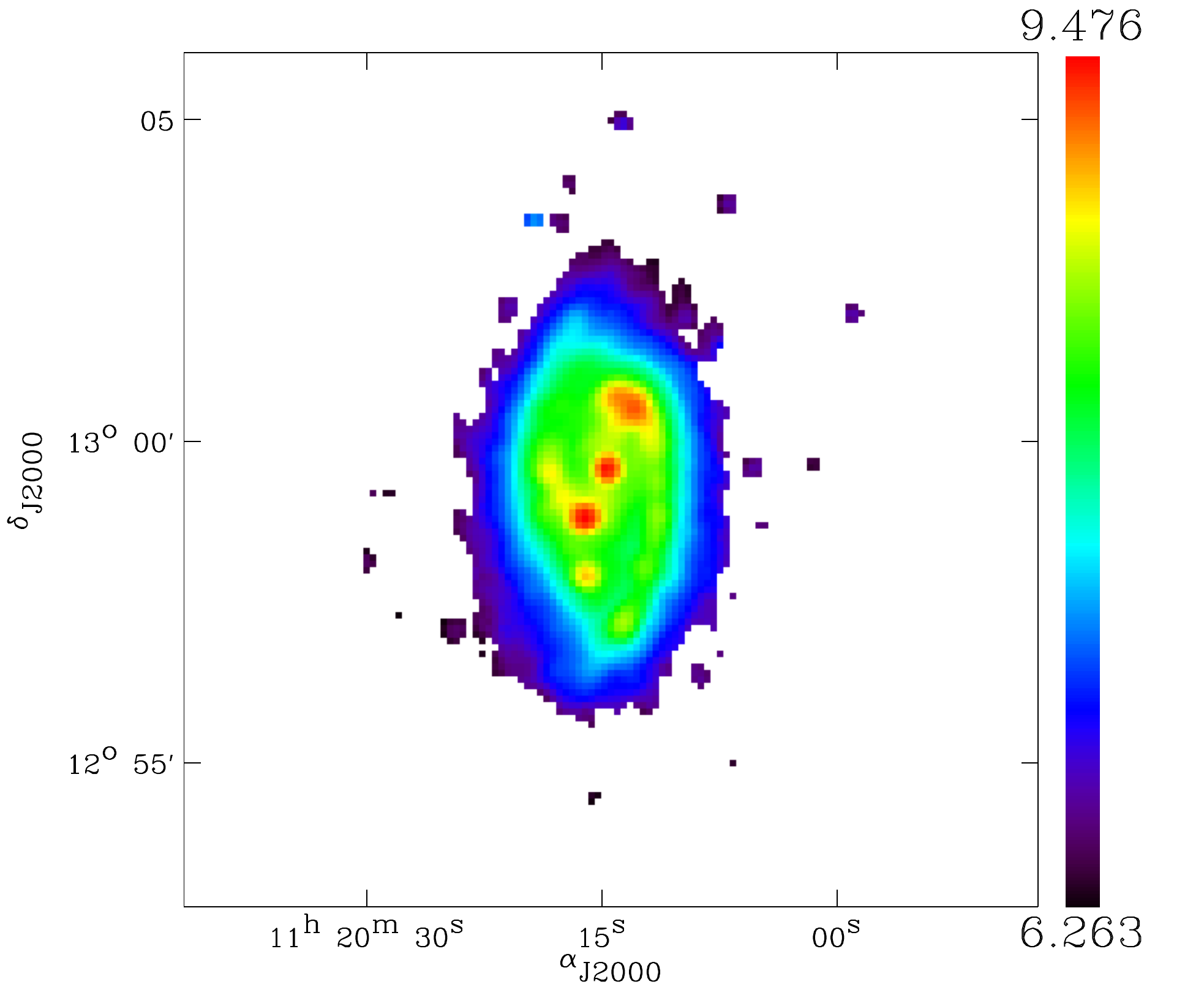} &
    \includegraphics[width=5.8cm]{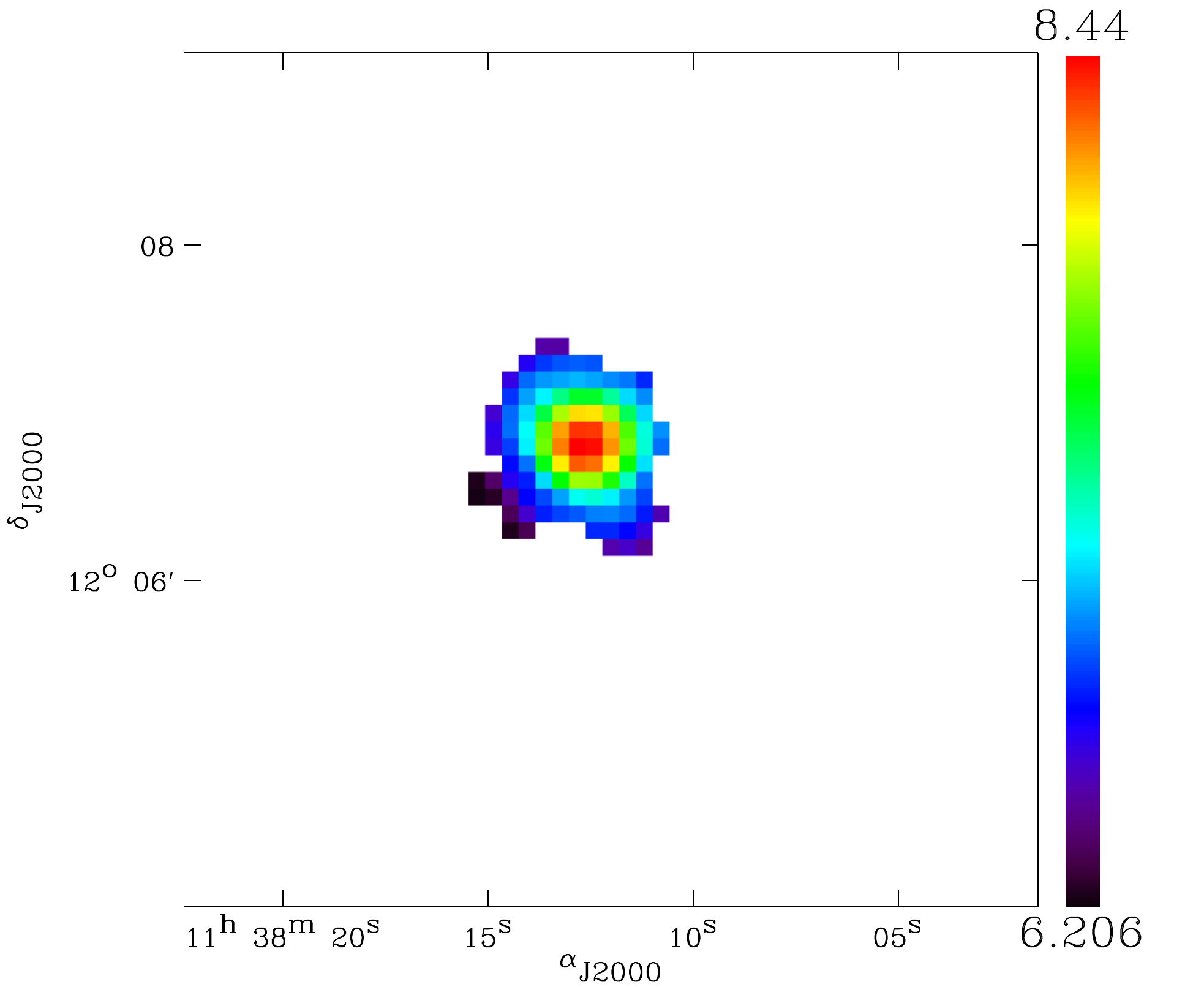} &
    \includegraphics[width=5.8cm]{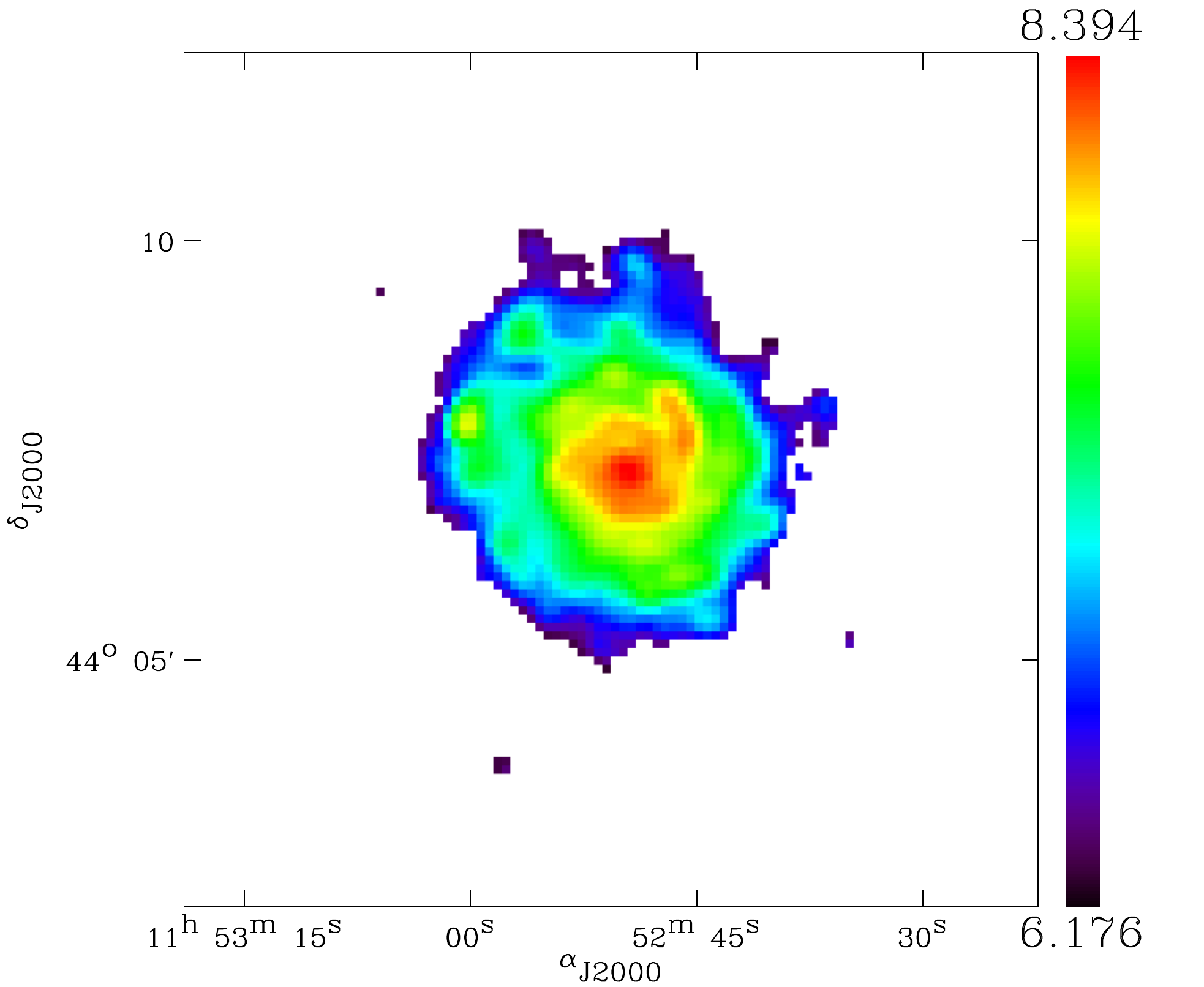} \\
    &&\\    
     {\large NGC 4236}~(4.45 Mpc, SBdm) &
    {\large NGC 4254}~(14.4 Mpc, SAc) &
    {\large NGC 4321}~(14.3 Mpc, SABbc) \\
     \includegraphics[width=5.8cm]{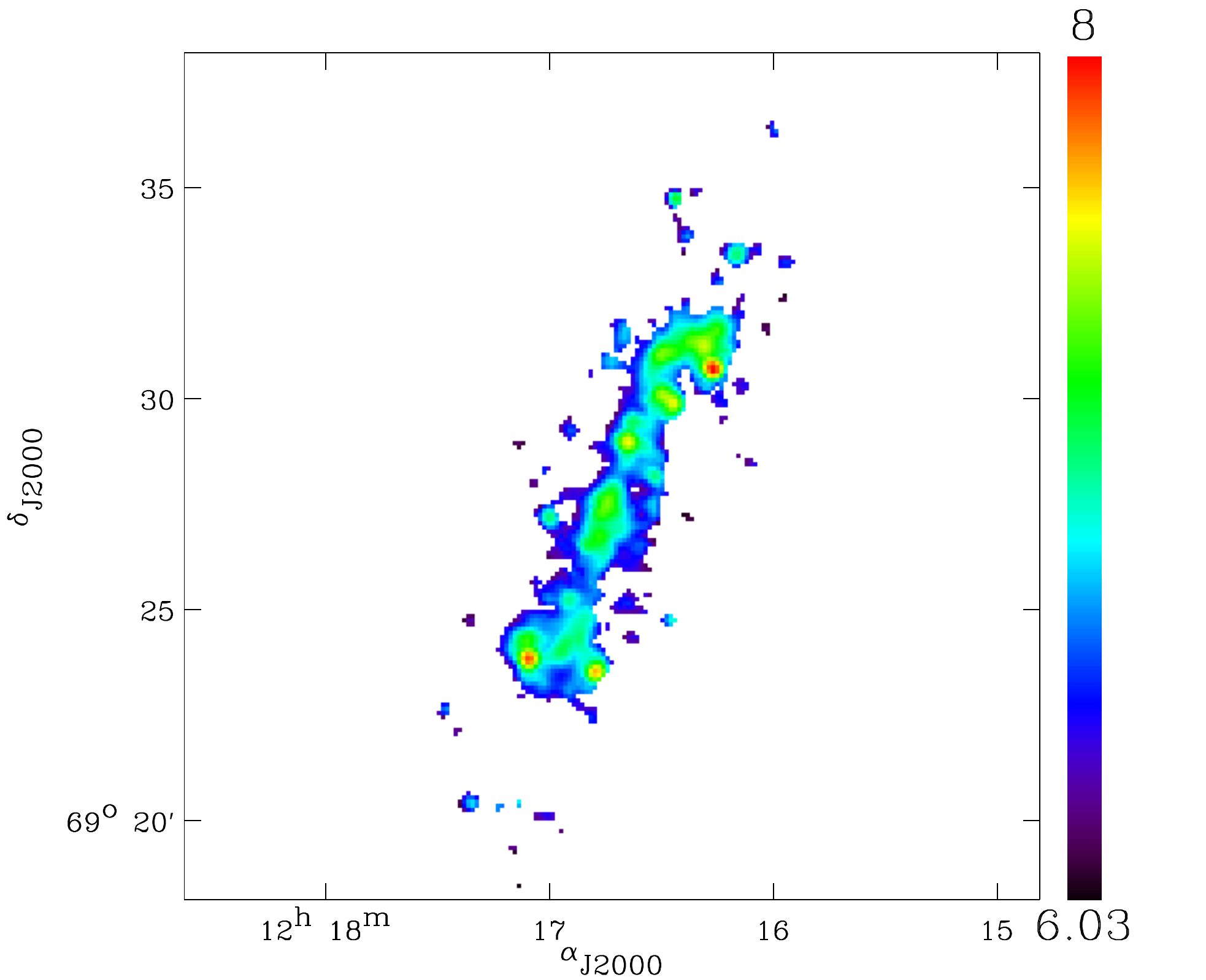} &
    \includegraphics[width=5.8cm]{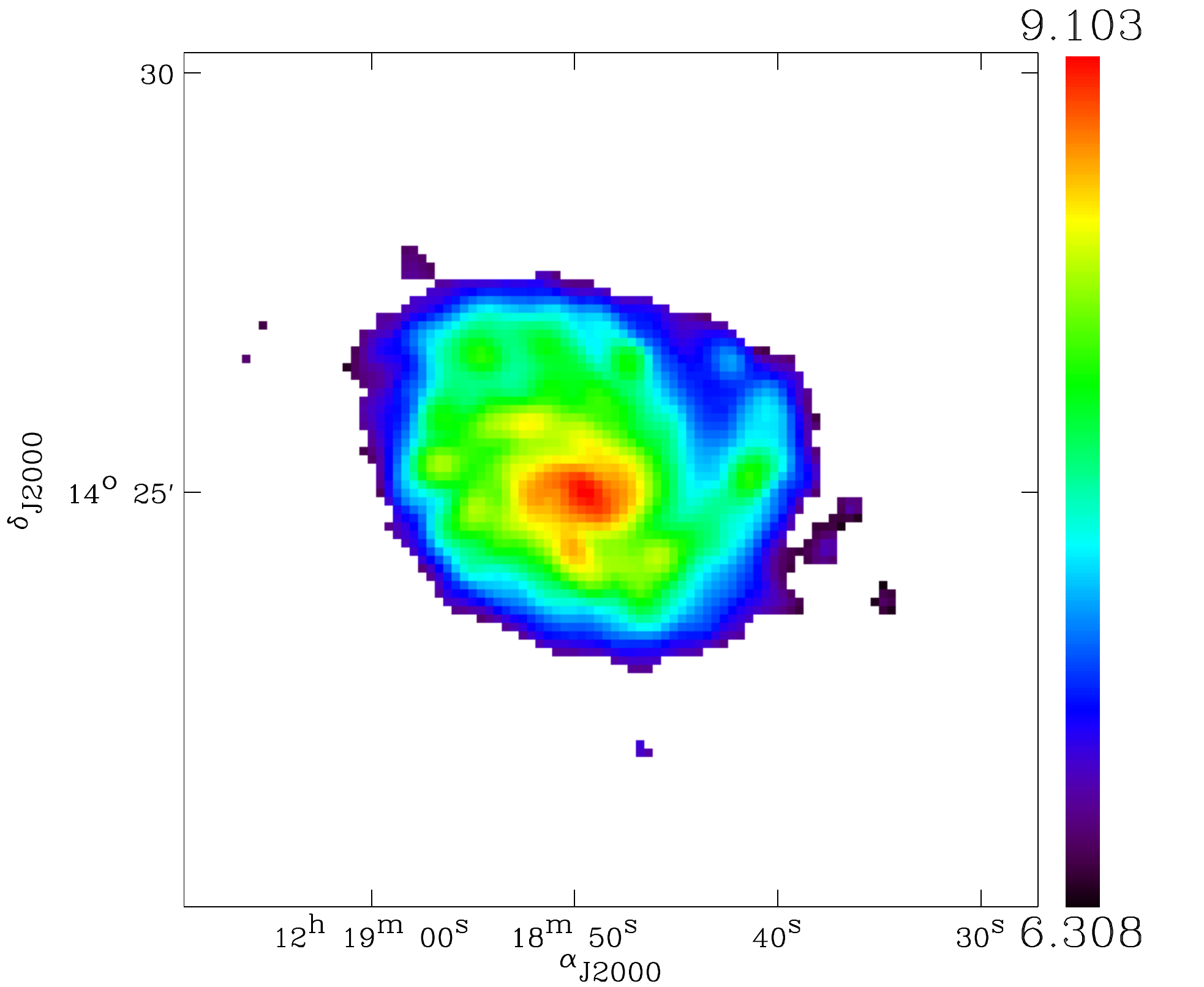} &
    \includegraphics[width=5.8cm]{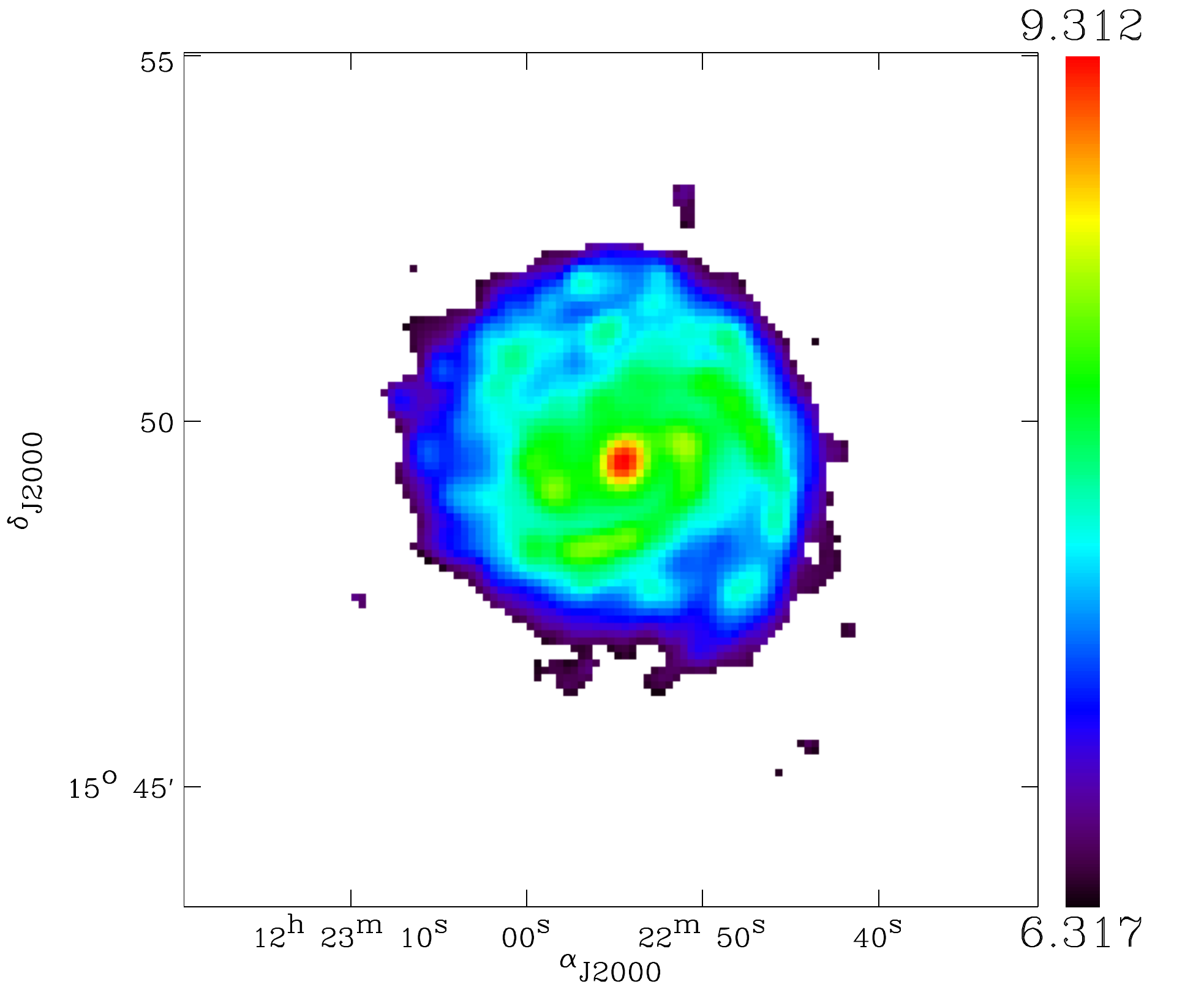} \\
    	\end{tabular}
		\caption{continued}
\end{figure*}

\newpage
\addtocounter {figure}{-1}

\begin{figure*}
    \centering
    \begin{tabular}{ p{5.8cm}p{5.8cm}p{5.8cm} }    
    {\large NGC 4536}~(14.5 Mpc, SABbc) &
    {\large NGC 4559}~(6.98 Mpc, SABcd) &
        {\large NGC 4569}~(9.86 Mpc, SABab) \\
    \includegraphics[width=5.8cm]{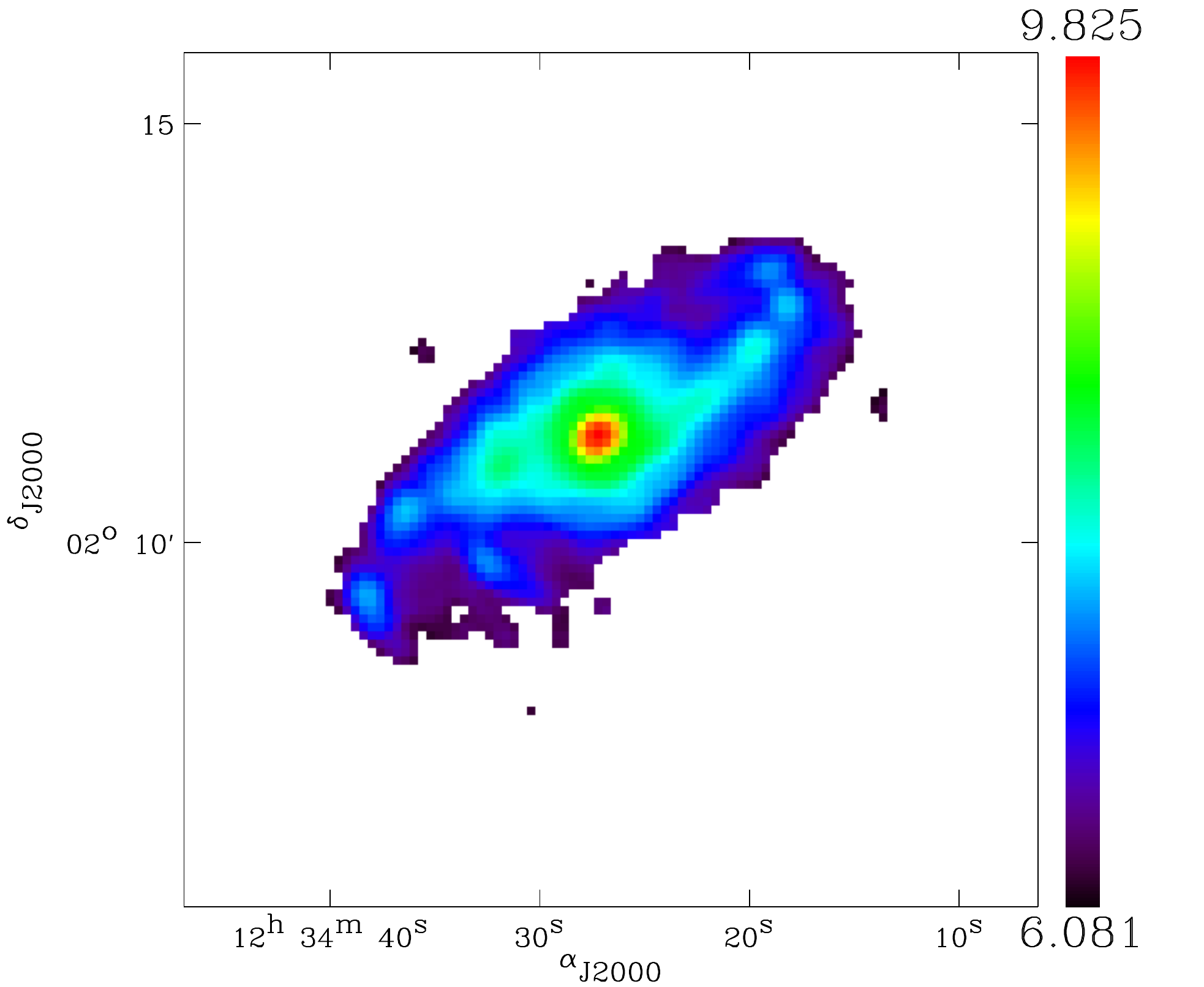} &    
    \includegraphics[width=5.8cm]{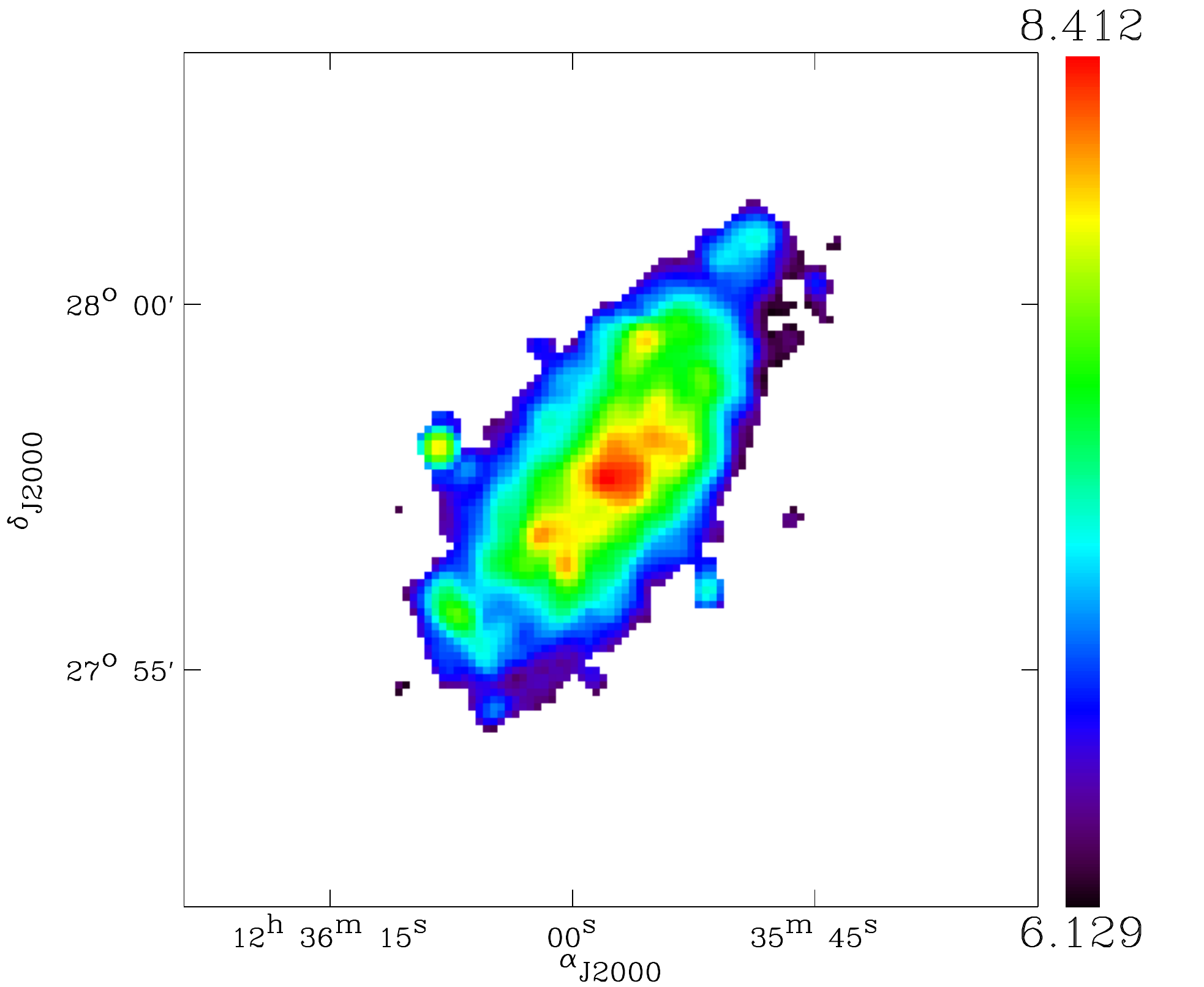} &
    \includegraphics[width=5.8cm]{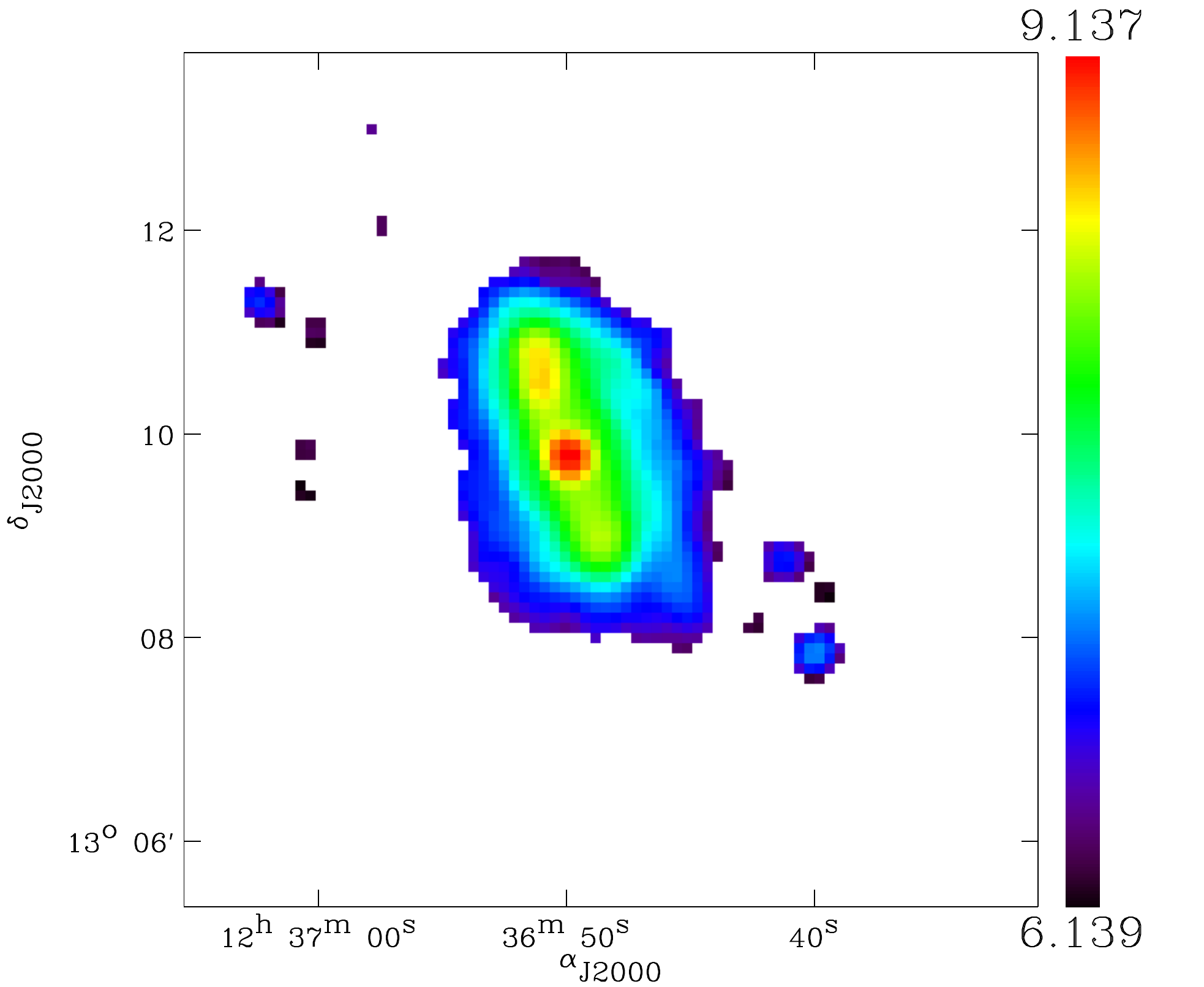} \\
    &&\\    
            {\large NGC 4579}~(16.4 Mpc, SABb) &
    {\large NGC 4594}~(9.08 Mpc, SAa) &
    {\large NGC 4625}~(9.3 Mpc, SABmp) \\    
   \includegraphics[width=5.8cm]{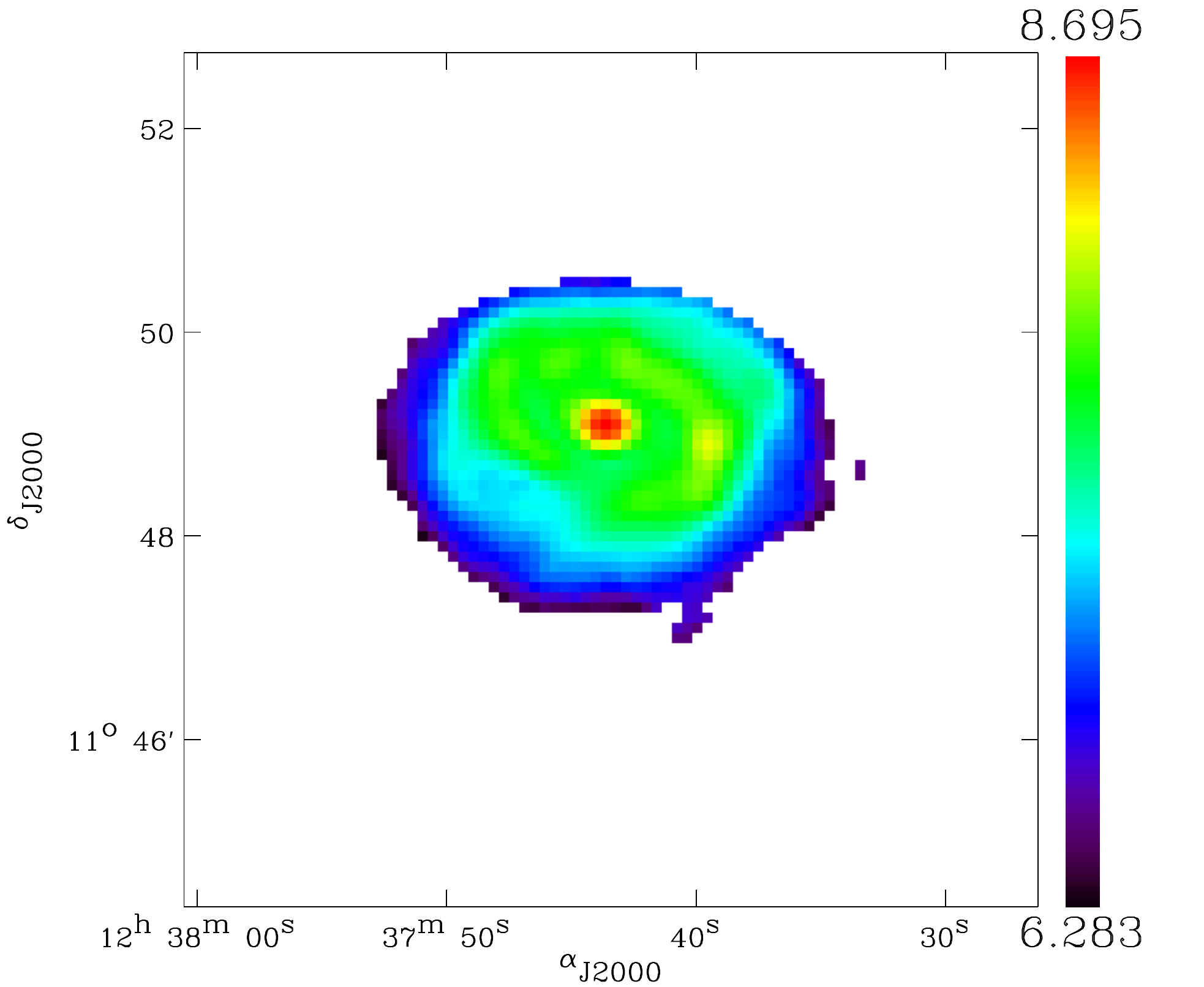} &  
    \includegraphics[width=5.8cm]{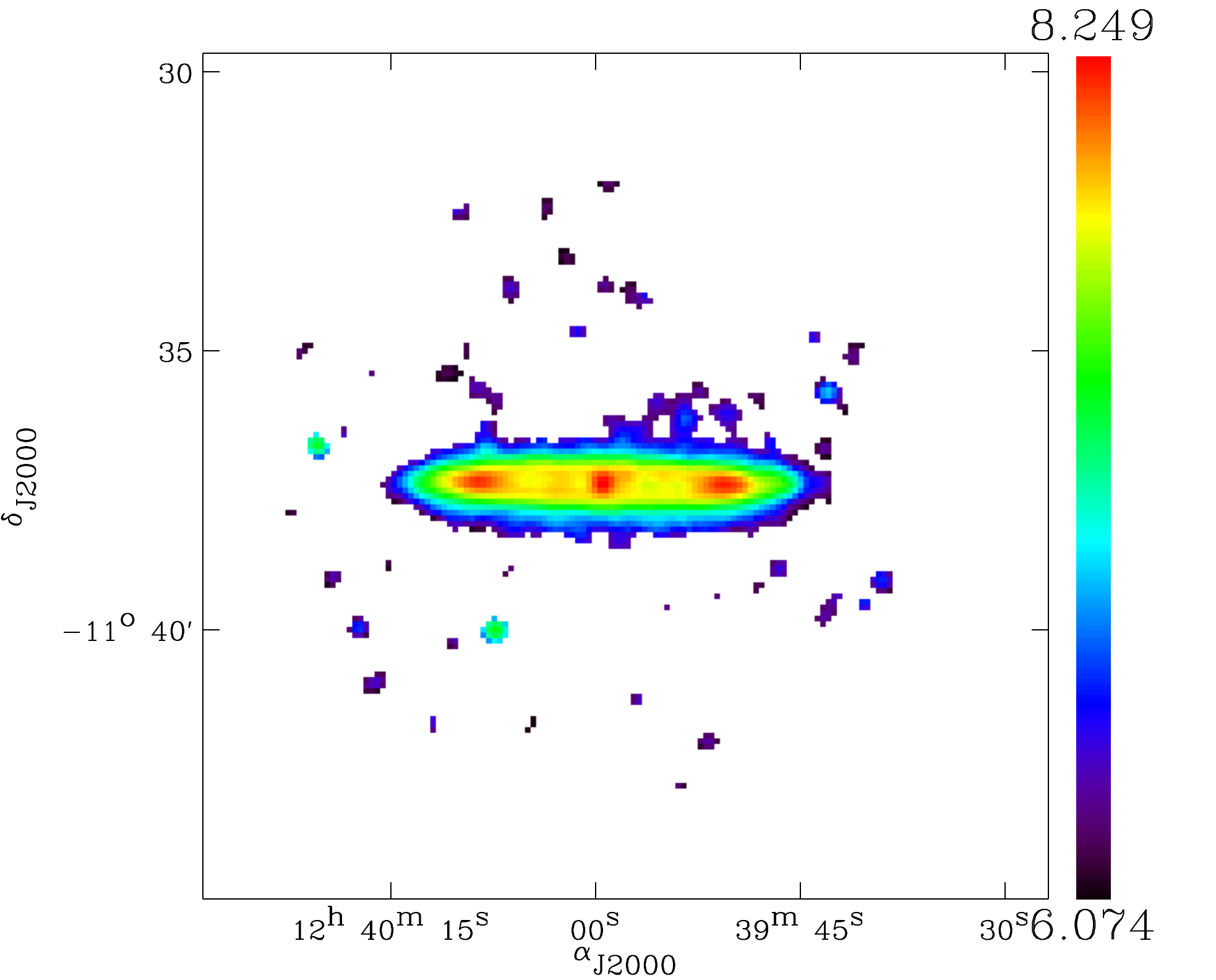} &
    \includegraphics[width=5.8cm]{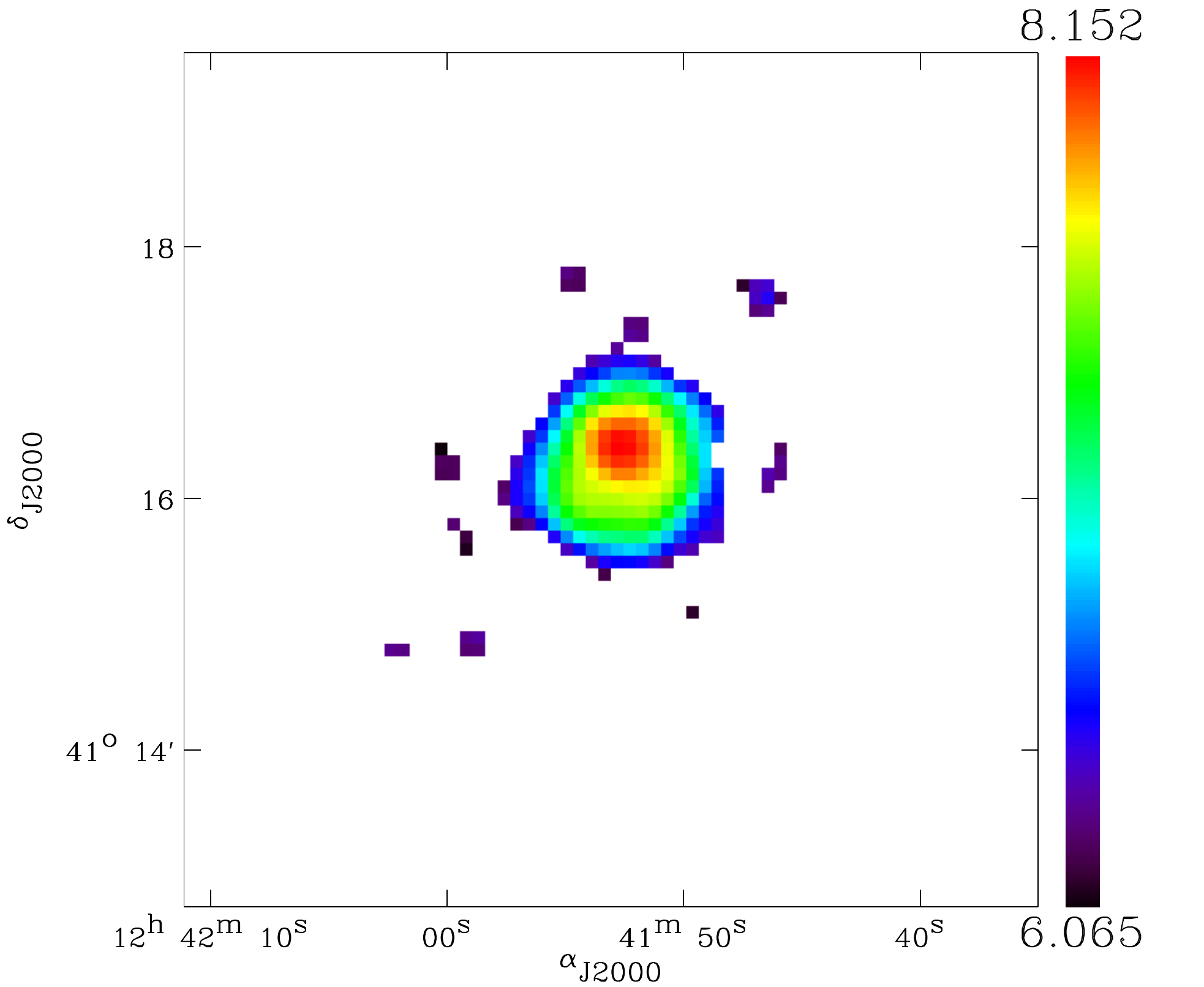} \\
    &&\\    
    {\large NGC 4631}~(7.62 Mpc, SBd) &   
    {\large NGC 4725}~(11.9 Mpc, SABab) &
    {\large NGC 4736}~(4.66 Mpc, SAab) \\ 
    \includegraphics[width=5.8cm]{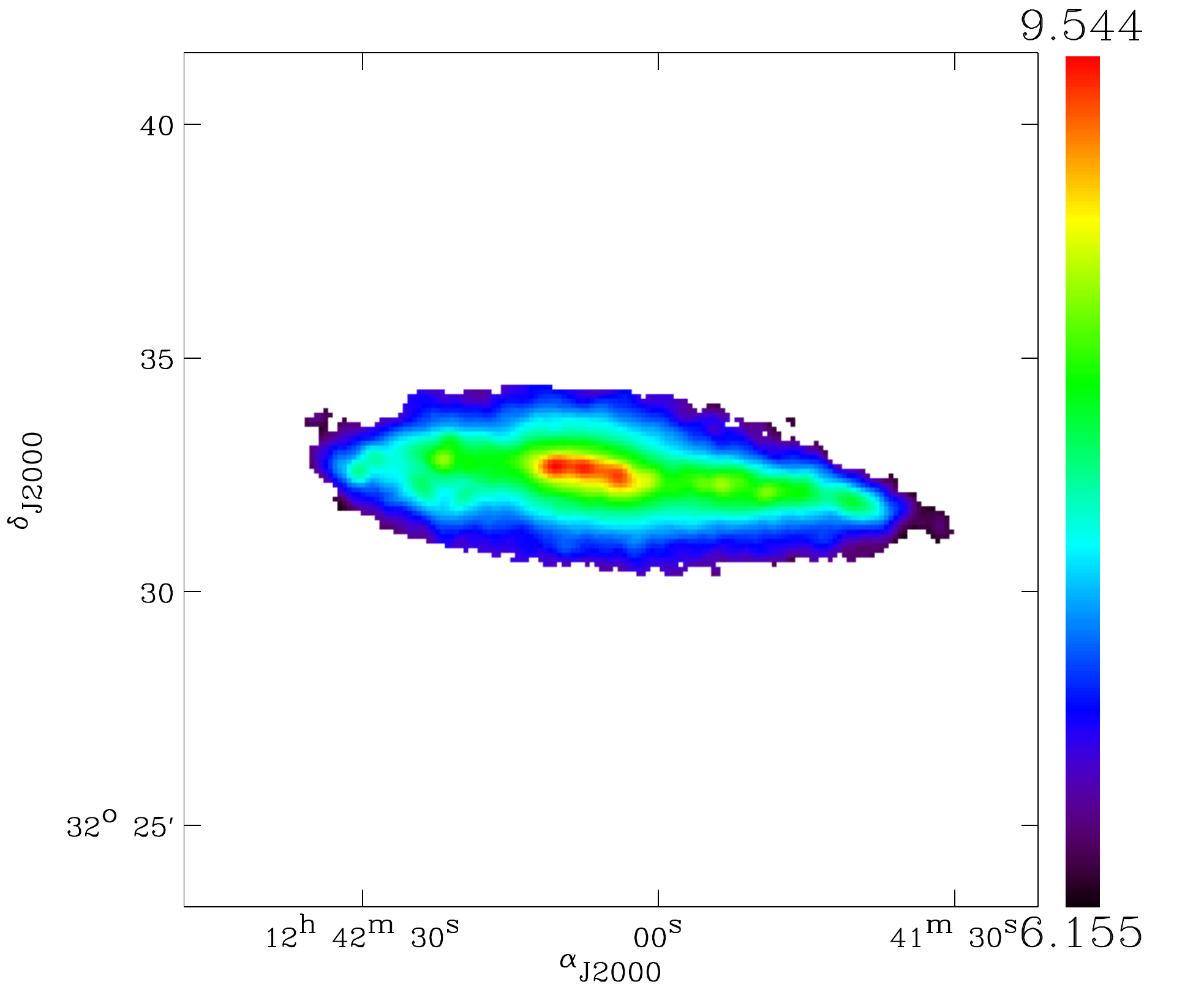} &
    \includegraphics[width=5.8cm]{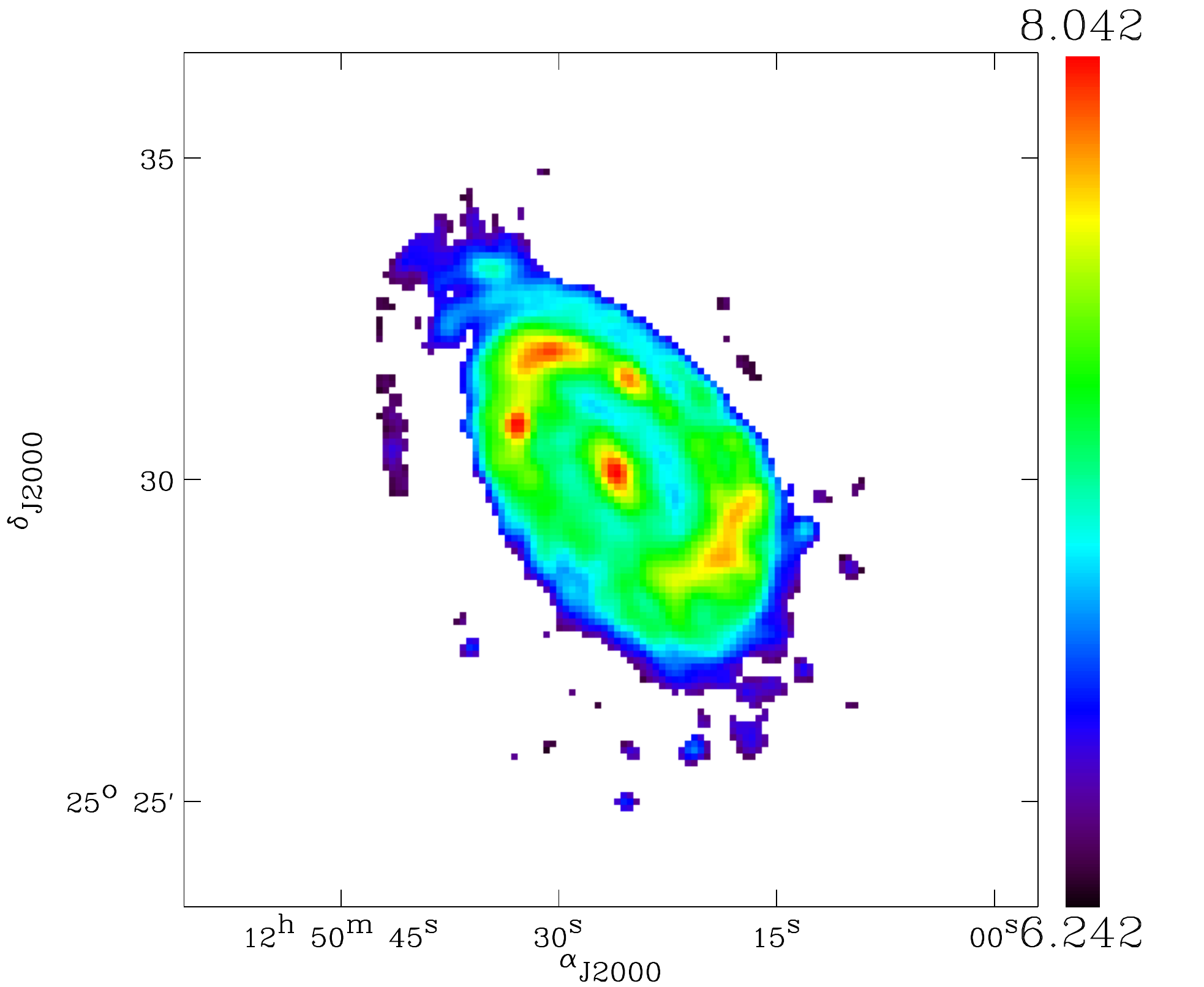} &
    \includegraphics[width=5.8cm]{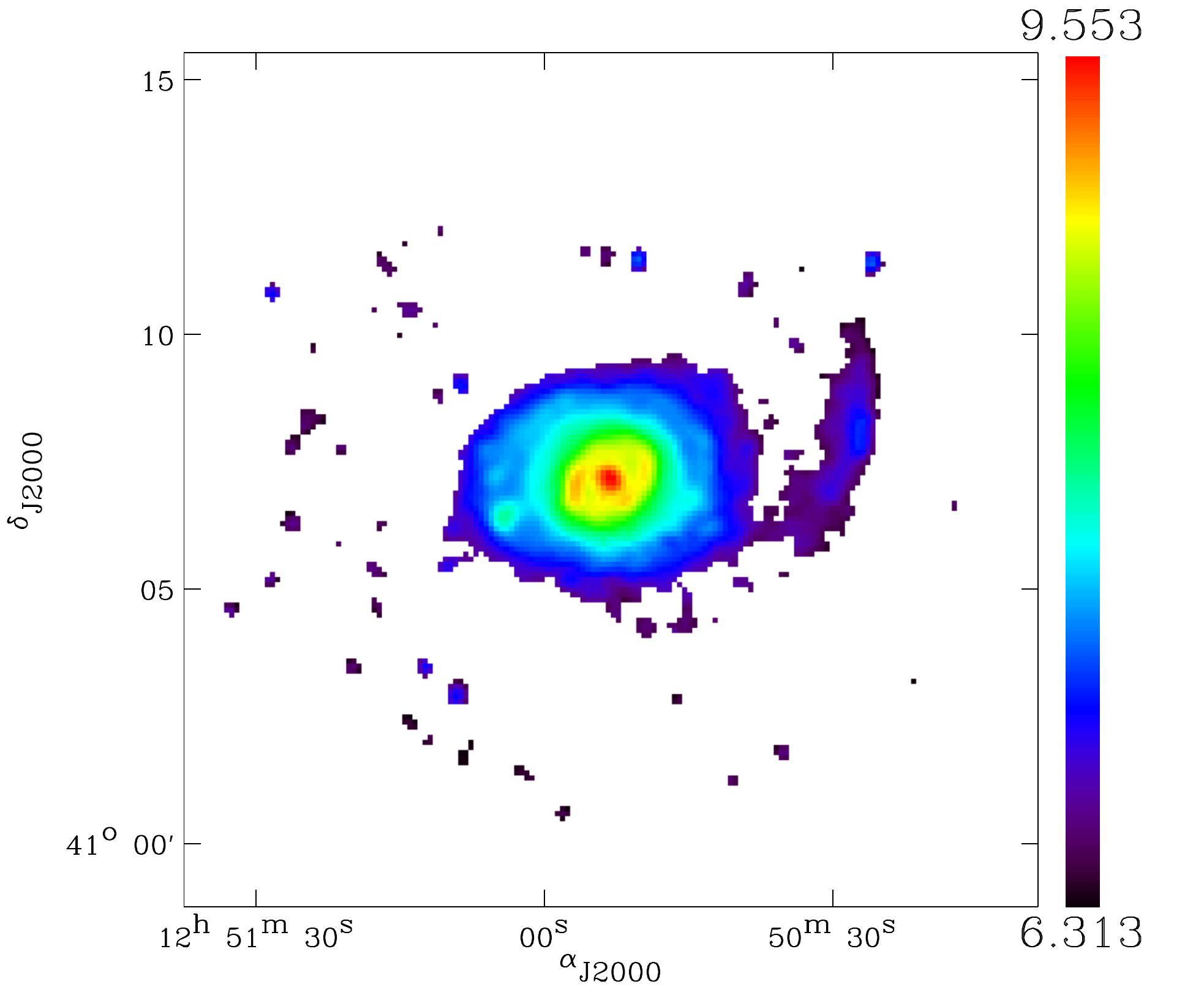} \\
    &&\\    
    {\large NGC 4826}~(5.27 Mpc, SAab)&    
    {\large NGC 5055}~(7.94 Mpc, SAbc) &
    {\large NGC 5398}~(7.66 Mpc, SBdm) \\
    \includegraphics[width=5.8cm]{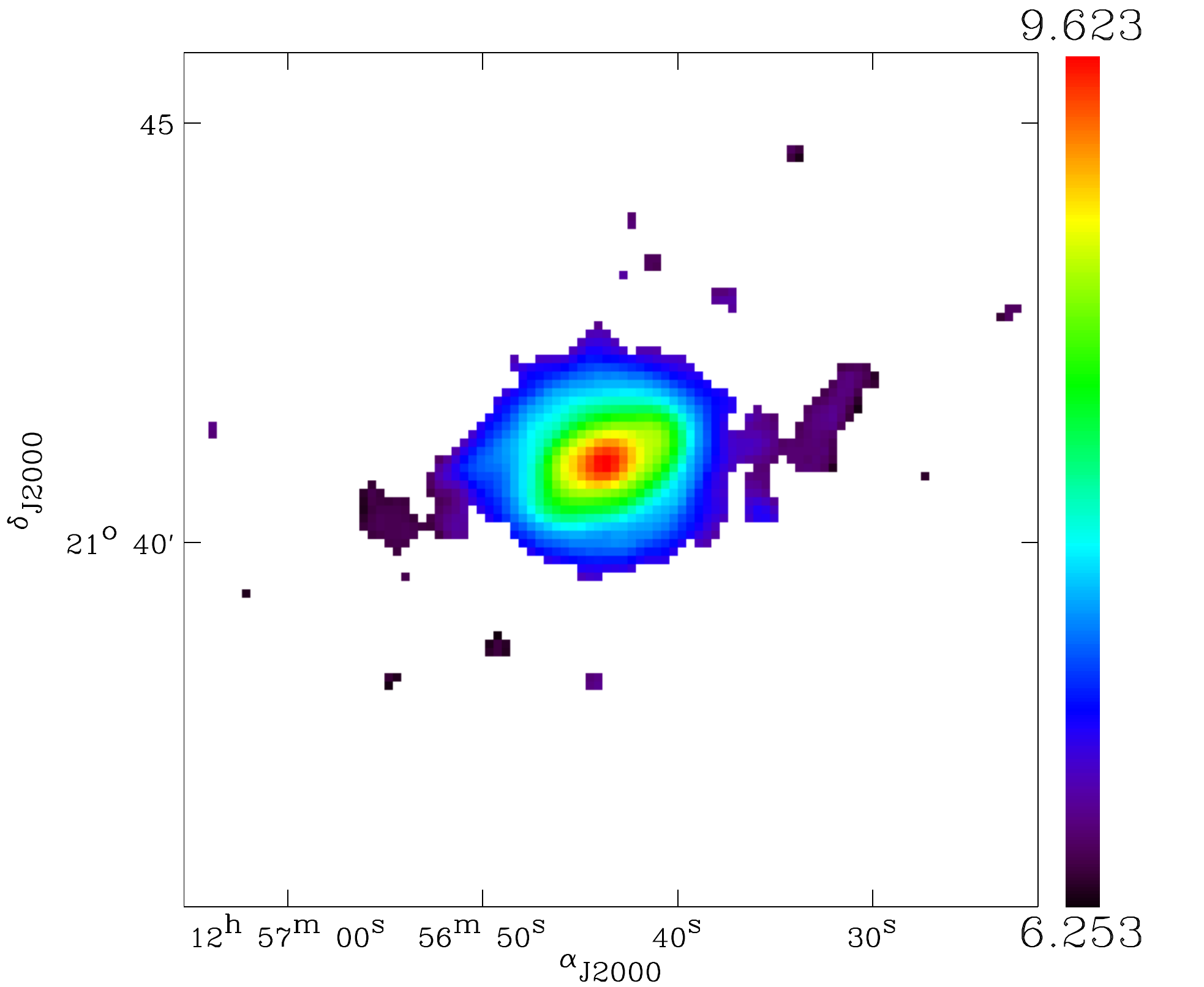} &
    \includegraphics[width=5.8cm]{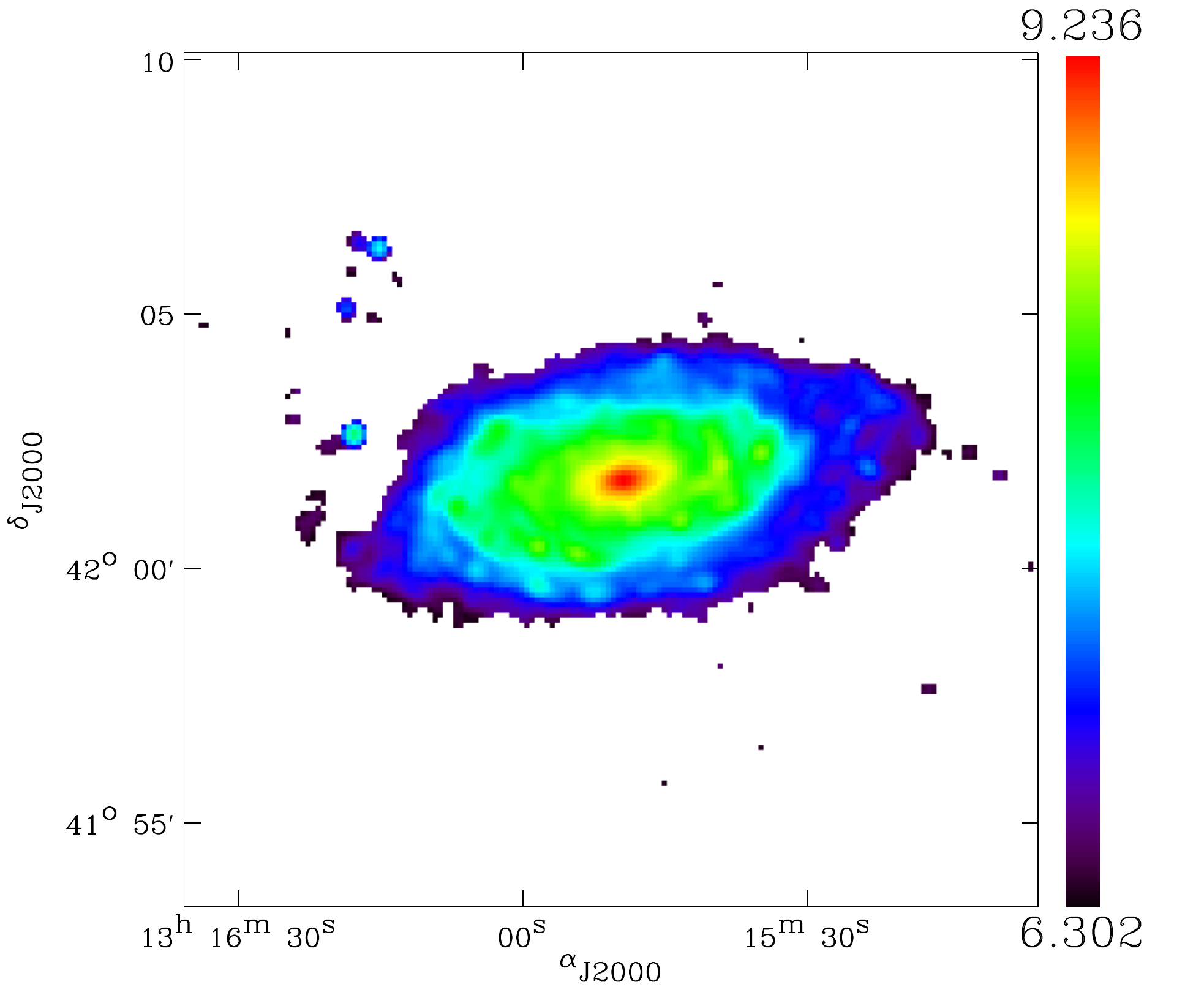} &
    \includegraphics[width=5.8cm]{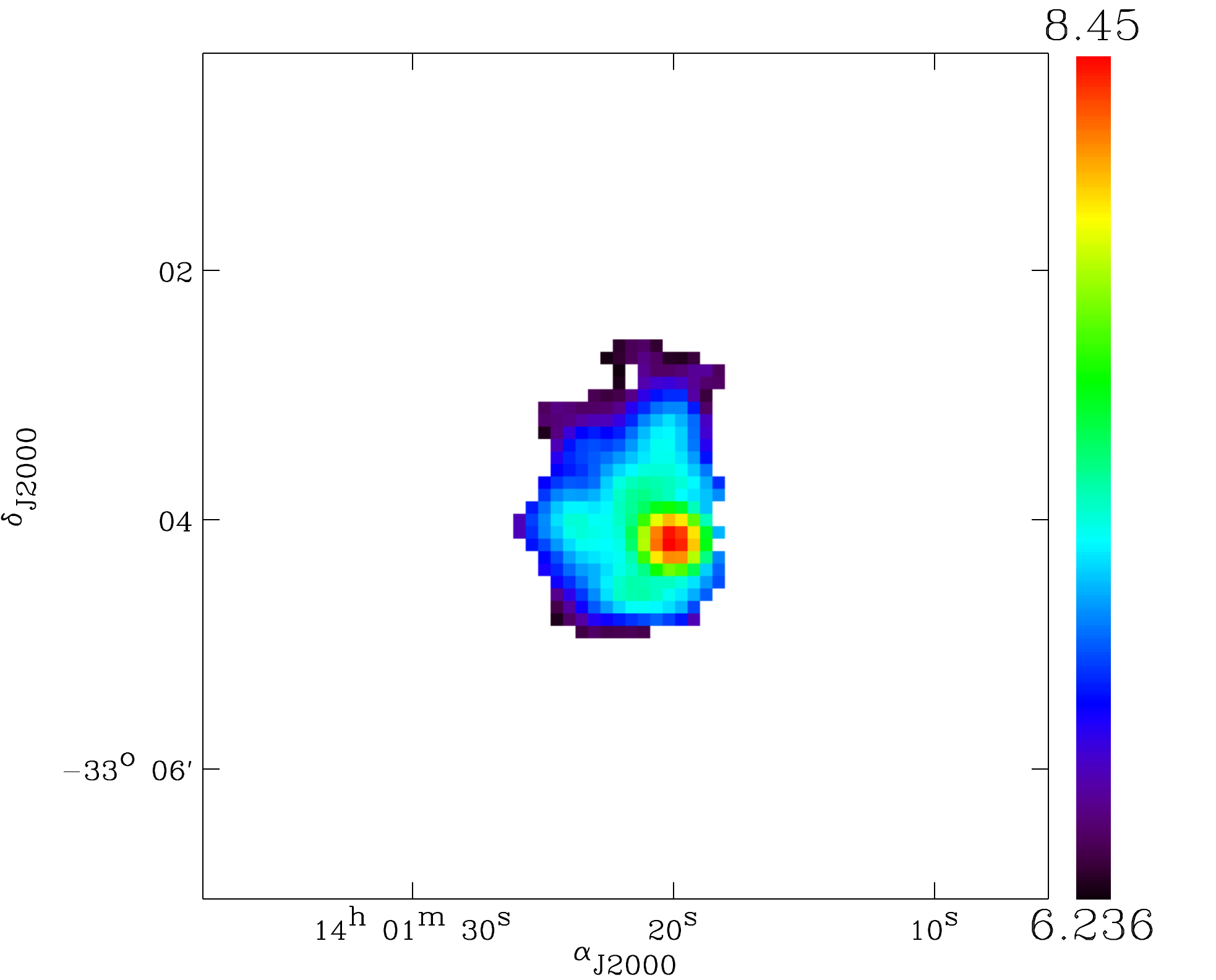} \\
    	\end{tabular}
		\caption{continued}
\end{figure*}

\newpage
\addtocounter {figure}{-1}

\begin{figure*}
    \centering
    \begin{tabular}{ p{5.8cm}p{5.8cm}p{5.8cm} }   
    {\large NGC 5408}~(4.8 Mpc, IBm) &    
    {\large NGC 5457}~(6.7 Mpc, SABcd) &
    {\large NGC 5474}~(6.8 Mpc, SAcd) \\
    \includegraphics[width=5.8cm]{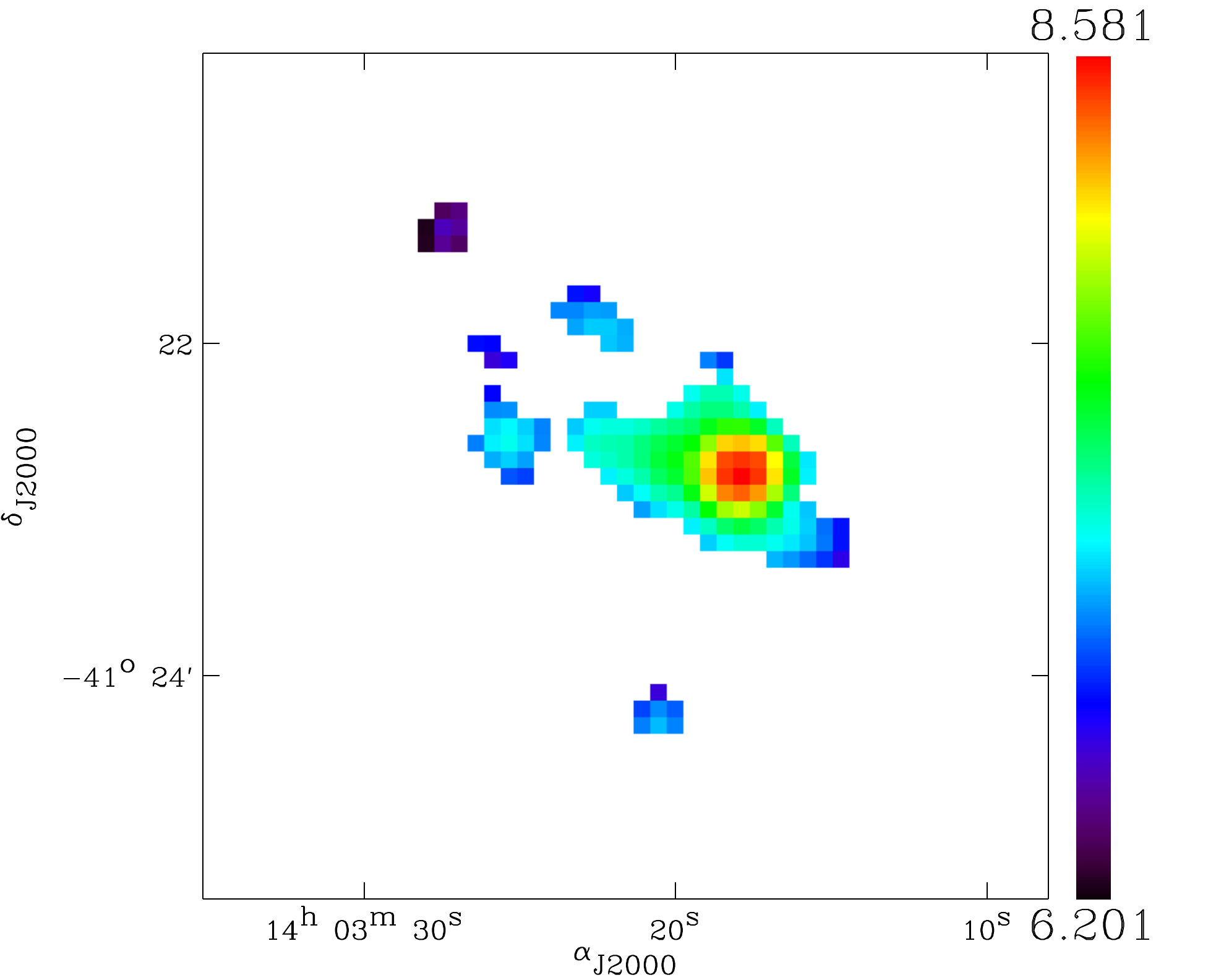} &      
    \includegraphics[width=5.8cm]{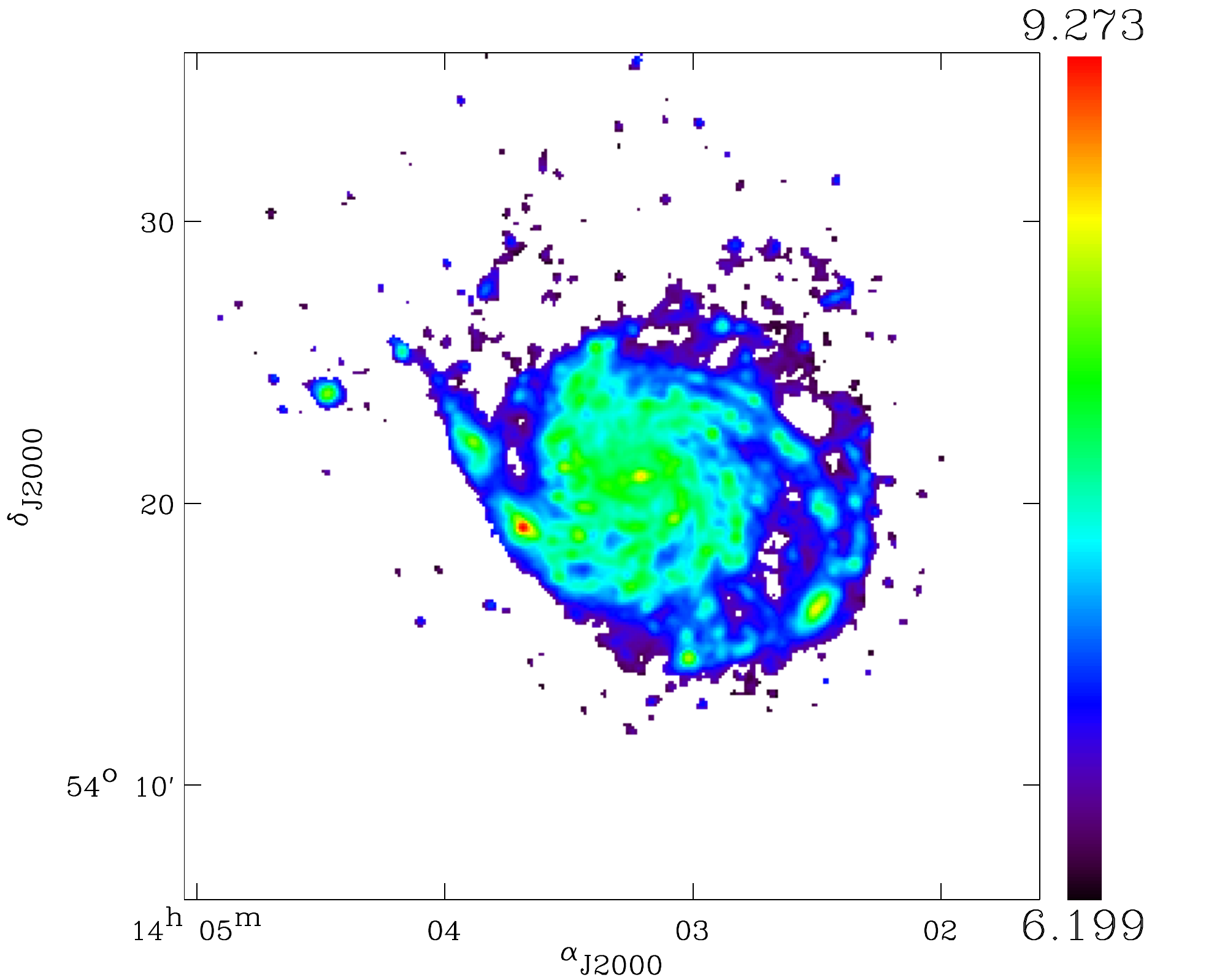} &
    \includegraphics[width=5.8cm]{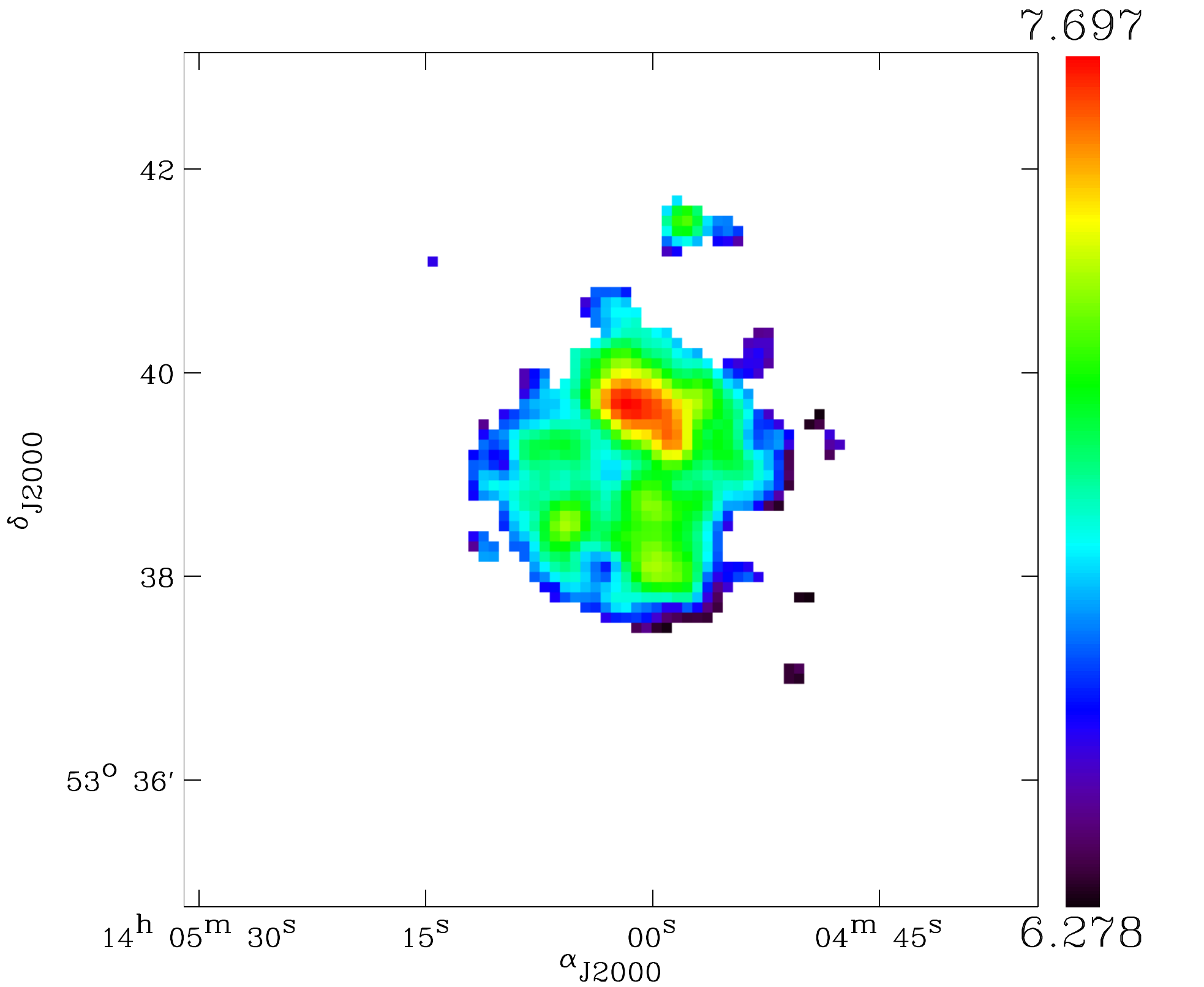} \\
    &&\\    
    {\large NGC 5713}~(21.4 Mpc, SABbcp) &      
   {\large NGC 5866}~(15.3 Mpc, S0) &
    {\large NGC 6946}~(6.8 Mpc, SABcd) \\
    \includegraphics[width=5.8cm]{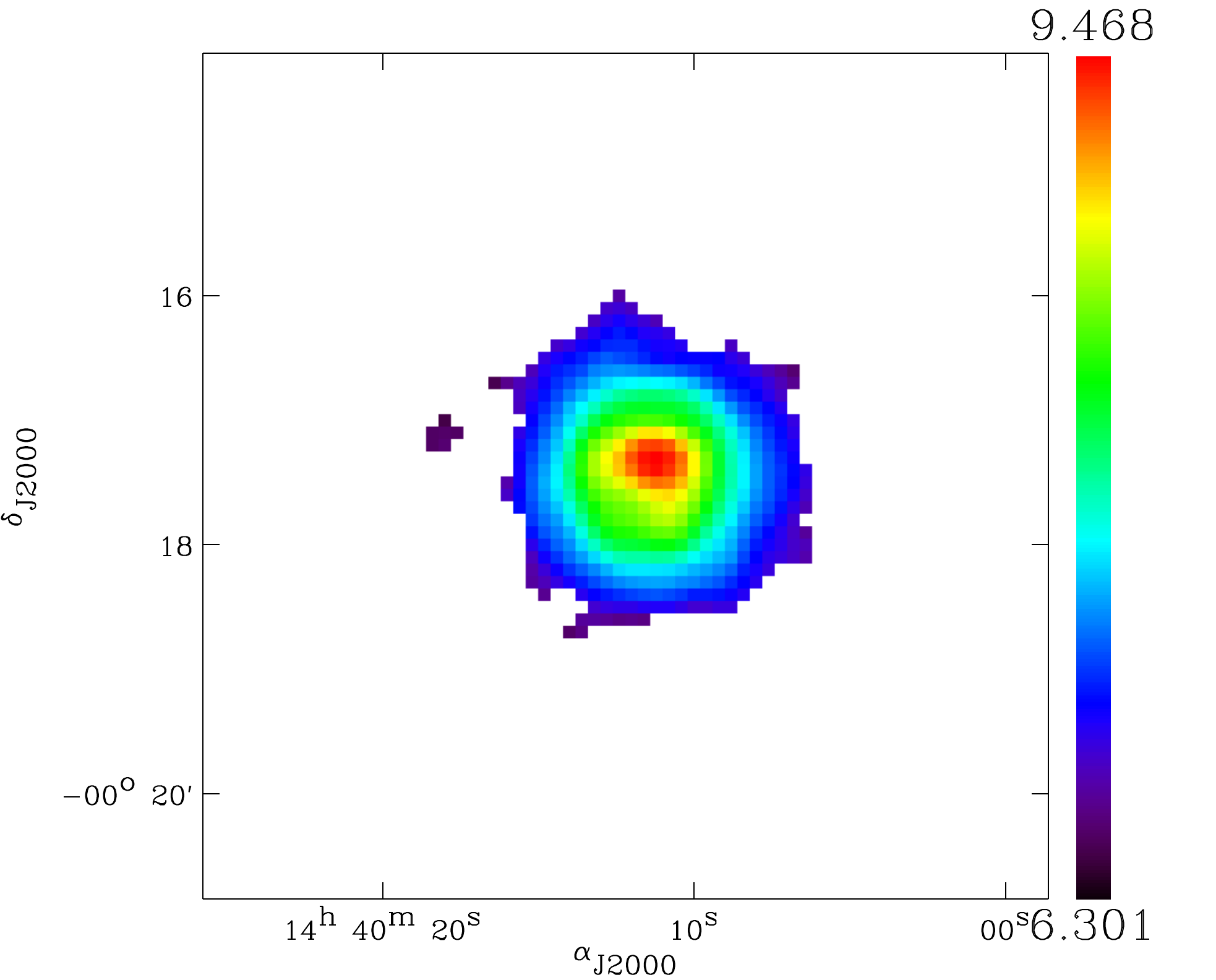} &
   \includegraphics[width=5.8cm]{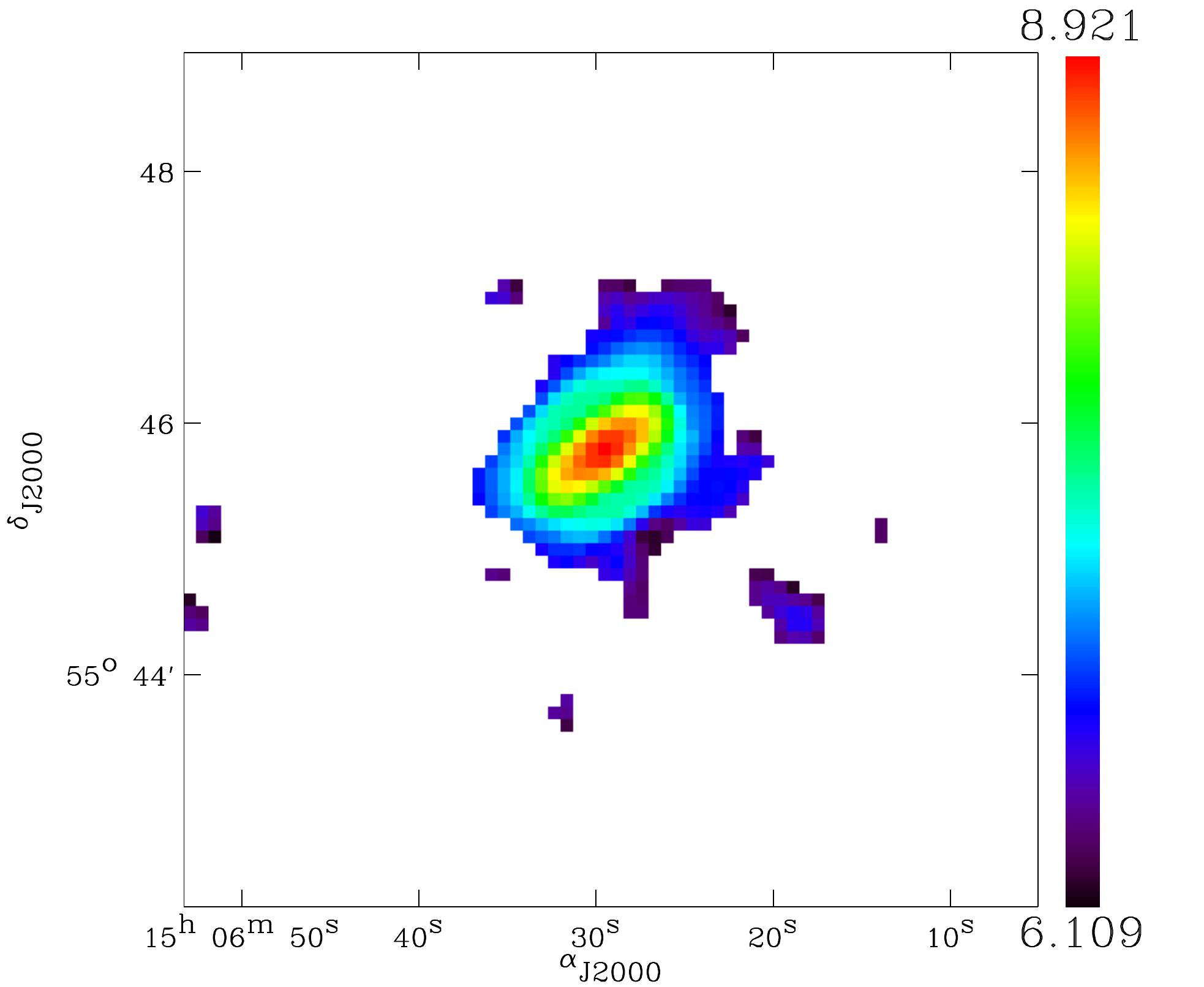} &
   \includegraphics[width=5.8cm]{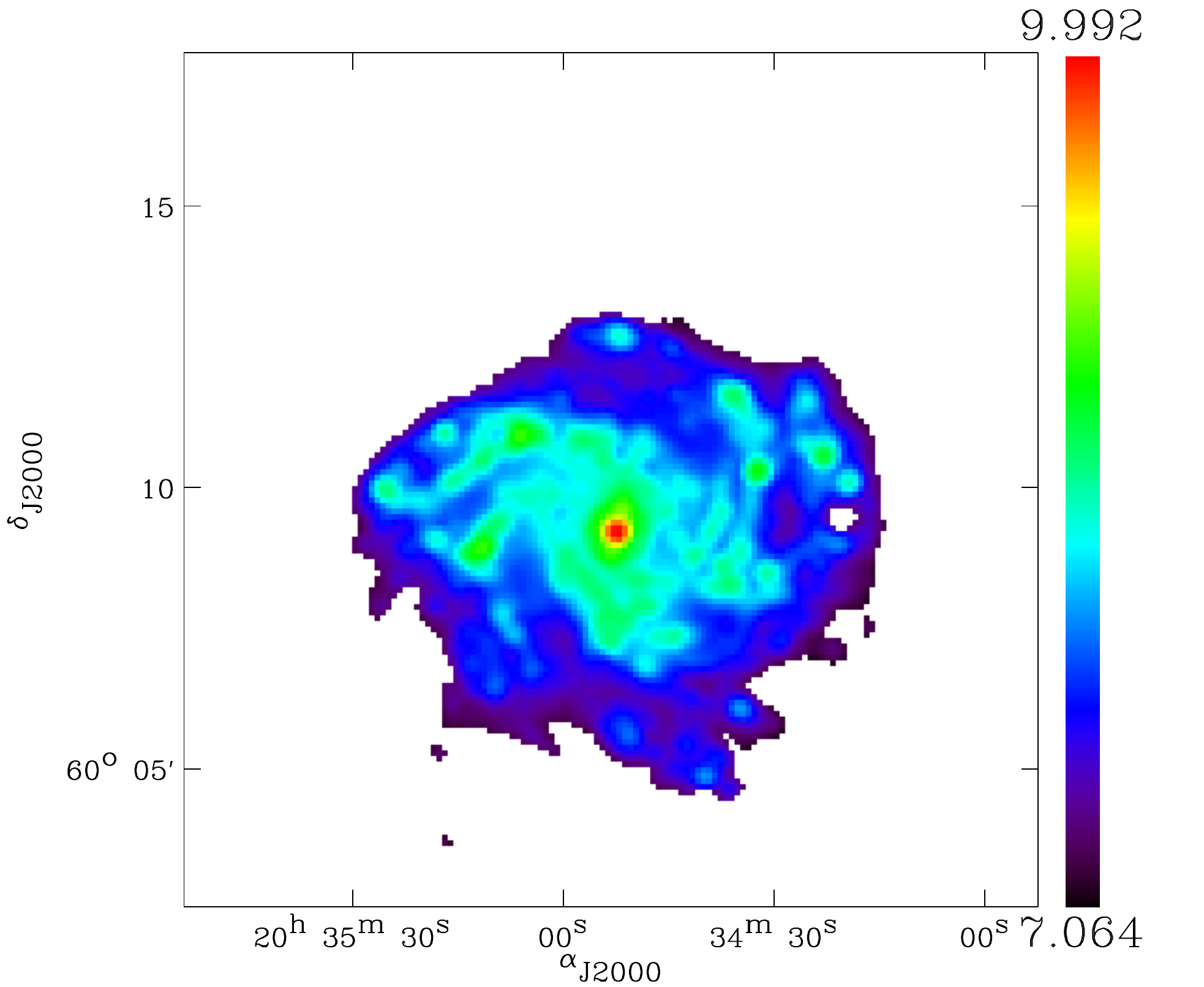} \\
    &&\\    
     {\large NGC 7331}~(14.5 Mpc, SAb) &   
   {\large NGC 7793}~(3.91 Mpc, SAd) &\\
     \includegraphics[width=5.8cm]{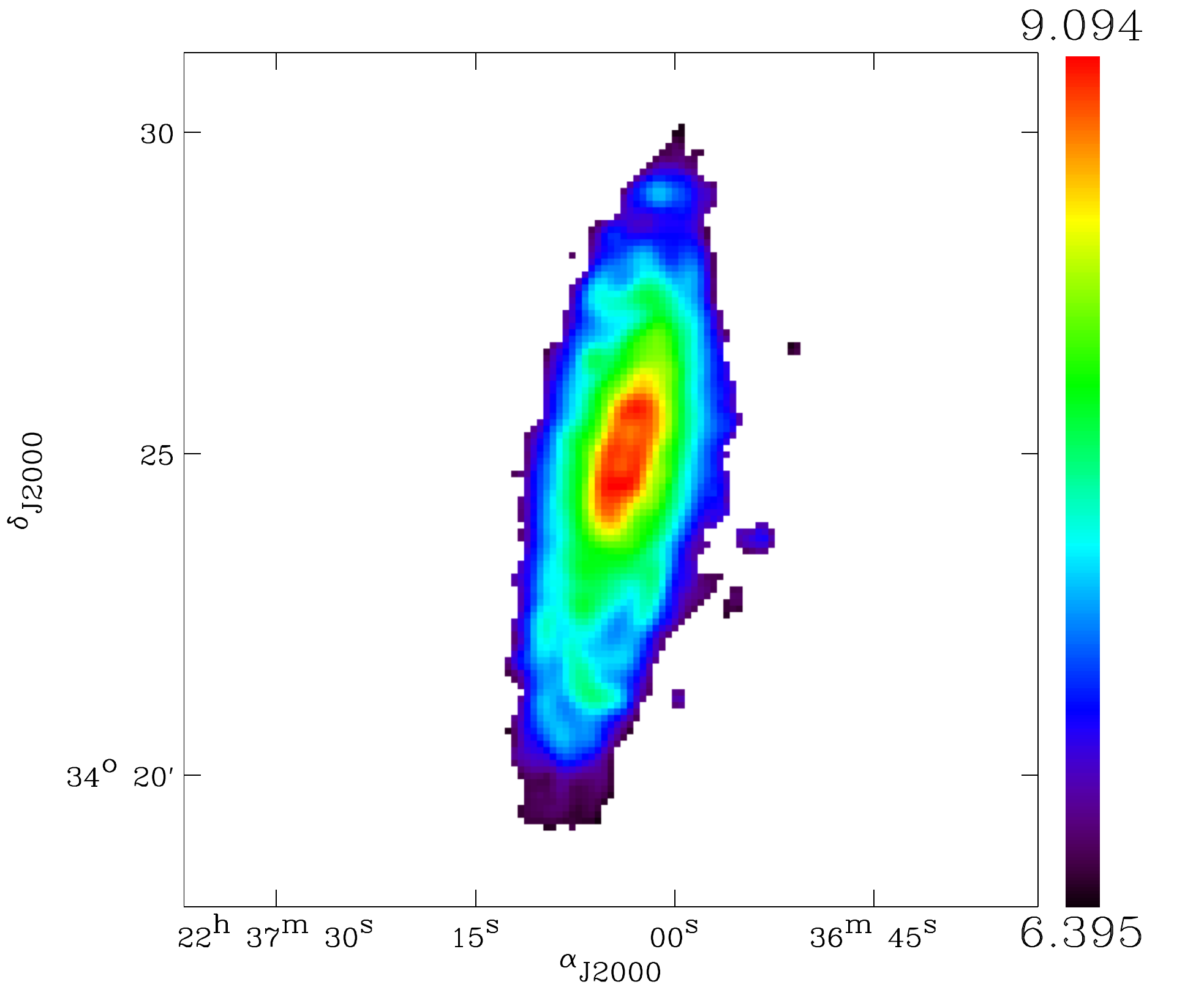} &       
      \includegraphics[width=5.8cm]{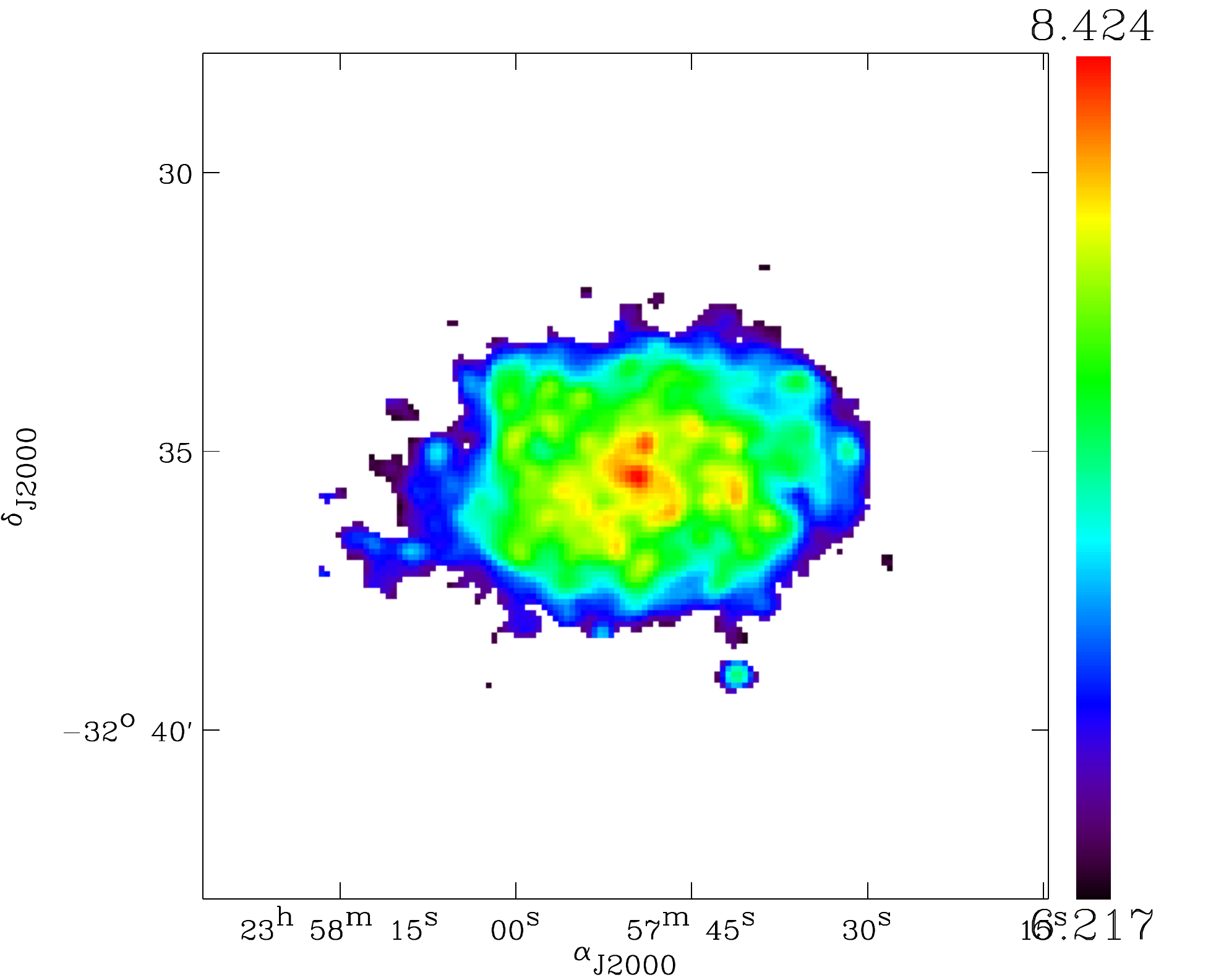} & \\
  	\end{tabular}
		\caption{continued}
 	\label{LTIR_Maps}
 \end{figure*}

\end{document}